\documentstyle[twoside,epsf]{article}

\catcode`\@=11
\long\def\@makefntext#1{
\protect\noindent \hbox to 3.2pt {\hskip-.9pt  
$^{{\eightrm\@thefnmark}}$\hfil}#1\hfill}		

\def\thefootnote{\fnsymbol{footnote}}
\def\@makefnmark{\hbox to 0pt{$^{\@thefnmark}$\hss}}	
	
\def\ps@myheadings{\let\@mkboth\@gobbletwo
\def\@oddhead{\hbox{}
\rightmark\hfil\eightrm\thepage}   
\def\@oddfoot{}\def\@evenhead{\eightrm\thepage\hfil
\leftmark\hbox{}}\def\@evenfoot{}
\def\sectionmark##1{}\def\subsectionmark##1{}}

\oddsidemargin=\evensidemargin
\renewcommand{\thefootnote}{\fnsymbol{footnote}}

\newcounter{sectionc}\newcounter{subsectionc}\newcounter{subsubsectionc}
\renewcommand{\section}[1] {\vspace{12pt}\addtocounter{sectionc}{1} 
\setcounter{subsectionc}{0}\setcounter{subsubsectionc}{0}\noindent 
	{\tenbf\thesectionc. #1}\par\vspace{5pt}}
\renewcommand{\subsection}[1] {\vspace{12pt}\addtocounter{subsectionc}{1} 
	\setcounter{subsubsectionc}{0}\noindent 
	{\bf\thesectionc.\thesubsectionc. {\kern1pt \bfit #1}}\par\vspace{5pt}}
\renewcommand{\subsubsection}[1] {\vspace{12pt}\addtocounter{subsubsectionc}{1}
	\noindent{\tenrm\thesectionc.\thesubsectionc.\thesubsubsectionc.
	{\kern1pt \tenit #1}}\par\vspace{5pt}}
\newcommand{\nonumsection}[1] {\vspace{12pt}\noindent{\tenbf #1}
	\par\vspace{5pt}}

\newcounter{appendixc}
\newcounter{subappendixc}[appendixc]
\newcounter{subsubappendixc}[subappendixc]
\renewcommand{\thesubappendixc}{\Alph{appendixc}.\arabic{subappendixc}}
\renewcommand{\thesubsubappendixc}
	{\Alph{appendixc}.\arabic{subappendixc}.\arabic{subsubappendixc}}

\renewcommand{\appendix}[1] {\vspace{12pt}
        \refstepcounter{appendixc}
        \setcounter{figure}{0}
        \setcounter{table}{0}
        \setcounter{lemma}{0}
        \setcounter{theorem}{0}
        \setcounter{corollary}{0}
        \setcounter{definition}{0}
        \setcounter{equation}{0}
        \renewcommand{\thefigure}{\Alph{appendixc}.\arabic{figure}}
        \renewcommand{\thetable}{\Alph{appendixc}.\arabic{table}}
        \renewcommand{\theappendixc}{\Alph{appendixc}}
        \renewcommand{\thelemma}{\Alph{appendixc}.\arabic{lemma}}
        \renewcommand{\thetheorem}{\Alph{appendixc}.\arabic{theorem}}
        \renewcommand{\thedefinition}{\Alph{appendixc}.\arabic{definition}}
        \renewcommand{\thecorollary}{\Alph{appendixc}.\arabic{corollary}}
        \renewcommand{\theequation}{\Alph{appendixc}.\arabic{equation}}
        \noindent{\tenbf Appendix \theappendixc #1}\par\vspace{5pt}}
\newcommand{\subappendix}[1] {\vspace{12pt}
        \refstepcounter{subappendixc}
        \noindent{\bf Appendix \thesubappendixc. {\kern1pt \bfit #1}}
	\par\vspace{5pt}}
\newcommand{\subsubappendix}[1] {\vspace{12pt}
        \refstepcounter{subsubappendixc}
        \noindent{\rm Appendix \thesubsubappendixc. {\kern1pt \tenit #1}}
	\par\vspace{5pt}}

\newcommand{\textlineskip}{\baselineskip=13pt}
\newcommand{\smalllineskip}{\baselineskip=10pt}

\def\eightcirc{
\begin{picture}(0,0)
\put(4.4,1.8){\circle{6.5}}
\end{picture}}
\def\eightcopyright{\eightcirc\kern2.7pt\hbox{\eightrm c}} 

\newcommand{\copyrightheading}[1]
	{\vspace*{-2.5cm}\smalllineskip{\flushleft
	{\footnotesize International Journal of Modern Physics B, #1}\\
	{\footnotesize $\eightcopyright$\, World Scientific Publishing
	 Company}\\
	 }}

\newcommand{\publisher}[2]{{\begin{center}\footnotesize\smalllineskip 
	Received #1\\
	Revised #2
	\end{center}
	}}

\def\abstracts#1#2#3{{
	\centering{\begin{minipage}{4.5in}\baselineskip=10pt\footnotesize
	\parindent=0pt #1\par 
	\parindent=15pt #2\par
	\parindent=15pt #3
	\end{minipage}}\par}} 

\def\keywords#1{{
	\centering{\begin{minipage}{4.5in}\baselineskip=10pt\footnotesize
	{\footnotesize\it Keywords}\/: #1
	\end{minipage}}\par}}

\renewenvironment{thebibliography}[1]			
	{\frenchspacing
	 \ninerm\baselineskip=11pt
	 \begin{list}{\arabic{enumi}.}
	{\usecounter{enumi}\setlength{\parsep}{0pt}
	 \setlength{\leftmargin 22pt}{\rightmargin 0pt}   
	 \setlength{\itemsep}{0pt} \settowidth
	{\labelwidth}{#1.}\sloppy}}{\end{list}}

\newcounter{itemlistc}
\newcounter{romanlistc}
\newcounter{alphlistc}
\newcounter{arabiclistc}

\newcommand{\fcaption}[1]{
        \refstepcounter{figure}
        \setbox\@tempboxa = \hbox{\footnotesize Fig.~\thefigure. #1}
        \ifdim \wd\@tempboxa > 5in
           {\begin{center}
        \parbox{5in}{\footnotesize\smalllineskip Fig.~\thefigure. #1}
            \end{center}}
        \else
             {\begin{center}
             {\footnotesize Fig.~\thefigure. #1}
              \end{center}}
        \fi}

\newcommand{\tcaption}[1]{
        \refstepcounter{table}
        \setbox\@tempboxa = \hbox{\footnotesize Table~\thetable. #1}
        \ifdim \wd\@tempboxa > 5in
           {\begin{center}
        \parbox{5in}{\footnotesize\smalllineskip Table~\thetable. #1}
            \end{center}}
        \else
             {\begin{center}
             {\footnotesize Table~\thetable. #1}
              \end{center}}
        \fi}

\def\@citex[#1]#2{\if@filesw\immediate\write\@auxout
	{\string\citation{#2}}\fi
\def\@citea{}\@cite{\@for\@citeb:=#2\do
	{\@citea\def\@citea{,}\@ifundefined
	{b@\@citeb}{{\bf ?}\@warning
	{Citation `\@citeb' on page \thepage \space undefined}}
	{\csname b@\@citeb\endcsname}}}{#1}}

\newif\if@cghi
\def\cite{\@cghitrue\@ifnextchar [{\@tempswatrue
	\@citex}{\@tempswafalse\@citex[]}}
\def\citelow{\@cghifalse\@ifnextchar [{\@tempswatrue
	\@citex}{\@tempswafalse\@citex[]}}
\def\@cite#1#2{{$\null^{#1}$\if@tempswa\typeout
	{IJCGA warning: optional citation argument 
	ignored: `#2'} \fi}}

\def\pmb#1{\setbox0=\hbox{#1}
	\kern-.025em\copy0\kern-\wd0
	\kern.05em\copy0\kern-\wd0
	\kern-.025em\raise.0433em\box0}

\def\fnt#1#2{\footnotetext{\kern-.3em
	{$^{\mbox{\scriptsize #1}}$}{#2}}}

\def\fpage#1{\begingroup
\voffset=.3in
\thispagestyle{empty}\begin{table}[b]\centerline{\footnotesize #1}
	\end{table}\endgroup}

\def\runninghead#1#2{\pagestyle{myheadings}
\markboth{{\protect\footnotesize\it{\quad #1}}\hfill}
{\hfill{\protect\footnotesize\it{#2\quad}}}}
\font\tenrm=cmr10
\font\tenit=cmti10 
\font\tenbf=cmbx10
\font\bfit=cmbxti10 at 10pt
\font\ninerm=cmr9
\font\nineit=cmti9
\font\ninebf=cmbx9
\font\eightrm=cmr8

\def\qed{\hbox{${\vcenter{\vbox{			
   \hrule height 0.4pt\hbox{\vrule width 0.4pt height 6pt
   \kern5pt\vrule width 0.4pt}\hrule height 0.4pt}}}$}}

\renewcommand{\thefootnote}{\fnsymbol{footnote}}	

\def\bsc{{\sc a\kern-6.4pt\sc a\kern-6.4pt\sc a}}	
\def\bflatex{\bf L\kern-.30em\raise.3ex\hbox{\bsc}\kern-.14em 
T\kern-.1667em\lower.7ex\hbox{E}\kern-.125em X} 

\begin{document}

\runninghead{C. J. P\'{e}rez, A. Corral, A. D\'{\i}az-Guilera,
K. Christensen, and A. Arenas} 
{Self-Organized Criticality and Synchronization in Lattice Models}

\normalsize\textlineskip
\thispagestyle{empty}
\setcounter{page}{1}

\copyrightheading{}			

\vspace*{0.88truein}

\fpage{1}
\centerline{\bf ON SELF-ORGANIZED CRITICALITY AND
SYNCHRONIZATION} 
\vspace*{0.035truein}
\centerline{\bf IN LATTICE MODELS OF COUPLED DYNAMICAL SYSTEMS}
\vspace*{0.37truein}
\centerline{\footnotesize CONRAD J. P\'{E}REZ\footnote{E-mail
address: conrad@ulyses.ffn.ub.es},\, 
\'{A}LVARO
CORRAL, ALBERT D\'{I}AZ-GUILERA}
\vspace*{0.015truein}
\centerline{\footnotesize\it Departament de F\'{\i}sica Fonamental, 
Universitat de Barcelona} 
\baselineskip=10pt
\centerline{\footnotesize\it Diagonal 647, 08028 Barcelona, Spain} 
\vspace*{10pt}
\centerline{\footnotesize KIM CHRISTENSEN}
\vspace*{0.015truein}
\centerline{\footnotesize\it Department of Physics, 
University of Oslo, P.O.  Box 1048}
\baselineskip=10pt
\centerline{\footnotesize\it Blindern, N-0316 Oslo 3, Norway}
\vspace*{10pt}
\centerline{\footnotesize ALEX ARENAS}
\vspace*{0.015truein}
\centerline{\footnotesize\it Departament d'Enginyeria 
Inform\`{a}tica, Universitat Rovira i
Virgili}
\baselineskip=10pt
\centerline{\footnotesize\it Carretera Salou s/n, 
E-43006 Tarragona, Spain }
\vspace*{0.225truein}
\publisher{(received date)}{(revised date)}

\vspace*{0.21truein}
\abstracts{Lattice models of coupled dynamical systems lead to a
variety of complex behaviors. Between the individual motion of
independent units and the collective behavior of members of a
population evolving synchronously, there exist more complicated
attractors. In some cases, these states are identified with
self-organized critical phenomena. In other situations, with
clusterization or phase-locking.  The conditions leading to such
different behaviors in models of integrate-and-fire oscillators and
stick-slip processes are reviewed.}{}{}

\vspace*{10pt}


\sloppy{

\vspace*{1pt}\textlineskip	
\section{Introduction}
\vspace*{-0.5pt}
\noindent
In nature, there are many examples of systems with complex collective
behavior. There is a big effort to identify and understand the
underlying mechanisms leading to such behavior, but it is difficult
to find general rules which could give, {\em a priori}, information
about spatio-temporal complexity.  The analysis of the time evolution
of the magnitudes which describe an isolated dynamical system is a
first step.  However, the features of a big system consisting of a
large number of individual units (where each unit is a dynamical
system itself) interacting according to a given criterion, can be
very difficult to determine, even if all the units are identical. The
nature of the interaction between units, the type of boundary
conditions, and the absence or existence of noise are some
ingredients which can change completely the dynamic behavior of a
coupled dynamical system.

\pagebreak

\textheight=7.8truein
\setcounter{footnote}{0}
\renewcommand{\thefootnote}{\alph{footnote}}

     One might think that complexity arises when the intrinsic
dynamics that govern the temporal behavior of each member of a
population as well as the interaction between them follow complicated
spatio-temporal rules.  Not necessarily.  It may arise as a result of
the continued local simple interactions between all parts in an
extended system.  On the other hand, extended systems with complex
local features may lead to a ''simple'' collective behavior. This is
observed in some biological systems where after some transient period
a regime characterized by a perfect synchrony in the temporal
activity of all the members is achieved.

     Between the trivial behavior of independent uncorrelated units,
which can be analyzed by considering the features of isolated
elements, and ''simple'' collective behaviors such as units evolving
in perfect synchrony, there is a wide spectrum of situations.  Some
of them manifest a large degree of complexity such as systems
displaying spatial fractal structures and temporal fractal behavior.
Fractal structures appear in a variety of physical
systems.\cite{Mand82,Feder88} They are geometrical objects, that look
alike on all length scales.  The universe consists of clusters of
galaxies, organized in clusters of clusters of galaxies and so on.
Earthquakes occur on structures of faults ranging from thousands of
kilometers to millimeters.  Likewise, the coast of Norway contains
fjords of all sizes, from very small ones to very big ones.  Temporal
fractal behavior occurs when the temporal fluctuations in a certain
quantity look alike on all time scales.  In general, the power
spectrum behaves like $1/f^{\varphi}$.  When $\varphi \approx 1$ we
talk about $1/f$ noise or flicker
noise.\cite{cmp7.103,rmp53.497,weissman} The fluctuations in light
intensity of quasars, the flow of the river Nile, and the current
flowing through a resistor are but a few examples of systems
displaying fluctuations without an intrinsic time scale.

     During the last years there has been an increasing interest to
understand the mechanisms that lead to other complex behaviors
closely related to fractals.  In particular, a lot of attention is
paid on conservative and nonconservative systems displaying
self-organized criticality (SOC).  Up to now, SOC has been observed
in models of cellular automata, coupled map lattices, and coupled
dynamical systems defined on a lattice.  It is our purpose in this
paper to review the mechanisms as well as the conditions required to
observe SOC and the corresponding transition to other collective
behaviors.

     The paper is organized as follows.  In Section II we review the
mechanisms that provoke the occurrence of SOC, in randomly driven as
well as in deterministically driven models.  In Section III we
analyze different models of oscillators focusing our interest on the
conditions required to observe a macroscopic degree of
synchronization.  We will discuss the effect of both short- and
long-range interactions.  Section IV is devoted to investigate the
circumstances in which a system develops the behaviors mentioned
above, their transitions, and their relations.  Finally, in Sect.  V
we present the conclusions of the present work along with the lines
we think deserve further investigation in order to deal with more
realistic physical problems.

\section{Self-Organized Criticality}
\noindent
The ubiquity of spatial fractal structures and temporal fractal
behavior observed in physical systems suggests a common underlying
dynamical origin.  The lack of a characteristic scale is the hallmark
of a critical process.  In equilibrium, critical processes require a
fine-tuning of a relevant physical control parameter such as the
temperature or the magnetic field to a critical point.  However,
nature does not, by itself, provide any fine-tuning of control
parameters, and with zero probability it sits at the critical point
by accident.  Thus it is very unlikely that the wide occurrence of
scale-invariance is due to critical processes in equilibrium systems.

     Recently, Bak, Tang, and Wiesenfeld (BTW) suggested that the
frequently observed scale-invariance in nature might be related to
the spontaneous organization (spatial complexity) and reorganization
(temporal complexity) in slowly driven, dissipative
systems.\cite{prl59.381,pra38.364} Internal mechanisms lead to
dissipation (heat generation) and quasi-stationary states are reached
in which the average fluxes of energy into and out of the systems are
equal.  The avalanches that occur when grains are dropped onto a pile
have been used to illustrate this idea.  When sprinkling grains of
sand onto a table, one after the other, a pile builds up.
Eventually, the pile ceases to grow and additional grains of sand
will ultimately fall off the pile.  The attractor of the dynamics is
a statistically stationary state with a fluctuating angle of repose.
If the slope is too steep, large system spanning avalanches would
make the pile to collapse to a more stable configuration.  If, on the
other hand, the slope is too shallow, only small avalanches would be
initiated.  The sandpile settles into a state in between with
avalanches of all sizes, power-law distributed.  This scale
invariance suggests that the system is critical in analogy with
equilibrium critical phenomena.  However, one deals with dynamical
nonequilibrium statistical properties and the system evolves
naturally to the critical state without any tuning of external
control parameters.  The criticality is an intrinsic property of the
dynamics of the system, and the phenomenon of self-organized
criticality may very well provide a connection between the occurrence
of fractal structures and $1/f$ noise, as well as being the physical
origin of these two phenomena.  The earth crust is another example of
a self-organized critical system: Along the boundary of tectonic
plates, a slow build-up of strain is relaxed through earthquakes of
all sizes; the energy-frequency relation of earthquake occurrence,
which is related to the Gutenberg-Richter law,\cite{GR44} is a
power-law distribution.

     A cellular automaton model for sandpile
dynamics\cite{prl59.381,pra38.364} and other numerical
models\cite{prl62.2632,prl68.1244,prl69.1629,prl71.4083} were shown
numerically to display SOC.  The dynamical rules in the BTW model ---
also known as the sandpile model --- at least intuitively resemble
the dynamics of a sandpile.  In the sandpile model, each site on a
lattice is characterized by an integer variable (local slope), and
this variable changes in time due to external perturbations (adding
grains of sand) and to interaction between different sites when an
avalanche propagates through the system.  With very simple rules, the
system reaches a state which is characterized by the lack of any
characteristic length or time scale.  The simplicity of the model
suggests that the phenomenon of SOC could be quite universal: Indeed,
it has been scrutinized by a host of researchers in fields spanning
statistical mechanics,\cite{prl68.205} condensed matter
physics,\cite{prl67.919,prl67.1334}
geophysics,\cite{prl68.1244,rmp66.657} biology,\cite{prl71.4083} and
economy.\cite{ricerche,p184a.127}

     The notion of self-organized criticality has initiated much
experimental work, especially on granular systems.  The main object
has been to address the question of SOC in real sandpiles.
Essentially, two distinct types of experiments have been performed:
Rotating a semi-cylindrical drum partially filled with grains at a
low, constant
velocity,\cite{prl62.40,pra43.2720,prl69.2431,pre47.2229} or dropping
single grains at a low rate on a conical sandpile resting on a
circular platform.\cite{prl65.1120,ajp61.329,pre47.1401,prl73.537} In
most of these studies, the flow over the rim of the system (the drop
number) was recorded.  In a drum experiment, this quantity was found
to be nearly periodic in time.\cite{prl62.40} Similar behavior was
seen for large conical piles: they displayed relaxation oscillations
when a system size of 30--40 particle diameters was
exceeded.\cite{prl65.1120} In some of the latter experiments,
power-law distributions has been reported for small conical piles.
However, recently it has been shown, that these distributions are
more likely to be stretched exponentials, that is, a characteristic
size for the drop number appears.\cite{kim.nature,newff} The observed
behavior may be attributed to inertial effects, which seem to play an
important role in all the
experiments.\cite{pre47.1401,kim.nature,p186a.82,pra45.665} Inertial
effects lead to a nonlocal process, whereas in the numerical models
only the local geometry determines the dynamics.  Furthermore, we
notice, that the relationship between the drop number measured in
experiments and the avalanche size measured in numerical models is
unclear.  For an alternative view on the dynamics of granular
material see the work by Mehta and Barker\cite{rpp57.383} which
reviews the progress that has been achieved experimentally and
theoretically.

     In a recent experiment, the internal dissipated energy due to
avalanches in a slowly driven one-dimensional ($1D$) rice pile, was
recorded\cite{kim.nature} and it was shown that the occurrence of SOC
depends on details in the grain-level dissipation mechanism.  With
spherical grains, a stretched-exponential distribution was observed
implying a characteristic scale which is inconsistent with SOC.  The
spherical grains typically accumulated kinetic energy when moving
down the slope.  However, with more elongated grains, the dynamics
was dominated by sliding grains.  This induced a higher effective
friction which suppressed the inertial effects, and a power-law
distribution of avalanche sizes appeared.  This provides the first
experimental evidence of self-organized critical behavior in slowly
driven granular systems.

\subsection{Randomly driven models}
\noindent
First, we briefly discuss the $1D$ BTW sandpile model. Although the
behavior of the $1D$ model is trivial, it illustrates the basic
concepts of the more complex behavior displayed by the $2D$ BTW
sandpile model.

     The $1D$ sandpile model\cite{pra38.364} is a cellular automaton
where an integer variable $h_i$ gives the height of the pile at site
$i$.  We define the local slope $z_i$ at site $i$ as 
\begin{equation}
z_i = h_i - h_{i+1}.
\end{equation} 
The addition of a grain of sand on a randomly chosen site
$i$ $(h_i \rightarrow h_i + 1)$ results in the following changes in
the slopes 
\begin{equation} 
\left.  \begin{array}{l} z_i \rightarrow
z_i + 1 \\ z_{i-1} \rightarrow z_{i-1} - 1.  \end{array} \right.
\label{adding}
\end{equation} 
We proceed by dropping grains at random sites until one site reaches
a slope which exceeds a critical value, $z_i > z_{c}$, and the site
topples by transferring one grain to its neighboring site on the
right, that is, 
\begin{equation} 
\left.  \begin{array}{l} z_i
\rightarrow z_i - 2
\\ z_{i \pm 1} \rightarrow z_{i \pm 1} + 1 \end{array} \right.
\label{toppling} 
\end{equation} 
unless at the rightmost site where the sand grains fall off the pile.
The neighbors that are affected by the toppling may topple in turn
generating a chain reaction or {\em avalanche}.  During the
avalanche, no grains are added to the pile thus separating the two
time scales involved in the dynamic evolution of the pile, a slow
time scale for the addition of the grains and a fast time scale for
the relaxation processes.  The avalanche stops when the system
reaches a stable state with $z_i \leq z_{c} \:\: \forall i$ and
another grain is added following (\ref{adding}) until a new avalanche
is initiated and so on.  After a transient period, whose duration
depends on the initial conditions, the system reaches a critical
state in which $z_i = z_{c} \:\: \forall i$.  This state is a fixed
point of the dynamics, since after any perturbation the system
relaxes returning to the stable state; the added grain will tumble
down the slope and simply fall off the pile.  The fixed point is an
attractor for the dynamics, however, this state has no spatial
structure, and correlation functions are trivial.

     In the $2D$ sandpile model (on a square
lattice),\cite{prl59.381} an integer $z_{i,j}$ is assigned to each
lattice site $(i,j)$, where $i,j = 1, \ldots , L$.  The integer
$z_{i,j}$ represents an appropriate dynamical variable (e.g. height
of a column of sand, mechanical stress, heat, pressure, or
energy\cite{prl59.381,p173a.22}) on site $(i,j)$ in a spatially
extended system.

     We perturb the system (add sand to it) by choosing at random a
position and increasing the dynamical variable with one unit, that
is, $z_{i,j}
\rightarrow z_{i,j} + 1$.  Whenever the dynamical variable on site
$(i,j)$ exceeds a threshold value $z_{i,j} > z_{c}$, the site
topples,
\begin{equation} 
z_{i,j} \rightarrow z_{i,j} - 4,
\label{topplea} 
\end{equation} 
\begin{equation} 
z_{nn} \rightarrow z_{nn} + 1 \label{toppleb} 
\end{equation} 
where $z_{nn}$ denotes the height at nearest neighboring sites.  As a
result one or more neighbors may exceed the threshold value in which
case they have to relax and an avalanche will propagate through the
system.  The toppling rule conserves the amount of $z$-values
whenever an interior site topples.  Dissipation only occurs when a
boundary site topples assuming open boundary conditions (bc).

     The $2D$ sandpile will evolve into a statistically stationary
state where, on the average, the rate of flow into the system equals
the rate of flow out of the system across the boundary.  Whenever the
average slope becomes too large, system spanning avalanches can
occur, which transfer grains to the boundary.  On the other hand,
when the average slope is too small, avalanches tend to be small, and
the pile builds up.  The stationary state is no longer a fixed point
as in $1D$ but a complex attractor.  Dhar\cite{prl64.1613} has
calculated the exact number of states in the attractor.  When $z_{c}
= 3$, only $(3.210 \cdots)^{L^2}$ of the $4^{L^2}$ stable states are
allowed.  Furthermore, the dynamics which takes the system from one
allowed state to another is ergodic in the sense that all the allowed
states occur with equal probability.\cite{prl64.1613}

     One way of characterizing the dynamics in a system of linear
size $L$, is to measure the distribution of avalanche sizes defined
as the total number of topplings.  Let $P(s,L)$ denote the
probability of initiating an avalanche of size $s$ in a system of
size $L$.  In the stationary state, an added particle will eventually
fall off the boundary of the system.  The average distance a particle
has to flow in order to reach the boundary is proportional to $L$ due
to the random deposit.  The average avalanche size $\langle s \rangle
= \int s P(s,L)\,ds$ must be infinite in the thermodynamic limit $L
\rightarrow \infty$.  Indeed, averaged over a large number of
perturbations (after the transient period), the avalanche size 
distribution is a power law 
\begin{equation} 
P(s,L) \propto s^{1-\tau}
\label{exponent-tau} 
\end{equation} 
limited only by the size of the system, see Fig.\ \ref{btwdist}(a).
From these results the power-law exponent $\tau$ is $2.15 \pm 0.1$.
Since the cutoff is a finite size effect, the average avalanche size
tends to infinity when $L \rightarrow \infty$.  A data collapse for
different system sizes $L$ is obtained when plotting
$L^{\beta}\,P(s,L)$ or $s^{\tau-1}\,P(s,L)$ against the rescaled
variable $s/L^{\nu}$, see Figs.\ \ref{btwdist}(b-c). 
Thus we can write 
\begin{equation} 
P(s,L) = L^{- \beta} F(s/L^{\nu}),
\label{fss} 
\end{equation} 
or, alternatively,
\begin{equation} 
P(s,L) = s^{1-\tau} G(s/L^{\nu}), 
\label{fssb} 
\end{equation} 
with $G(x) = x^{\tau -1}\,F(x)$. The exponent $\beta = \nu
({\tau-1})$, where $\tau$ is the power-law exponent and $\nu$ a
critical index expressing how the cutoff scales with system size.
The scaling function $F$ approaches a power law (or, equivalently,
$G$ a constant) when $s/L^{\nu} \rightarrow 0$ since the avalanche
size distribution becomes independent of the system size when $L
\rightarrow \infty$ and decays very quickly for $s \gg L^{\nu}$.
 
\begin{figure}[htbp] 
\centerline{\epsfxsize=4.0truein \epsffile{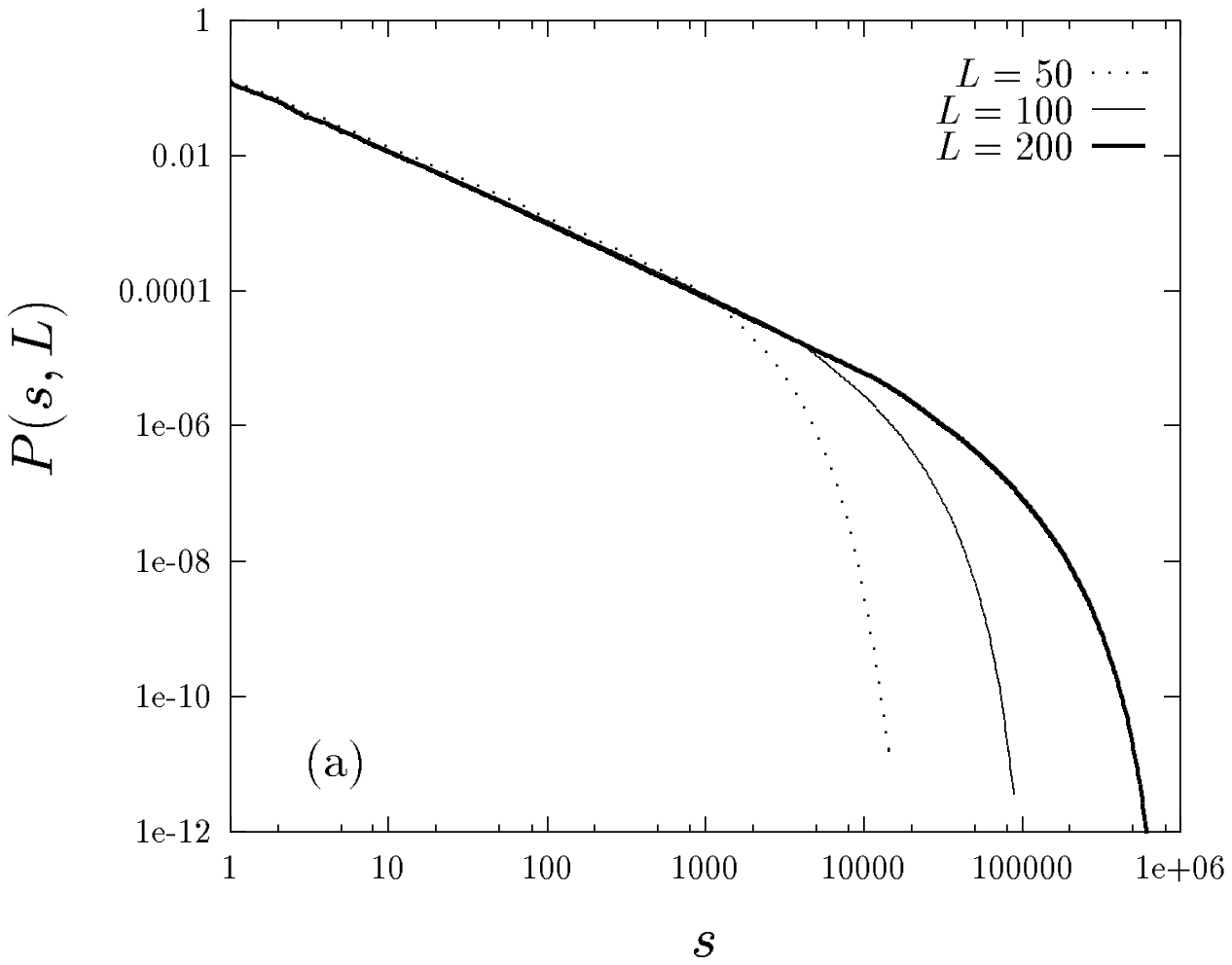}}
\vspace*{0.2cm}
\centerline{\epsfxsize=4.0truein \epsffile{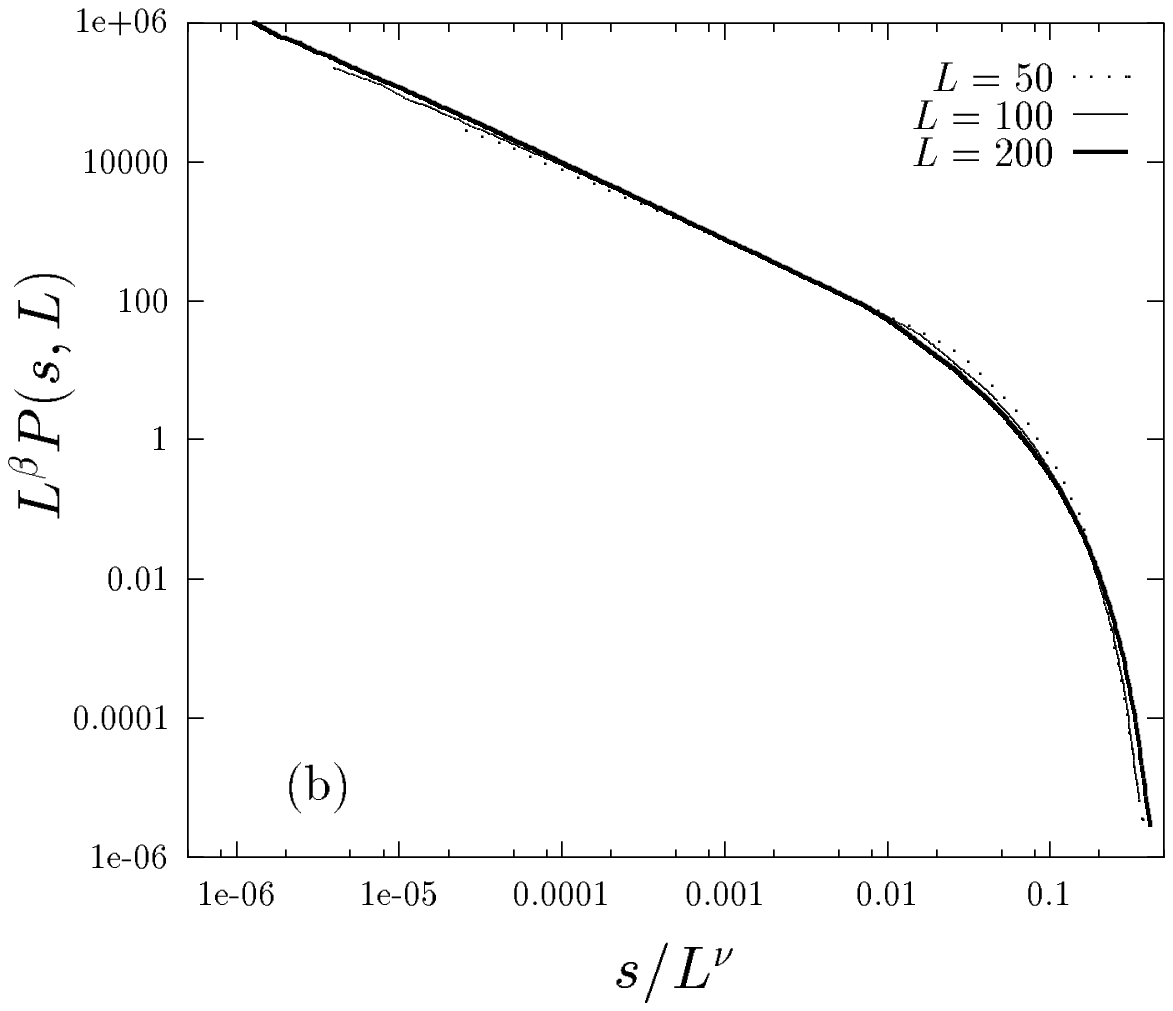}}
\vspace*{0.3cm}
\fcaption{The $2D$ Bak, Tang, and Wiesenfeld sandpile model for
system sizes $L = 50, 100,$ and $200$ with open bc.  (a) The
probability of initiating an avalanche of size $s$ in a system of
size $L$, $P(s,L)$ decreases algebraically with $s$.  The cutoff is a
finite-size effect.  The displayed distribution functions are
averaged over exponentially increasing bins.  (b) Using Eq.
({\protect\ref{fss}}), a reasonable data collapse is obtained with
$\tau = 2.15 \pm 0.1$ and $\nu = 2.7 \pm 0.1$.  (c) However, the
alternative data collapse suggested by Eq. (\ref{fssb}) shows the bad
scaling of the model for ''small'' system sizes.}
\label{btwdist} 
\end{figure}

\addtocounter{figure}{-1}
\begin{figure}[htbp]
\centerline{\epsfxsize=4.0truein \epsffile{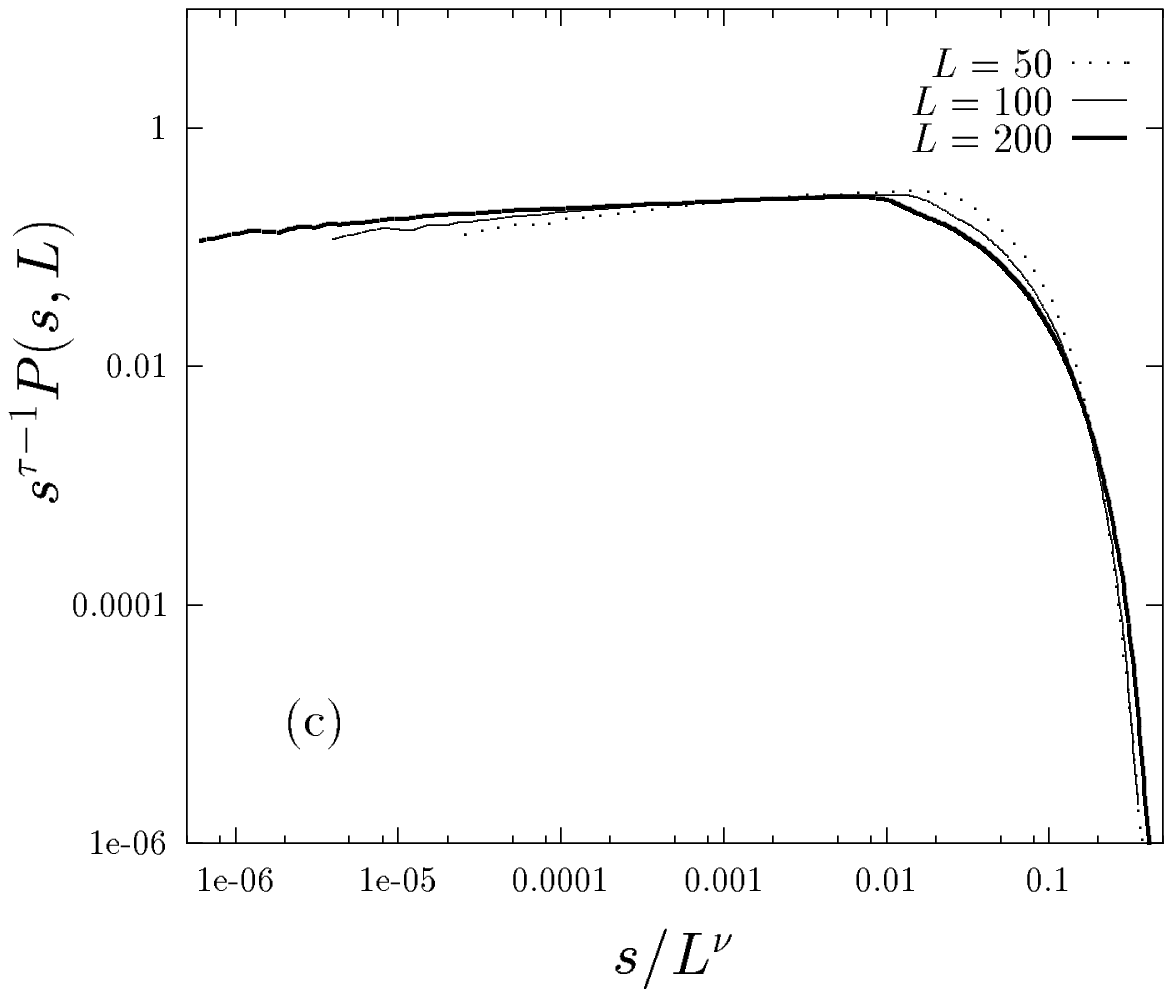}}
\vspace*{0.5cm}
\fcaption{({\em continued})}
\end{figure}

     The BTW relaxation rules ({\ref{topplea}}) and ({\ref{toppleb}})
conserve the dynamical variable $z$ except at the boundary.
Introducing nonconservative dynamics in the interior of the system
will leave the BTW model noncritical.  Changing rule
({\ref{topplea}}) to, say, $z_{i,j} \rightarrow z_{i,j} - 5$ while
({\ref{toppleb}}) remains unchanged, one grain of sand will dissipate
for every toppling.  In the stationary state, the average rate of
dissipation equals the average rate of flow into the system.  Thus
$\langle s \rangle = 1/P_{z_{c}}$ where $P_{z_{c}}$ is the
probability of adding a grain to a site with $z_{c}$ units, thereby
initiating an avalanche.  This probability will approach a constant
value when $L \rightarrow \infty$, that is, $\langle s \rangle$ will
approach a finite value.  Such a system cannot display SOC.  The
avalanche size distribution decays exponentially with a
characteristic avalanche size, that does not depend on the system
size, and the avalanches are ''localized''.  However, a system with
globally conservative but locally nonconservative dynamical rules
(changing rule ({\ref{topplea}}) to $z_{i,j} \rightarrow z_{i,j} - 4
+ \theta$, where $\theta \in \{-3,...,3\}$ is an annealed random
variable) displays SOC.\cite{jsp22.923}

     Memory effects can be introduced into the BTW model in the sense
that a site which has relaxed cannot receive any amount from
neighboring sites during the ongoing avalanche.  This kind of memory
effects makes the dynamics nonconservative and hence destroys the SOC
behavior.

     In order to model the early experiments on SOC in slowly driven
granular piles, inertia effects were included in a BTW-like
model.\cite{pra45.665} It was observed that when the system size
exceeded a certain size, characteristic system spanning avalanches
appeared due to the inertia effects in agreement with experiments on
real sandpiles.\cite{prl62.40,pre47.1401,p186a.82,pra45.665} Thus the
presence of inertia effects also destroy the SOC behavior.

     An apparently different stochastically driven continuous-energy
model was introduced by Zhang.\cite{prl63.470} In this model, a real
number $E_{i,j}$ is assigned to each site $(i,j)$ on a square
lattice.  The system is perturbed randomly by adding an amount
$\delta E$ to a randomly chosen site, that is, $E_{i,j} \rightarrow
E_{i,j} + \delta E$.  If the value at any site exceeds the threshold
value $E_{i,j} > E_{c}$, the system relaxes according to
\begin{equation} 
\left.  \begin{array}{l} E_{nn} \rightarrow E_{nn} +
\varepsilon E_{i,j} \\ E_{i,j} \rightarrow 0, \end{array} \right.
\label{sbrelax} 
\end{equation} 
with $\varepsilon = 0.25$. The system is conservative and dissipation
occurs only at the boundary, where the number of nearest neighbors is
less than 4.  The model was studied with $E_{c} = 1$ and $\delta E$
chosen uniformly in the interval $[0,0.5],$\cite{p173a.22,prl63.470}
in which case it is believed to be in the same universality class as
the $2D$ BTW model, that is, the values of critical exponents, such
as the power-law exponent $\tau$ and scaling index $\nu$, are
identical.\cite{el26.177,pre51.1711}

     With $\varepsilon < 0.25$, the dynamics in the Zhang model is
nonconservative.  When $\delta E$ is chosen in the range $[0.1,1]$,
the nonconservative relaxation rule introduces a characteristic
avalanche size.  In this case, the Zhang model is also noncritical in
the presence of nonconservation.  Furthermore, the introduction of
the memory effects mentioned above leads to a characteristic
avalanche size inconsistent with the hypothesis of SOC.  Similar
models have been considered in the context of
fracturing.\cite{pra45.2211} However, additional rules where
introduced in order to make the dynamics globally conservative which
introduces a kind of inertia effects.  Indeed, the general behavior
of such systems is the periodic generation of system spanning
avalanches coexisting with power-law distributed small avalanches.

     The microscopic rules of the lattice models can be written in
the form of a set of coupled equations, one for each site.  For a
continuous version of the stochastically driven BTW model we
have\cite{granada}
\begin{equation} 
E_{i,j}(t+1) = E_{i,j}(t) - \Theta (E_{i,j}(t) - E_{c})\,
E_{c} + \sum_{nn}\Theta (E_{nn}(t) - E_{c})\, E_{c}/4 + \eta_{i,j}(t),
\label{discretebtw} 
\end{equation} 
where $\Theta(x) = 0$ when $x \leq 0$ and $\Theta(x) = 1$ when $x >
0$ is the Heaviside function.  The variable $t$ refers to lattice
updates.  However, in a slowly driven system, the physical time
between lattice updates is very short when avalanches are propagating
and very long when perturbing the system.\cite{adg.fractals} The sum
takes into account a possible increase in the dynamical variable due
to toppling of nearest neighboring sites $nn$.  When an avalanche has
come to a halt, a random site is perturbed.  The random drive is
represented by
\begin{equation} 
\eta_{i,j}(t)= \eta\, \delta_{i,n}
\delta_{j,m} \prod_{k,l} \left[1-\Theta (E_{k,l}(t)-E_{c})\right],
\label{discretenoise}
\end{equation}
$n, m$ being two random integers between $1$ and $L$ chosen once for
each lattice update, and $k,l$ two index running over all the lattice
sites.  In the original BTW-model $\eta = 1.$ The Heaviside function
$\Theta$ that appears in the noise terms represents the separation of
time scales.  The models are driven slowly in the sense that no
perturbation takes place while avalanches evolve.  On the other hand,
for the conservative Zhang model, where the toppling variable is
reset to zero, the corresponding equations are 
\begin{equation} 
E_{i,j}(t+1) = [1-\Theta (E_{i,j}(t) -
E_{c})]\, E_{i,j}(t) +\sum_{nn}\Theta (E_{nn}(t) - E_{c})\,
E_{nn}(t)/4 + \eta_{i,j}(t).  
\label{discretezhang} 
\end{equation}
The noise $\eta_{i,j}(t)$ is again given by (\ref{discretenoise}).
In the original Zhang model, $\eta$ is chosen uniformly in the
interval $[0,0.5]$.  In order to deal analytically with these
equations, one usually neglects the Heaviside function in the noise
term.  This fact creates some difficulties, since the notion of
avalanches in such models is not well-defined.\cite{prl74.2511}

     Equations (\ref{discretebtw}) and (\ref{discretezhang}) can be
coarse-grained in order to obtain a continuum equation for the
effective $E(\vec{r},t)$.  Using the prescriptions for the temporal
derivative and the Laplace operator one gets, after a rescaling of
the energy $E - E_c \rightarrow E$\cite{23DRG}
\begin{equation} 
\frac{\partial E(\vec{r},t)}{\partial t}= \alpha \nabla ^2
[(Z E(\vec{r},t) + E_c) \Theta(E(\vec{r},t))] +\eta(\vec{r},t),
\label{continuumexact} 
\end{equation} 
where $Z=1$ for the Zhang model and $0$ for the continuous BTW model.
The noise $\eta(\vec{r},t)$ takes into account the effective external
noise as well as the internal noise that appears due to the
elimination of internal degrees of freedom.

     On the other hand the $\Theta$-function representing the
threshold dynamics is modeled as the limit of a smooth function,
whereupon dynamical renormalization group calculations can be
applied.\cite{el26.177,adg.fractals} This fact also changes the
problem intrinsically.  Left to itself, without adding noise, the model
would relax to a unique flat ground state with $E(\vec{r},t) = 0$.
The concept of stable states (or memory) is removed.  Some other
authors have built nonlinear stochastic differential equations
according to the symmetries of the discrete
models.\cite{prl62.1813,prl64.1927,pra45.7002} One should
distinguish, however, between these approaches, in which the main
goal is to study the "generic scale invariance" by means of dynamical
renormalization group theory, and the lattice models.

\subsection{Continuously driven models}
\noindent
A continuously driven dynamical system was introduced by Olami,
Feder, and Christensen (OFC) in the context of earthquake
modeling.\cite{prl68.1244} Though the dynamics of earthquakes is very
complex there are two basic components which have to enter a model:
(a) earthquakes are generated by the very slow relative motion of
tectonic plates, (b) they occur as abrupt rupture events when the
fault can no longer sustain the stress, that is, the occurrence is
intermittent.  Hence, there are two time scales involved in the
process; one is related to the stress accumulation while the other,
which is orders of magnitude smaller, is associated to the duration
of the abrupt releases of stress.  A simplified spring-block model
introduced by Burridge and Knopoff\cite{BK67} includes the basic
components mentioned above.  The model consists of a $2D$ network of
blocks interconnected by springs.  (For a review on the work of $1D$
Burridge-Knopoff models of earthquake faults, see the work by
Carlson, Langer, and Shaw.\cite{rmp66.657}).
Each block is connected to the four nearest
neighbors.  Additionally, each block is connected to a single rigid
driving (tectonic) plate by another set of springs as well as
connected frictionally to a fixed rigid (tectonic) plate, see Fig.\
\ref{spmodel}.

\begin{figure}[htbp] 
\epsfxsize=4.9truein 
\hskip 0.15truein
\epsffile{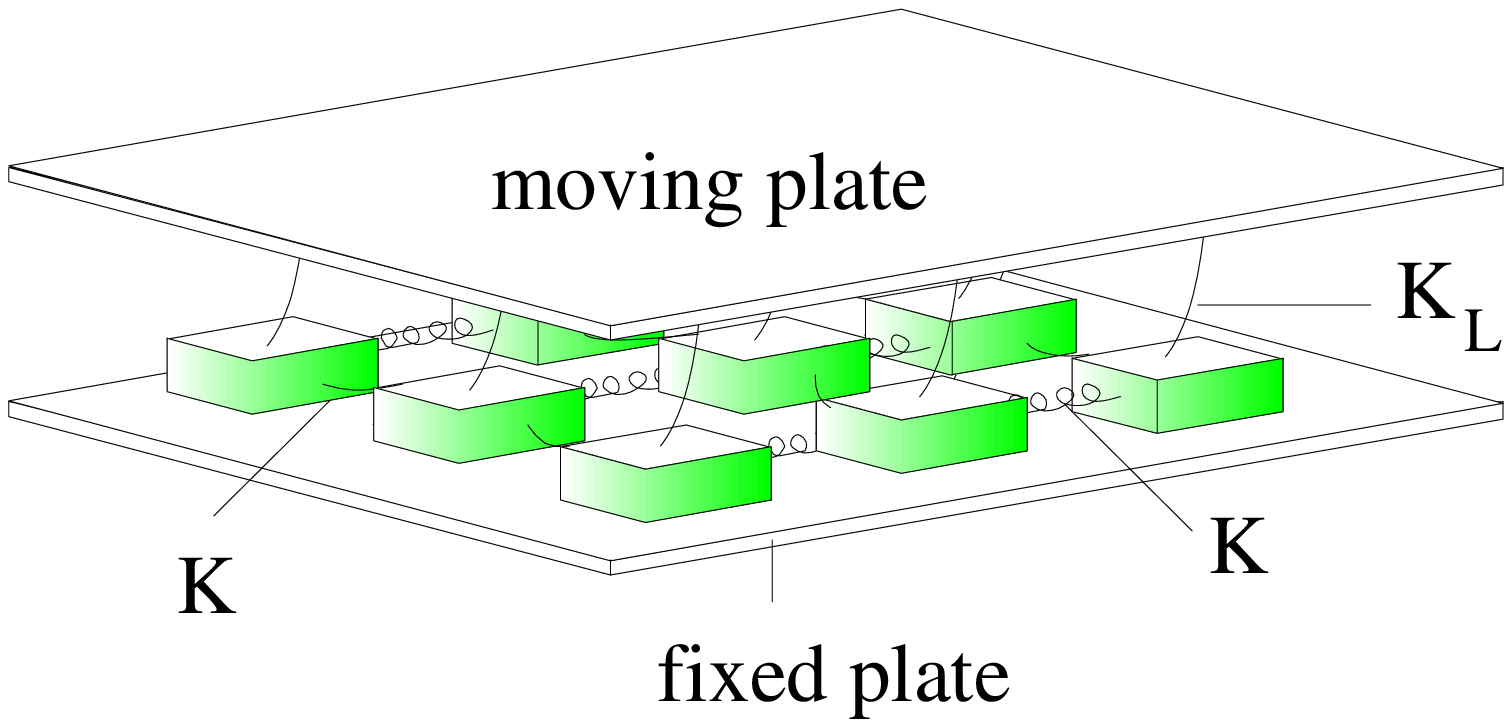} 
\fcaption{The $2D$ spring-block model of an earthquake fault. All
the blocks are connected to a moving plate by springs with spring
constant $K_L$.  The blocks are also connected to the four nearest
neighbors through springs of strength $K$. 
(After Olami, Feder, and Christensen\cite{prl68.1244})} 
\label{spmodel} 
\end{figure}

     Strain is accumulated uniformly across the system as the rigid
plates move with a constant relative velocity.  When the strain on
one of the blocks exceeds the static friction force, the block slips.
The released stress is transferred to the neighboring blocks, which
in turn may slip, and an earthquake can evolve.  This simple picture
can be mapped to a coupled map lattice model.\cite{pra46.1829} A
uniformly increasing stress is assigned to each lattice site $(i,j)$.
When a block exceeds the threshold stress $E_{c}$ it slips.  Simple
arguments lead to the relaxation rules identical to Eq.
(\ref{sbrelax}) with $\varepsilon = K/(4K+K_L)$, $K$ and $K_L$
denoting the spring constants in the model, see Fig.\ \ref{spmodel}.
The earthquakes are considered to be instantaneous, that is, the
loading of the system stops during an earthquake and the driving
recommences when the evolving earthquake has come to a halt.  The
size of an earthquake is defined as the total number of relaxations.

     When $\varepsilon = 0.25$, the model is a continuously driven
version of the Zhang model.\footnote{There is also a uniformly driven
continuous version of the $2D$ BTW sandpile model.\cite{pra38.364}
Choosing random initial conditions, this model will be in the same
universality class as the stochastically driven integer model.  In
the continuous model, the fractional part of the dynamical variable
takes, in some sense, the role of the random number generator.  Note
that this model and the OFC model are deterministic.  The inherent
randomness enters via the initial conditions.}\ 
Apparently, this model
has the same power-law exponent $\tau$ as the stochastically driven
models.  However, since $K_L \neq 0$, the generic situation
corresponds to the nonconservative case where $\varepsilon < 0.25$.

     In Fig.\ \ref{sbdist}(a) we plot the results of a simulation
with $\varepsilon = 0.20$ for different system sizes. A reasonably
good data collapse is obtained using the finite-size scaling ansatz
Eq.  (\ref{fss}), see Fig.\ \ref{sbdist}(b). Thus the system is
scale invariant; it displays SOC behavior even in the nonconservative
case.

\begin{figure}[htbp]
\centerline{\epsfxsize=4.0truein \epsffile{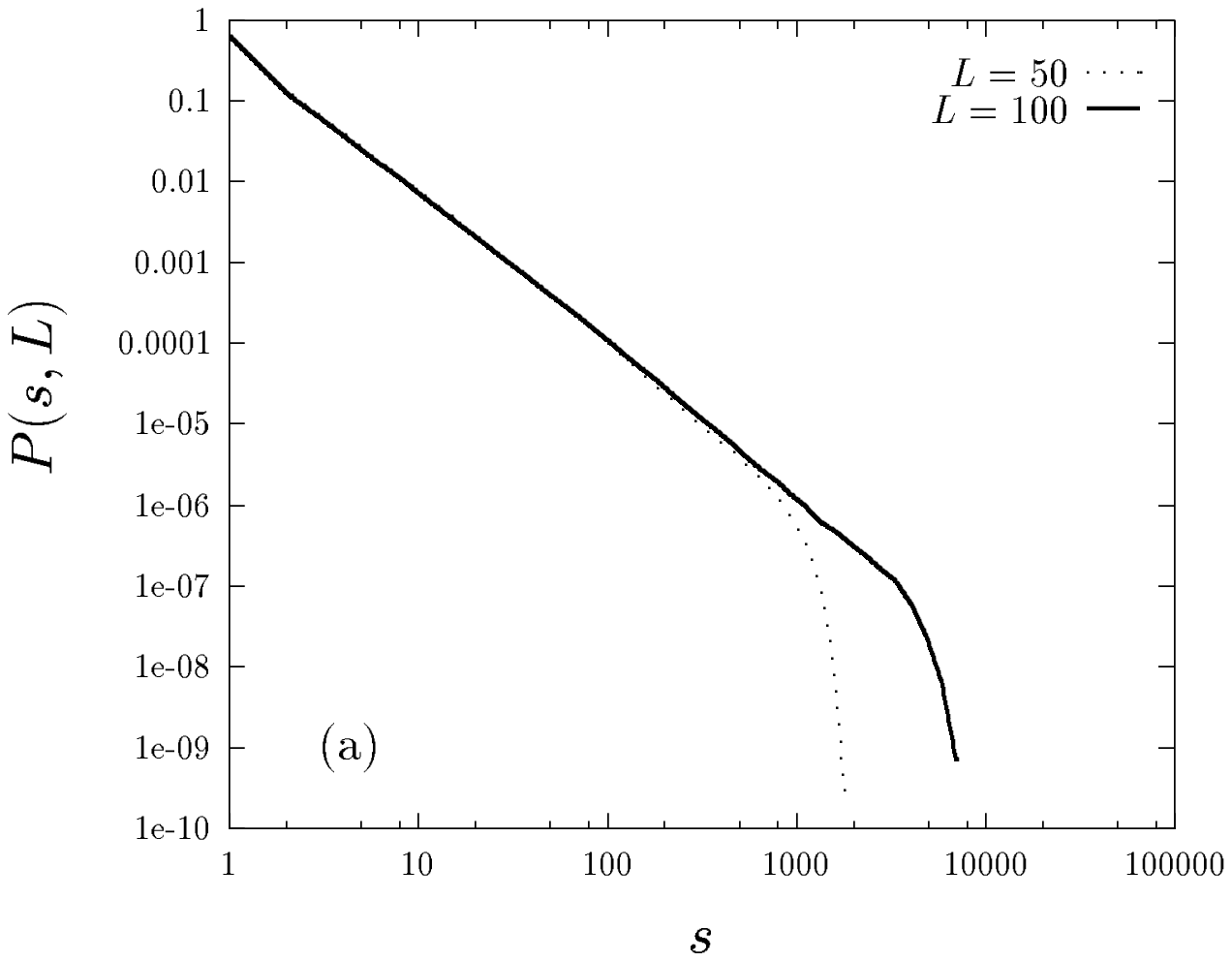}} 
\vspace{0.2cm}
\centerline{\epsfxsize=4.0truein\epsffile{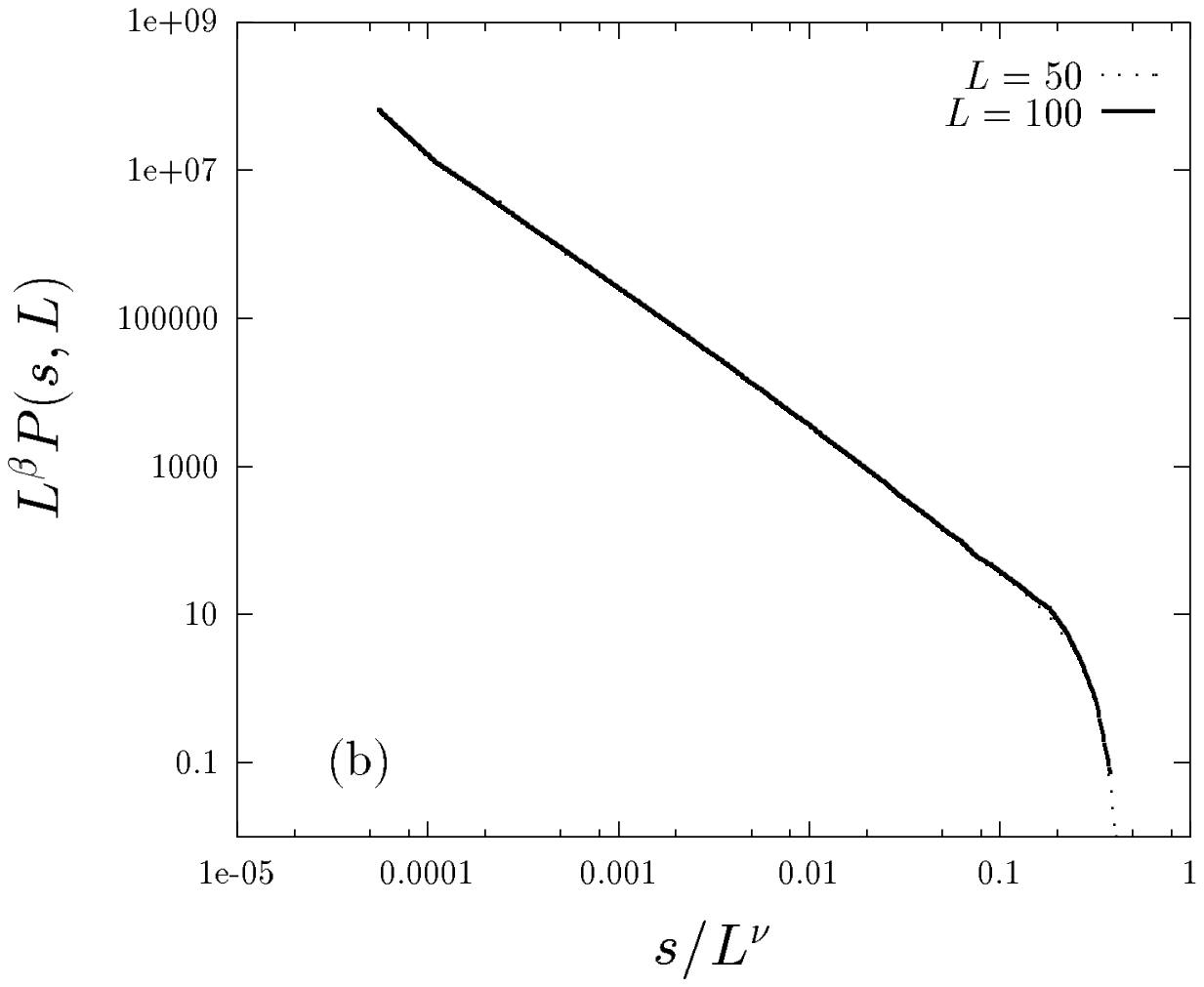}}
\vspace*{0.3cm}
\fcaption{(a) The probability density $P(s,L)$ of having an
earthquake of size $s$ when $\varepsilon = 0.20$, for the OFC model.
The different curves refer to different system sizes $L = 50$,
and  $100$.  The cutoff in energy distribution increases with system
size $L$.  (b) A finite-size scaling plot using Eq.
(\protect\ref{fss}) with $\tau = 2.9 \pm 0.1$ and $\nu = 2.15
\pm 0.1$\protect.\cite{comment.ofc,reply.kim}}  
\label{sbdist} 
\end{figure}

     When introducing nonconservative relaxation rules into the
stochastically driven BTW and Zhang models, the distribution of
avalanche sizes decay exponentially with a characteristic avalanche
size.  The systems are subcritical.  Thus the occurrence of
criticality in the nonconservative OFC model is very intriguing
since 
it suggests a different mechanism for the generation of the scale
invariance. Since the majority of natural phenomena are inherently
nonconservative, the SOC behavior of nonconservative systems is
probably much more generic than the corresponding behavior of
conservative systems.

     The microscopic rules for the continuously driven OFC model are
given by\cite{granada}
\begin{equation} 
E_{i,j}(t+1) = [1-\Theta (E_{i,j}(t) -
E_{c})]\, E_{i,j}(t) +\varepsilon \sum_{nn}\Theta (E_{nn}(t) -
E_{c})\, 
E_{nn}(t) + \eta(t) 
\end{equation} 
where 
\begin{equation}
\eta(t)=\eta\prod_{k,l} [1-\Theta (E_{k,l}(t)-E_{c})] 
\end{equation} 
is a global perturbation.  In simulations, one uses $\eta = E_c -
E_{max}(t)$ where $E_{max}(t)$ is the maximal value of all the
dynamical variables in the lattice after the avalanche has come to a
halt. To simulate this model a very efficient algorithm has been
proposed by Grassberger.\cite{pre49.2436} 

     The only difference between the Zhang model and the OFC model is
the way of driving. The uniform (global) driving must be a crucial
element in driving nonconservative models to
criticality.\cite{tesikim} To move continuously between
stochastically and uniformly driven systems, one can perturb the
system randomly by adding a fixed amount $\delta E$ to a randomly
chosen site $(i,j)$, that is, $E_{i,j} \rightarrow E_{i,j} + \delta
E$.  If the value at any site exceeds the threshold value $E_{i,j}
> E_{c}$, the system relaxes according to Eq.  (\ref{sbrelax}).
The limit of small $\delta E$ is the limit of continuous drive.
Figure\ \ref{zhang} shows the distribution function of avalanche
sizes $P(s,L)$ for $\varepsilon = 0.20$ and various values of $\delta
E$.
The avalanches are localized for ``large'' values of $\delta E$.
The distribution function is essentially an exponentially decreasing
function and no scale invariance is observed when changing the system
size.  Apparently a phase transition from a localized into a critical
system occurs between Fig.\ \ref{zhang}(c) and
\ref{zhang}(d).  For ``small'' values of $\delta E$ the distribution
function approaches a power-law behavior and the cutoff scales with
system size, implying that the avalanches are no more localized.

\begin{figure}[htbp] 
\centerline{\epsfxsize=4.0truein \epsffile{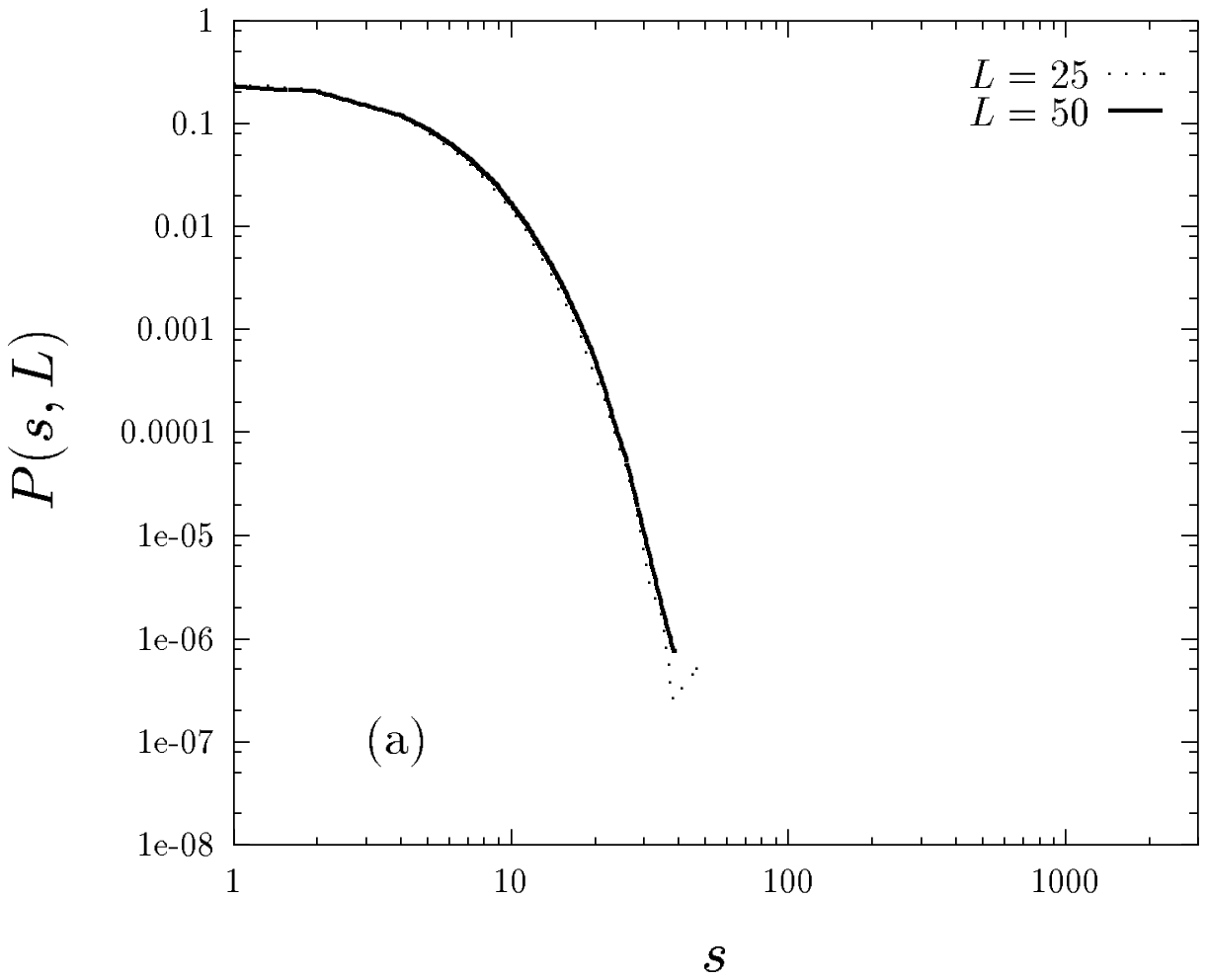}} 
\vspace*{0.2cm}
\centerline{\epsfxsize=4.0truein \epsffile{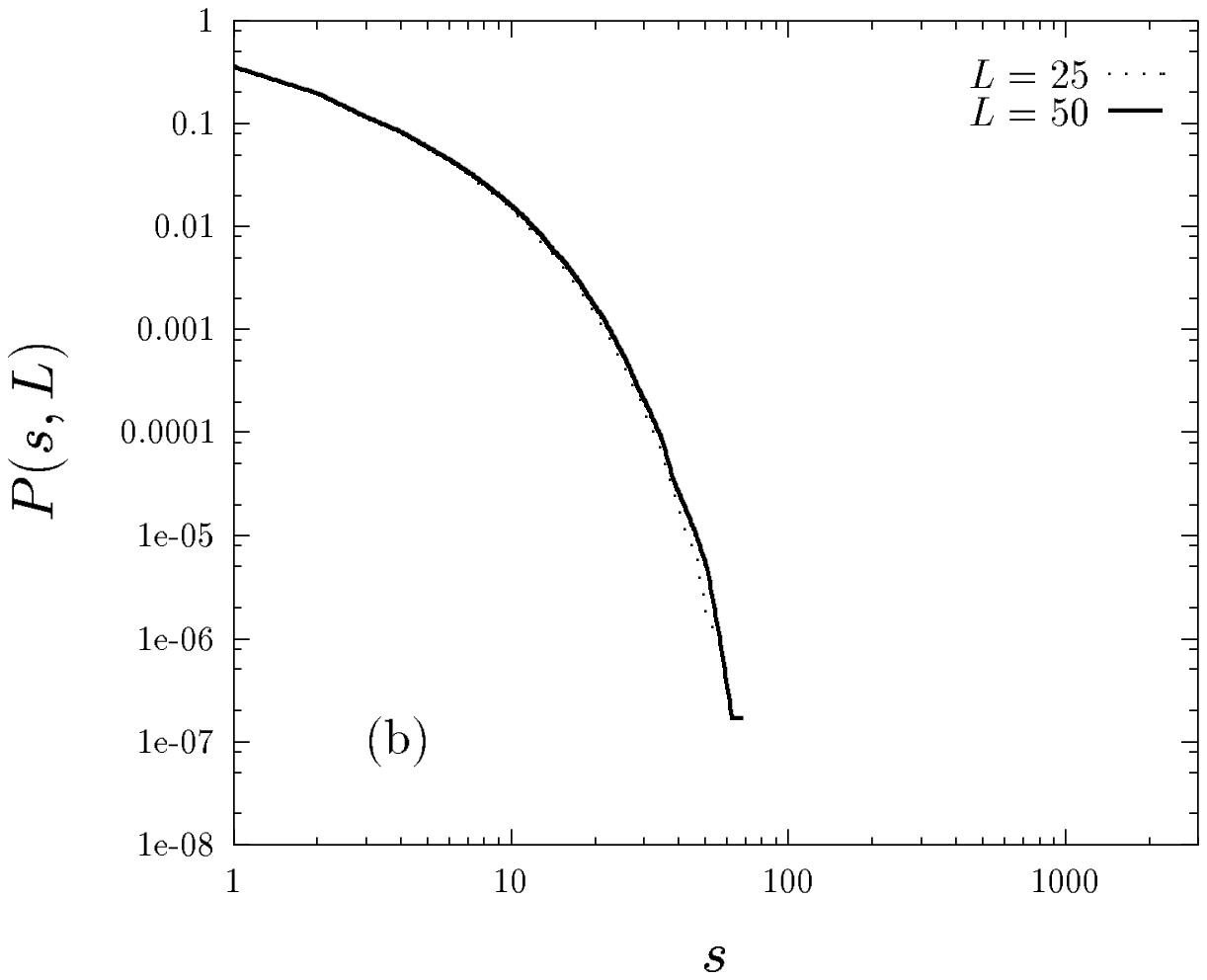}} 
\vspace*{0.3cm}
\fcaption{The nonconservative stochastically driven Zhang model with
$\varepsilon = 0.20$, $E_c = 1,$ and $L = 25, 50$ when (a) $\delta E
= 1$, (b) $\delta E = 0.2$, (c) $\delta E = 0.008$, and (d) $\delta E
= 0.00032$.} 
\label{zhang}
\end{figure}

\addtocounter{figure}{-1}
\begin{figure}[htbp]
\centerline{\epsfxsize=4.0truein \epsffile{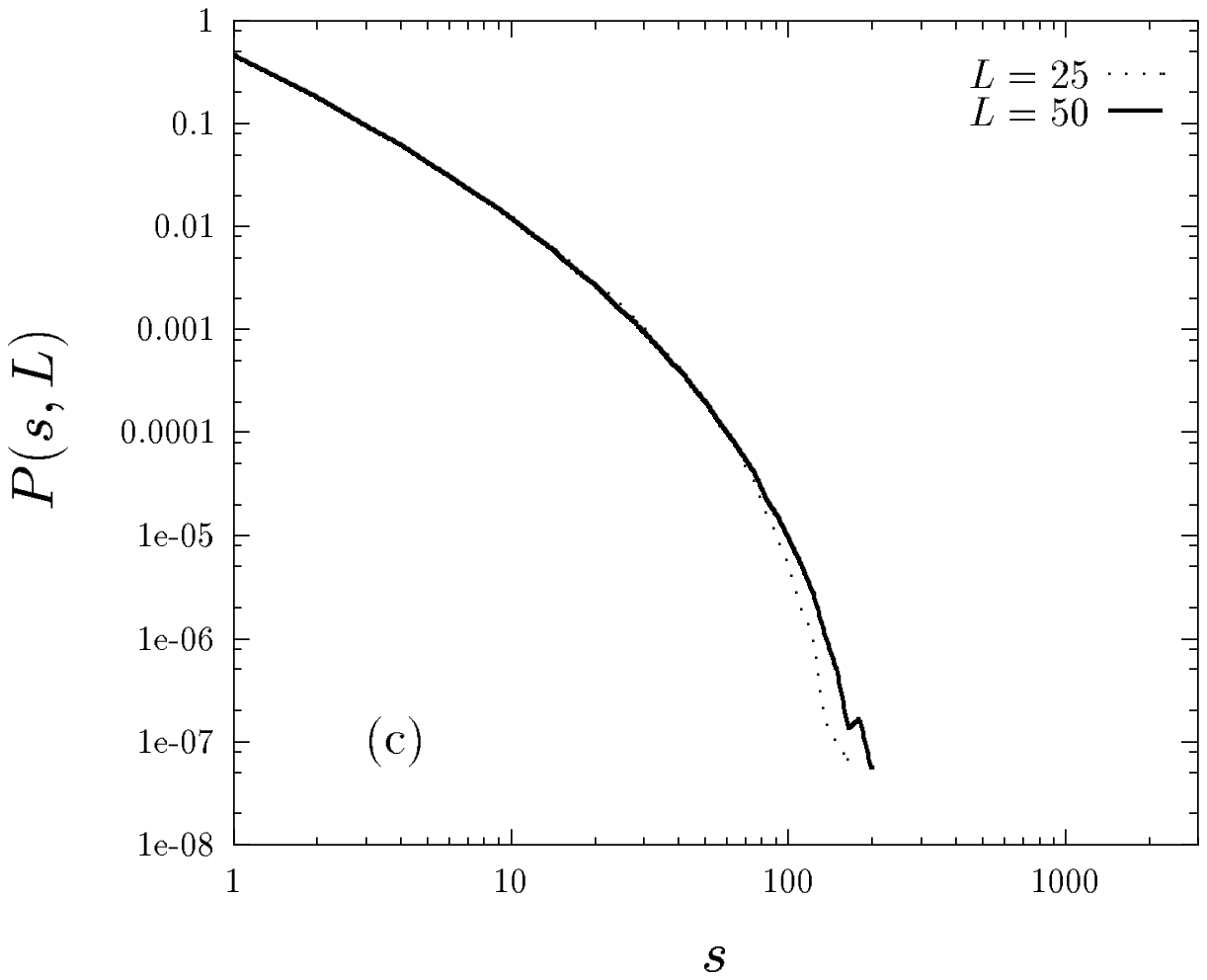}}
\vspace*{0.2cm}
\centerline{\epsfxsize=4.0truein \epsffile{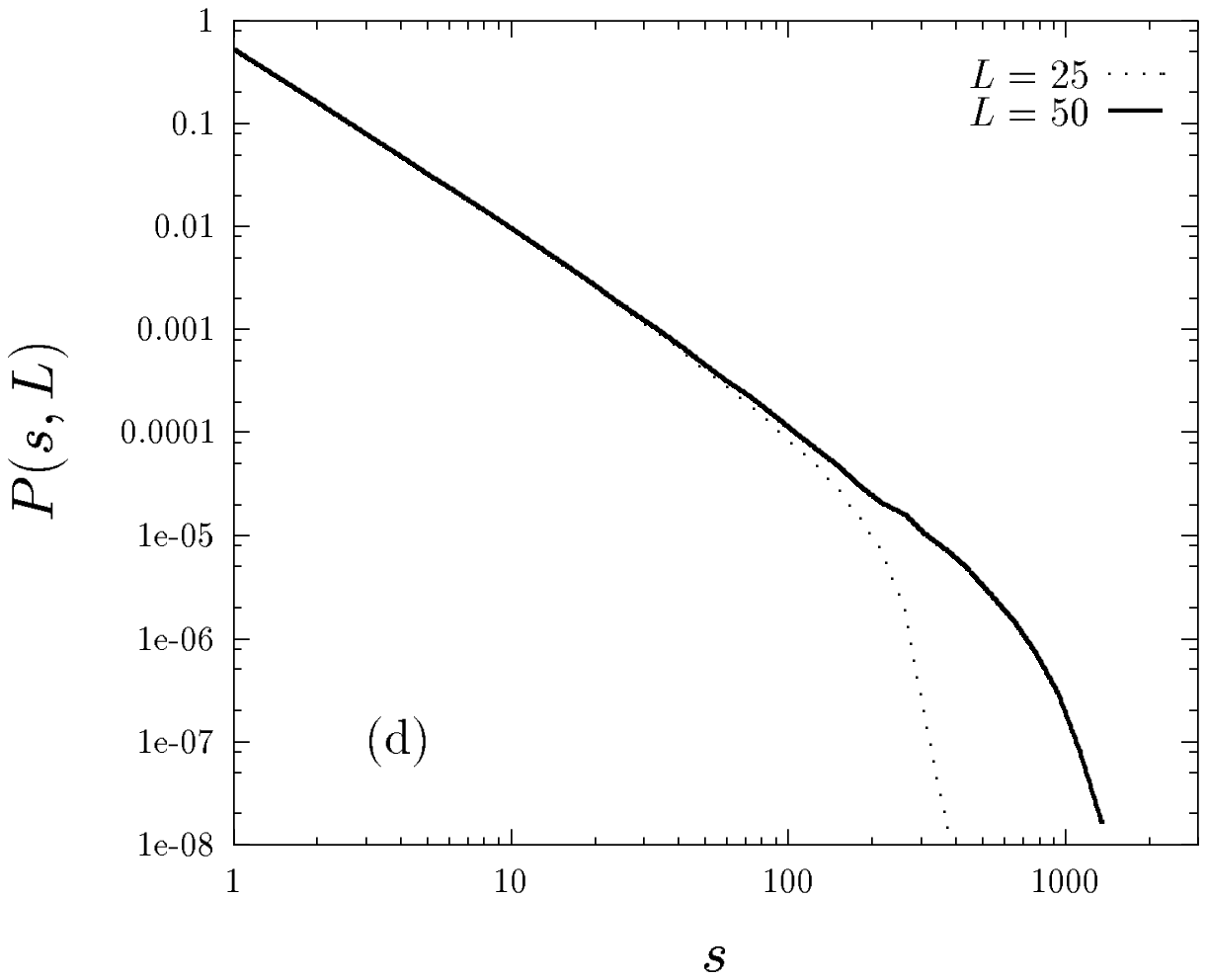}}
\vspace*{0.3cm}
\fcaption{({\em continued})}
\end{figure}

     The random drive represents an effective noise given by
$\sqrt{\frac{E_{c}-E_{max}(t)}{\delta E}} \delta E \propto
\sqrt{\delta E}$.  Obviously, large noise can destroy spatial
correlations in a coupled dynamical system,\footnote{This is seen for
example in the model discussed 
by Manna, Kiss, and Kert\'{e}sz,\cite{jsp22.923} where the
dynamical rule destroys the spatial correlations in the BTW model by
letting the toppling site $z_{i,j} \rightarrow z_{i,j} - 4 + \theta,$
where $\theta \in \{-3, \ldots , 3\}$ is an annealed random variable.
They obtain $\tau_s = 2.515 \pm 0.02$ in perfect agreement with the
analytical result $\tau_s = 5/2$ for models without spatial
correlations.\cite{tesikim,pre48.3361}}\,
but the $\delta E
\rightarrow 0$ represents the weak noise limit where long-range
correlations can appear.\cite{tesikim}
Analytically, the nonconservation introduces a length scale
which in principle one expects to break the scale invariance.
Additionally, the continuous driving is difficult to handle
analytically.  However, an appealing approach would be to treat the
OFC model in the limit of random weak noise driving since this is
conceptually equivalent to a continuous driving.

\section{Synchronization}
\noindent
Synchronization is a fascinating phenomenon observed in
biological, chemical, and physical systems that has captivated
the attention of many scientists in the last century. The
paradigmatic example is observed in some forests of South Asia
where, at night, myriads of fireflies are sitting in the trees.
At the beginning, the members of the population emit flashes of
light incoherently but after a short period of time the whole
swarm is flashing in unison creating one of the most striking
visual effects ever seen. But mutual coherence in the temporal
activity of a group of units has been observed in other
contexts. Some examples are the cell mitotic cycle, heart
pacemaker cells, circadian pacemakers (biological clocks),
neurons in the visual cortex, menstrual period of women,
Josephson junctions, or some chemical reactions, just to cite a
few.  Winfree's book\cite{winfree} is an excellent reference to
read about these systems.

     The relevance of synchronization has been stressed frequently
although not always fully understood. In the case of fireflies it may
help the courtship between male and female.\cite{qrb63.265,s82.151}
In the heart, the impulses coming through the vagus nerve trigger the
contraction of the heart only if they are properly
synchronized.\cite{pt.feb94.40} In other cases the relevance is still
a matter of discussion. There are some evidences which support that
coherent oscillations play an important role in sensory processing.
It has been suggested that the discrimination and segmentation of
objects, so often performed by living systems, can be explained by
analyzing the temporal firing patterns of different neurons in the
visual cortex.\cite{bc54.29,bc60.121,pnas86.1698,n388.334}
 
     Which mechanisms are capable of generating a collective
synchronous behavior? It is accepted that in order to observe a
global coherent activity, oscillatory interacting units are
required. 
The rhythmic activity of each element is provoked either by some
internal processes or by external sources (external stimuli or
forcing). The internal processes may have a physical or biochemical
origin of great complexity, which probably is different for any of
the systems considered above, but the essence of the phenomenon can
be explained in terms of basic principles which allow to create a
common framework. Within this environment, a simple and general
mathematical description can be formulated.

     Suppose that the rhythmic activity of each element can be
described in terms of a magnitude which evolves regularly in time.
When such variable reaches a certain threshold value the unit emits a
pulse (action potential for neurons) which is transmitted to all the
members of the population connected to it. Moreover, a resetting
mechanism initialize the state of the unit that has fired. Then, a
new cycle recommences. Essentially the behavior is analog to an
oscillator. Assuming that the period of the rhythm is $T$, it is
convenient to define the concept of phase, without loss of generality
defined in $[ 0,1 ]$, which is a periodic measure of the elapsed
time. In general, the phase is a nontrivial function of the state of
the considered element. These features define the so-called
relaxation or integrate-and-fire oscillators (IFO's).

     Essentially, the effect of the emitted pulse is to alter the
current state of the neighbors by modifying their periods. This
change is not uniform, sometimes the period is lengthened and
sometimes it is shortened. The perturbation depends on the current
state of the oscillator which receives the external impulse. The
modification induced by such a pulse can also be studied in terms of
a phase-shift, what has been called phase response curve (PRC).
Several different PRC's have been reported in the literature
depending on the particular system under
study.\cite{jmb12.13,bc43.157,jmb24.291} In general, the analysis of
the collective behavior of the system can be done in terms of a PRC
if the phase-shift provoked by an impulse is independent of the
number of impulses arriving within an interspike interval and if the
arrival of such an impulse affects the period of the current interval
but does not modify future intervals. This fact fostered a bunch of
papers published in the seventies and the beginning of the eighties
devoted to analyze the different patterns of entrainment between
pairs of pacemakers. However, a systematic treatment of the whole
population was still missing.

     There is another type of models where synchronization effects
have been studied extensively. In these models each member of the
population is modeled as a nonlinear oscillator moving in a globally
attracting limit cycle of constant amplitude.  These are the phase,
or limit cycle, oscillators. The units interact weakly to ensure that
any perturbation does not move any of them away from the limit cycle.
Then, only one degree of freedom is necessary to describe the dynamic
evolution of the model.  Perhaps, the best known example is the
Kuramoto model,\cite{Kurabook} where the phase $\varphi_i$ of each
oscillator obeys the following Langevin equation 
\begin{equation}
\dot{\varphi_i} = \omega_i + \sum_{j=1}^N J_{ij}\sin(\varphi_j -
\varphi_i) + \eta_i(t),
\end{equation}
where $\omega_i$ is the intrinsic frequency of each member, and
$\eta_i$ denotes a gaussian white noise of zero mean and correlation
given by
\begin{equation}
<\eta_i(t) \eta_j(t')>= 2D \delta_{ij}\delta (t-t').
\end{equation}
When there is no dispersion in the values of $\omega_i$ this model
can be transformed into the planar $XY$ model, well known in
statistical mechanics. However, in the context of oscillators
$\omega_i$ is picked up from a random distribution. Note that the
main difference between these phase-coupled oscillators and the
pulse-coupled IFO's explained before is that now the coupling in not
pulsatile but it acts continuously in time, depending only on the
phase difference between units.

     For ferromagnetic all-to-all interactions $(J_{ij} = J/N, J>0)$,
there is a critical value of the coupling strength $J$ for which a
phase transition occurs between a state where all the oscillators run
incoherently with its own frequency to another state where a certain
degree of synchronization appears
spontaneously.\cite{ptp79.223,ptp75.1319} The nature of such
transition depends on the distribution of
frequencies,\cite{jsp63.613,jsp67.313} on the features of the
coupling,\cite{prl68.1073,el26.79,ptp88.1213,ptp89.929} and on the
strength of the noise $D$.\cite{ptp79.39,jsp70.921} In this review
our interest will be focussed in the pulse-coupled model although
under certain conditions it is possible to convert a model of IFO's
into a model of phase-coupled oscillators as has been shown
by several authors.\cite{jpa23.3835,schuster,cn4.193} 

     In general, three different levels of synchronization can be
observed when considering populations. The strongest implies that all
the oscillators have the same phase.  When the members of a group do
not fire in unison but keep a fixed phase difference between them, a
''phase locking'' regime is achieved. Finally, the weakest situation
appears when several oscillators run with the same average frequency
but without keeping any fixed phase difference. This is called
''frequency locking''.

     A particularly interesting task was developed by Winfree who was
one of the first authors who considered the population as a whole and
who gave mathematical tools to tackle this problem.\cite{jtb16.15}
His ideas were picked up by other scientists who developed them more
deeply or proposed alternative mathematical techniques to solve
problems which remained open for decades.\cite{jmb22.1} Mean-field
models are sufficiently simple to be solved exactly and analyze under
which conditions any of the three levels of synchronization defined
above can be observed.

\subsection{Mean-field models}
\noindent
We will consider systems formed by an assembly of $N$ identical IFO's
with all-to-all couplings. Each member of the population is described
in terms of a phase variable $\varphi$ and a state variable $E$ which
evolve in time according to the following equations 
\begin{equation}
\frac{d \varphi_i}{dt}=1
\end{equation}
and
\begin{equation}
\frac{dE_i}{dt} = f(E_i) + g(E_i,t),
\label{fidete}
\end{equation}
where $f(E_i)$ is the driving rate, which gives the natural evolution
of the oscillator, and $g(E_i,t)$ is a function that accounts for the
effect of the coupling between unit $i$ and the rest. It may have a
complicated dependence on time. In this description self-interaction
is not considered but an explicit dependence of $g(E_i,t)$ on the
current state of the oscillator which receives the incoming input.
The dynamics follows the essential trends mentioned in the
introduction for relaxation oscillators.  When an oscillator reaches
the threshold it is reset $(E\rightarrow 0, 
\varphi\rightarrow 0)$ and the rest of
oscillators suffer a perturbation of their internal state.  The
features of this perturbation depend on $g(E_i,t)$. 
This description is
quite general and includes the majority of models of
integrate-and-fire oscillators studied so far. Moreover, it allows a
straightforward generalization to models with nonidentical periods.
We can see in this dynamics a first contact point with models which
display SOC behavior.

     A particularly interesting case was considered by Mirollo and
Strogatz (MS).\cite{sjam50.1645} Their work was inspired in the
analysis developed by Peskin\cite{peskin} for the cardiac 
pacemaker, though the main outcomes are general enough to be applied to a
wide range of models. MS consider systems whose intrinsic dynamics is
given by a driving rate $f(E)$ that satisfies
\begin{equation}
f(E) >0 \hspace{5 mm} \mbox{and} \hspace{5mm} f'(E)
= \frac{df(E)}{dE}<0, 
\hspace{5mm} \forall E,
\label{condri}
\end{equation}
that is, the state variable $E$ and the phase $\varphi$ are related
through a convex function $E(\varphi)$, hereafter called the driving
(see Fig.~\ref{figpes}). 

\begin{figure}[htbp]
\centerline{\epsfxsize=3.0truein\epsffile{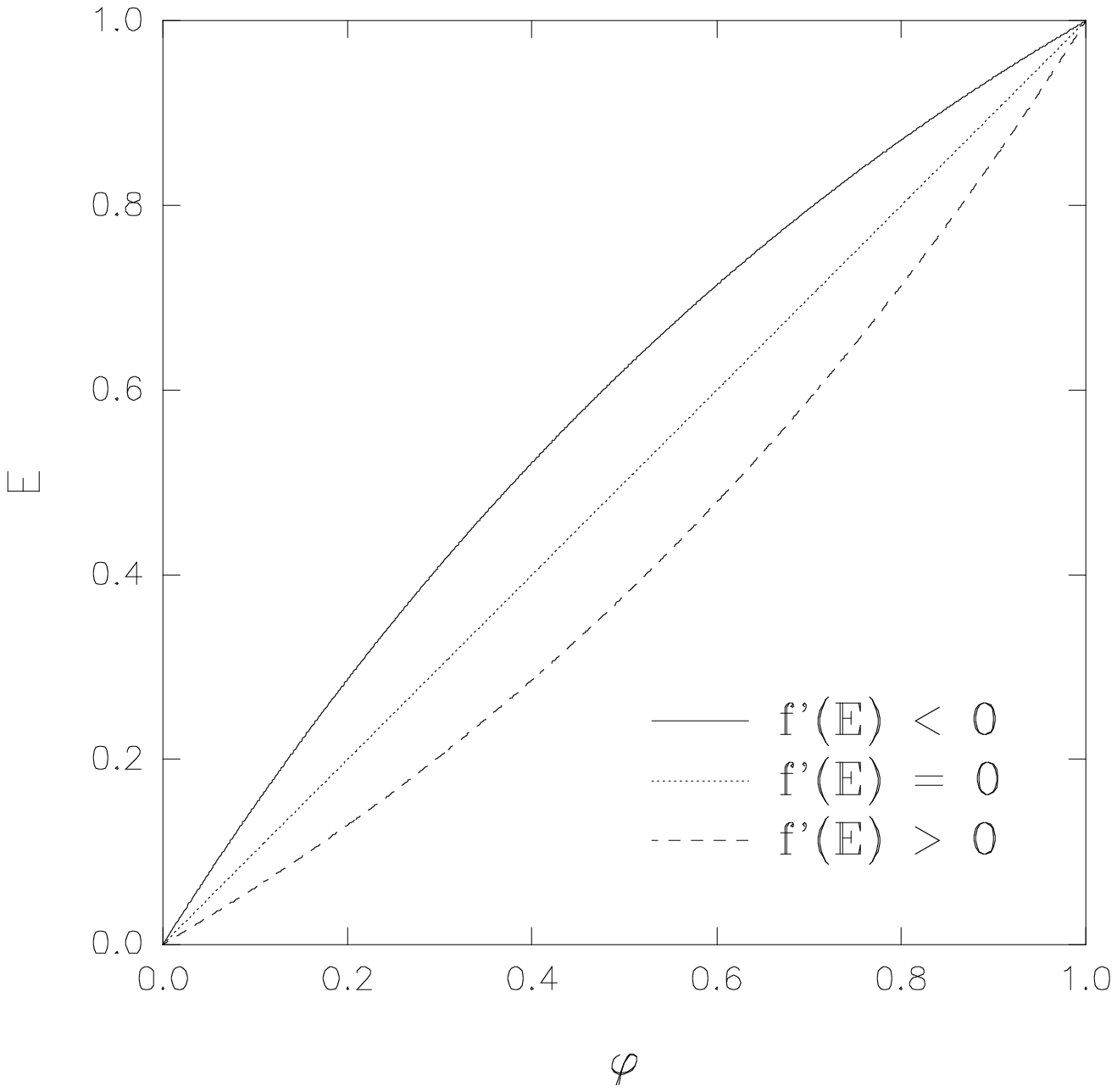}}
\vspace*{0.5cm}
\fcaption{Functional relation between the state of a unit $E$, and
its phase $\varphi$, for different driving rates corresponding to a
convex driving ($f'(E)<0$), linear driving ($f'(E)=0$), and concave
driving ($f'(E)>0$), respectively.}
\label{figpes}
\end{figure}

\noindent As an example, for the Peskin model\cite{peskin}
\begin{equation}
f(E) = \gamma (C-E),
\label{peskdri}
\end{equation}
where $\gamma$ is a constant that gives information about the
convexity of the driving and $C=\frac{1}{1-e^{-\gamma}}$ ensures that
the period is one when the threshold is one. This simplified model
accounts for the variation of the membrane potential $E$ (voltage
difference between the internal and external part of a cell), as a
consequence of the transport of Na, K and other ions across the
membrane channels. The system is simply represented as an $RC$
circuit, which implies $\gamma >0$.\cite{jppg9.620,Tuck}

On the other hand, in the MS model, 
the coupling between the oscillators is given by
\begin{equation}
g(E_i,t) = \varepsilon \sum_j \delta(t - t_j),
\label{coup}
\end{equation}
where the sum involves all the oscillators which have fired at a
given moment ($t_j$ denotes the time at which the $j$-th oscillator
fires). An equivalent description can be made in terms of the state
variable $E$
\begin{equation}
g(E_i,t) \propto \varepsilon \sum_j \delta(E_c - E_j),
\end{equation}
where now the sum runs over all the oscillators and $E_c$ is the
threshold.  The coupling is instantaneous, excitatory ($\varepsilon
>0$), and constant since there is no dependence on $E_i$, which means
that when a cluster of $m$ members reaches the threshold
simultaneously, the state of the other units is changed immediately
according to
\begin{equation} 
E \rightarrow E + m \varepsilon .
\label{additivity}
\end{equation}
One of the effects of such pulse-like interaction is that when one
oscillator fires other members can also reach the threshold. These
oscillators may in turn push other elements up triggering an
avalanche. In this case all the units which have fired merge in a
group which does not break up anymore. This effect is called
absorption. The concept of absorption implies the existence of a
refractory time, a period where the oscillator which has fired is not
sensitive to new incoming inputs. It has a crucial effect when the
interaction is modeled in terms of a delta function, as we will see
later. Notice that a coupling given by (\ref{coup}) implies the
existence of two time scales, a slow scale for the intrinsic dynamics
of the units and a fast one, indeed infinitely rapid when compared
with the other, for the interaction between IFO's. Once again we can
establish a connection with models displaying SOC. In terms of this
time scale separation we can say that the refractory time only acts
in the fast time scale.

     MS were able to prove that for any $\varepsilon > 0$ the
population of oscillators will be firing in synchrony in the
stationary state for almost any initial condition (except in a 
set of Lebesgue measure zero) 
provided the driving is convex, i.e., when Eq.
(\ref{condri}) holds. In fact, this result can be derived even with
more restrictive conditions given by
\begin{equation}
g(E_i,t) = \varepsilon \delta(t - t_j).
\label{notaddi}
\end{equation}
According to this description the strength of the pulse emitted by a
group of oscillators is independent of its size, that is, there is no
additivity in the coupling and therefore $m=1$ in Eq.
(\ref{additivity}). This result is very powerful because
(\ref{notaddi}) does not favor synchronization, just the opposite.
When additivity is taken into account, the process of entrainment is
accelerated and perfect synchrony can be found under broader
conditions than the given by the MS theorem.\cite{tesikim,prl74.4189}

     Note that there is no contradiction in these results because the
theorem only gives sufficient but not necessary conditions to
synchronization (Eq. (\ref{condri})). Figure
\ref{alltoall} illustrates this situation. 

\begin{figure}[htbp]
\centerline{\epsfxsize=4.0truein\epsffile{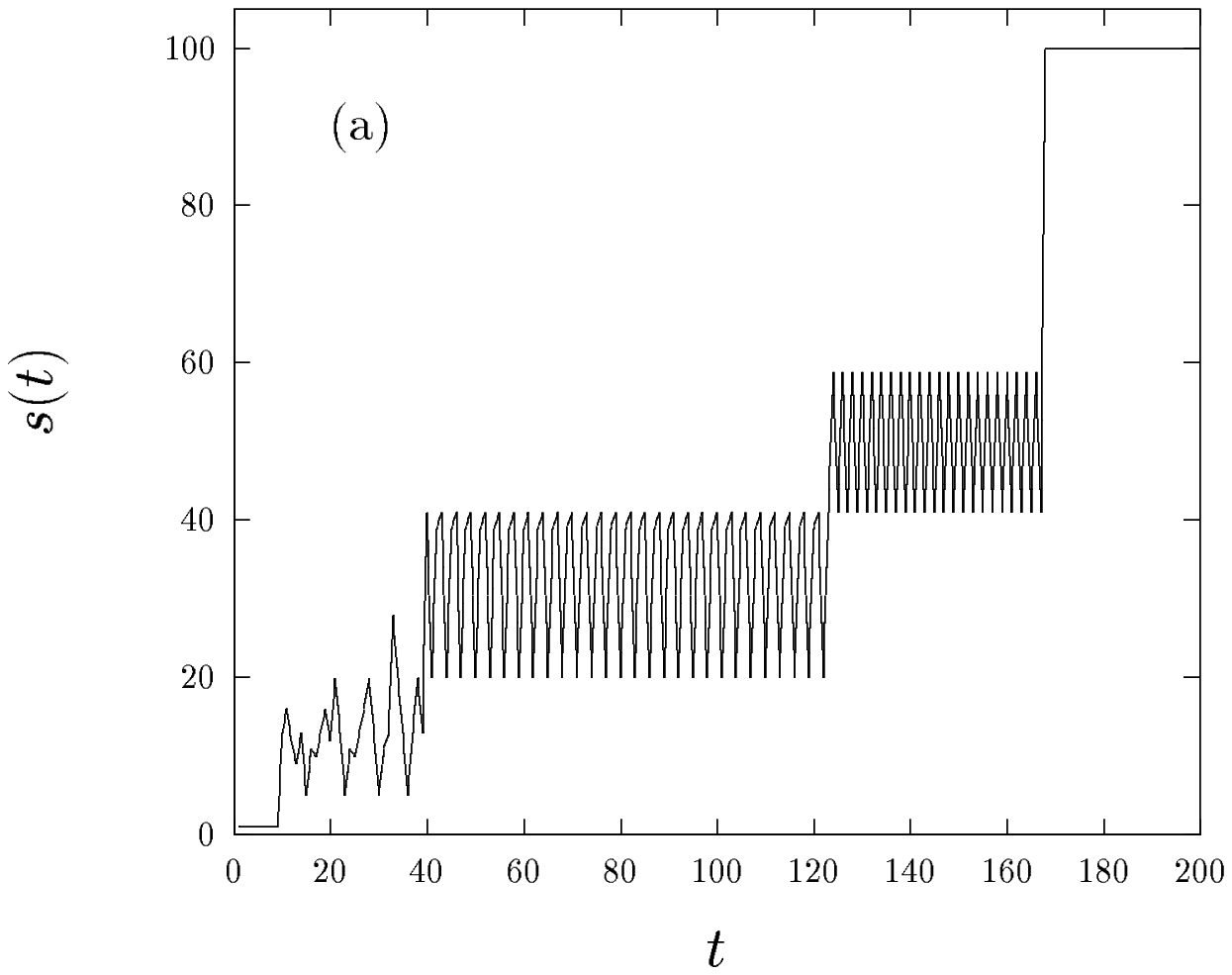}}
\vspace*{0.2cm}
\centerline{\epsfxsize=4.0truein\epsffile{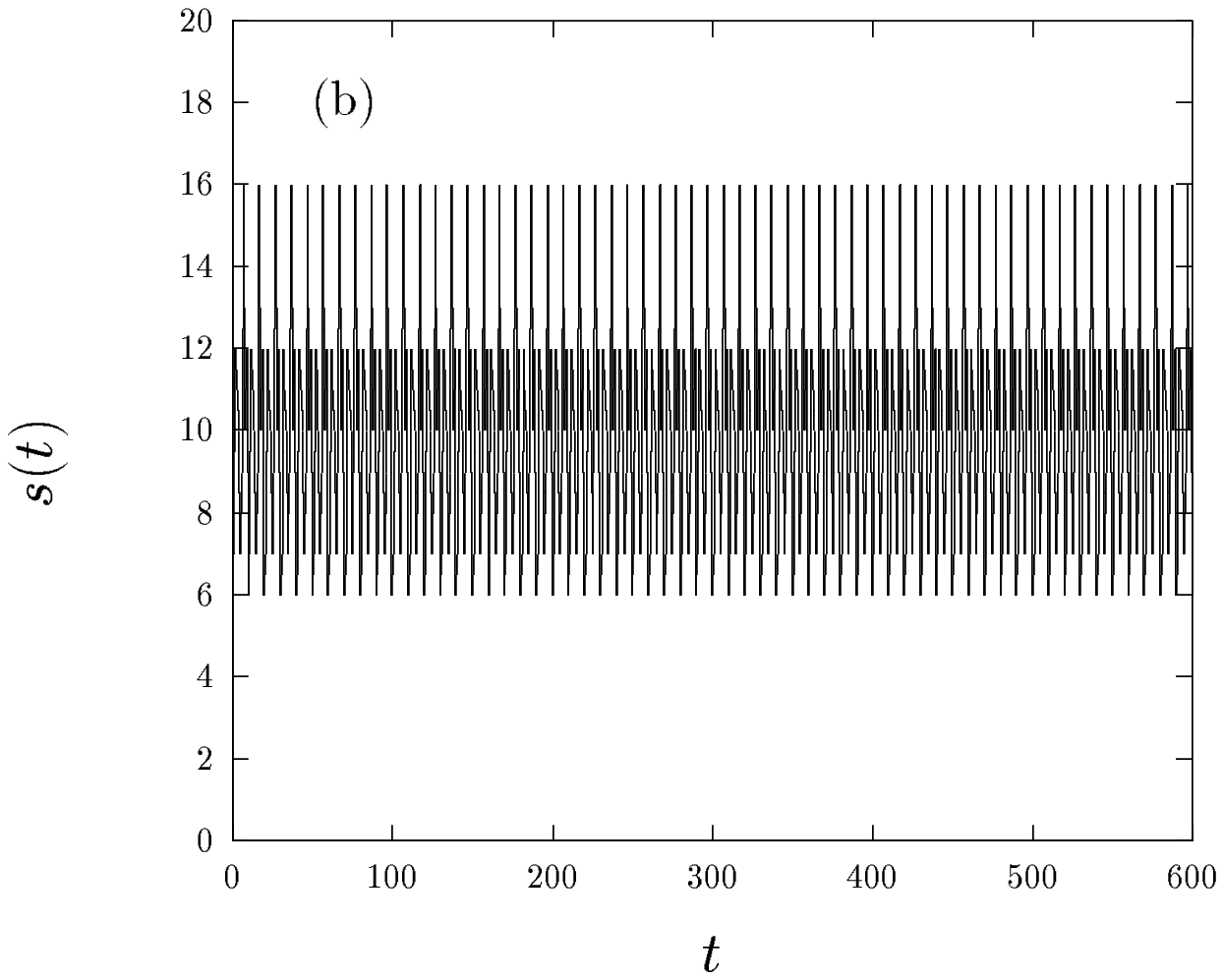}}
\vspace*{0.3cm}
\fcaption{Time evolution of the size of the "avalanches" 
for the Peskin
model with all-to-all coupling, $\varepsilon =0.1$, and $N=100$. (a)
A convex driving, $\gamma =0.1$, leads to synchronization, and (b)
with a concave driving, $\gamma =-0.1$, the system settles in a
periodic state which is very sensitive to initial conditions. In
these simulations we have considered the extreme situation given by
Eq.  (\protect\ref{notaddi}). Time is measured in number of
avalanches.}
\label{alltoall}
\end{figure}

     The MS model has been generalized to other situations. It has
been studied in presence of weak noise\cite{p50d.15} and with a
random distribution of thresholds.\cite{pre49.2668} Some studies have
emphasized the necessity to overcome some unrealistic features of the
model. For instance, the instantaneous character of the coupling does
not take into account neither the finite time of the synaptic
response nor the finite time associated with the transmission of
information propagating along the axons (delays). Both are relevant
in models of spiking neurons and make the collective behavior much
richer.\cite{pre48.1483,n4.259,prl71.1280} 
As an example, recent studies have
shown that under certain conditions inhibitory couplings rather than
excitatory lead to
synchronization\cite{pre51.738,jcn1.313,prl74.1570} 
in
agreement with some experimental observations.\cite{nc4.84} These
results contrast with the MS model for inhibitory coupling
($\varepsilon < 0$). A straightforward revision of their proof shows
that the sufficient conditions to find synchronization in this case
are $f(E)>0$ but $f'(E)>0$, $\forall E$.

     Another underlying hypothesis that restricts the range of
interest of the MS model concerns the strength of the coupling
$\varepsilon$ which does not depend on the state of the oscillator
which receives the pulse.  As we have mentioned above, in many
studies of biological pacemakers it is assumed that the response of a
cell due to an external stimulus is a phase shift (given by the PRC)
which induces an energy-shift $\varepsilon(E)$ in Eqs. (\ref{coup})
or (\ref{notaddi}), hereafter called the energy response curve (ERC),
that, in general, is a nonconstant function of $E$. It is possible
to generalize the MS theorem by considering a general ERC as well as
an arbitrary driving rate, with the only restriction of having $f(E)
> 0$. Then, a sufficient condition to find a stable synchronous
regime is given by\cite{PRL2}
\begin{equation} 
\frac{f'(E)\varepsilon (E)}{f(E)} < \varepsilon '(E),
\hspace{5mm} \forall E.
\label{conditions} 
\end{equation} 
This means that, in general, the knowledge of the intrinsic dynamics
(i.e. the driving) of an individual oscillator is not enough to
predict the final state of the population. Information about the
response of a given unit in presence of a external stimulus is also
required.

\subsection{Short range models}
\noindent
The question whether the conclusions extracted from the MS theorem
can also be applied to lattice models with short-range connectivity
is of great importance. First, notice that in mean-field models
synchronization emerges in an absorption process, where clusters of
oscillators of increasing size merge with each other and never break
up. However, in a lattice model, big assemblies of oscillators with
the same phase, which eventually may break up, are generated by means
of avalanches that start at a given point and propagate through the
lattice. When they sweep the whole system we call them relaxation
oscillations. Therefore, due to the different nature of both
mechanisms one would expect that the conditions required to find
synchronization/relaxation oscillations are different. In general, it
is not easy to get analytical results for short-range models.  Only
for very special bc and for systems of low dimensionality
(essentially $1D$) exact results have been derived.\cite{ptp92.1039}
Therefore, in the majority of situations the main outcomes rely on
computer simulations. As we will see later, they show that under
certain conditions mean-field arguments can be extrapolated to
systems with finite connectivity.  

     Let us consider a lattice model of IFO's with local rules given
by (\ref{fidete}) and (\ref{coup}) (but now the sum runs only over
nearest neighbors). An important point concerns the way in which an
avalanche is propagated through the lattice. Suppose that a given
unit, the seed, has fired and as a consequence some of its neighbors
reach the threshold.  In turn, these oscillators will fire,
interacting with their neighbors being the seed among them. The
question is what happens with the seed.  According to the strategy
applied in the mean-field version of the MS model such unit should
remain at the reset point during the whole avalanche. This is
important because such effect implies the existence of a refractory
time which persists until all the oscillators are below the
threshold. Furthermore, all the members which have fired will have
zero phase when the avalanche is over. 
We can describe this refractory time by an ERC that verifies
$\varepsilon (0)=0$.
Under such conditions computer
simulations show that relaxation oscillations appear when $f'(E)<0$
and $\varepsilon$ is a definite positive constant, in agreement with
a conjecture proposed by MS.\cite{sjam50.1645} For a more general
function $\varepsilon(E)$ (but $\varepsilon(0)=0$), it has been 
shown\cite{PRL2} 
that (\ref{conditions}) is a sufficient condition to
observe relaxation oscillations. Moreover, in this case relaxation
oscillations lead to a perfect synchronous regime.\footnote{Note that
for open bc and $\varepsilon(0) \neq 0$, after a relaxation
oscillation of the size of the system not all the units have the same
phase. For instance, the cells located at the boundaries remain at a
different state with respect to those located in the bulk because the
number of neighbors in each case is different.}\, Figure \ref{prl2}
gives numerical evidence of this fact. 

\begin{figure}[htbp]
\centerline{\epsfxsize=3.0truein\epsffile{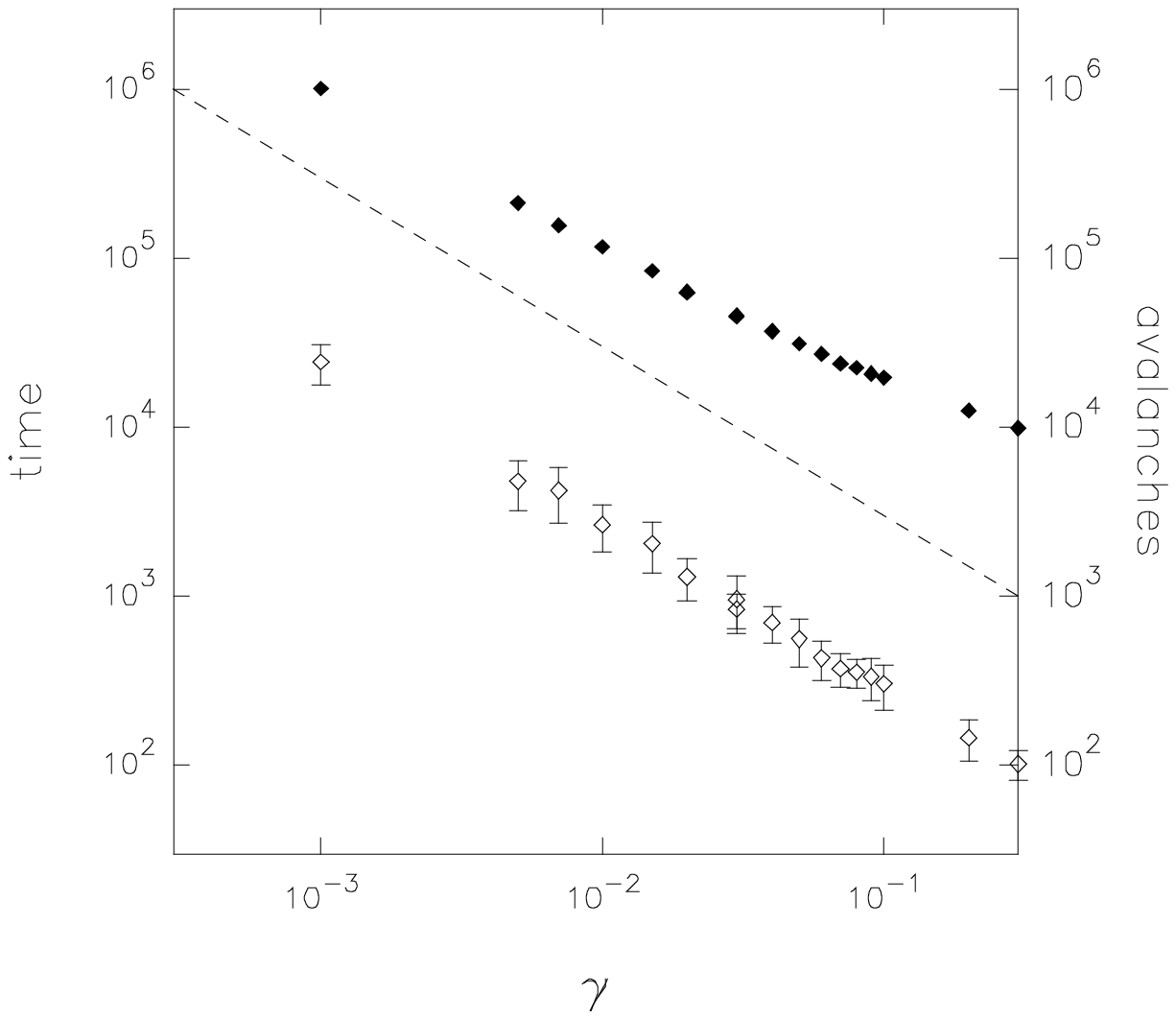}}
\vspace*{0.5cm}
\fcaption{Number of avalanches (filled) and time in period units 
(hollow) required for a $32 \times 32$ lattice of Peskin's
oscillators with open bc to observe complete synchronization as a
function of $\gamma$ in a $\log -\log$ scale. The straight line shows
the $\gamma^{-1}$ behavior. The value of $\varepsilon$ is
constant and equal to $0.1$.
Similar results are obtained for periodic bc and other driving
rates.}
\label{prl2}
\end{figure}
The most appealing consequence of the previous results is that for
coupled dynamical systems like those considered before, for open or
periodic bc, mean-field results seem to be still valid and represent
sufficient conditions to ensure the existence of the synchronized
regime, provided one considers the refractory time.  

     When the driving is linear, the phase and the energy become
identical. In this case, for a constant ERC, 
it is very simple to show that starting from
a random initial configuration the system evolves towards a periodic
state formed by clusters of oscillators with the same phase.
The phase difference, after a cycle, remains constant in time 
once all the units have 
fired, and consequently the system does not synchronize in 
agreement with
Fig. \ref{prl2}. On the other hand, 
a concave driving may give rise to the
appearance of stable complex phase-locked spatio-temporal
structures\cite{PRL2,chinchon,physd} as, for instance, a chessboard-like
configuration. This sort of ''antisynchronized'' states are exclusive of
the short-range models since there is no counterpart in the
mean-field case.

     On the other hand, we could suppose in the previous discussion
that the seed was updated as any other neighboring cell, that is,
$\varepsilon(0)
\ne 0$, so the refractory time is removed. In that case the coupling
between neighbors is governed, when $E_{ij} \ge E_c$, by 
\begin{equation} 
\begin{array}{l}
E_{nn} \rightarrow  E_{nn} +  \varepsilon   \\ 
E_{i,j} \rightarrow  0.
\end{array}
\label{ff}
\end{equation}
These rules are not new. They were introduced in a stick-slip model
by Feder and Feder (FF)\cite{prl66.2669} in the context of earthquake
dynamics, with the restriction of having a constant driving rate
and open bc.

     As in all the models of IFO's discussed so far, since the
relaxing site is reset to zero and a fixed amount is transferred to
the neighbors, the FF model does not have any memory of the initial
configuration. After an avalanche, all the sites that have fired have
an energy which is a multiple of $\varepsilon$. When the threshold
$E_c$ and $\varepsilon$ are commensurable, that is $E_c/\varepsilon$
is an integer $M$, the system eventually reaches a state with $M$
distinct values of the dynamical variables. Hence, after an
avalanche, $E_{i,j} = n \varepsilon, n = 0
\ldots M-1$, and since the driving is uniform, $E_{i,j} = n
\varepsilon, n = 1 \ldots M$ when initiating an avalanche. When $E_c$
is incommensurable with $\varepsilon$, that is, when
$E_c/\varepsilon$ is not an integer, the system is not able to reach
a ''partially synchronized'' state.  Figure \ref{epsilonfig}(a)
displays the average number of totally synchronized sites as a
function of time for $E_c/\varepsilon = 4, 100/22,$ and $5$.  When
$E_c$ is commensurable with $\varepsilon$, the time the system
needs to settle down into such a state increases as
$\varepsilon$ decreases.  
Note that the case of $M = 4$, was originally studied
by Feder and Feder.\cite{prl66.2669} Their results are quite
different. This will be explained in detail in the next section.
Figure\ \ref{epsilonfig}(b) displays the corresponding distributions
of avalanche sizes. When $M = 4$ and $5$ only large avalanches are
seen. For $M = 100/22$, avalanches of all sizes occur, but they are
not power-law distributed.

\begin{figure}[htbp]
\centerline{\epsfxsize=4.0truein\epsffile{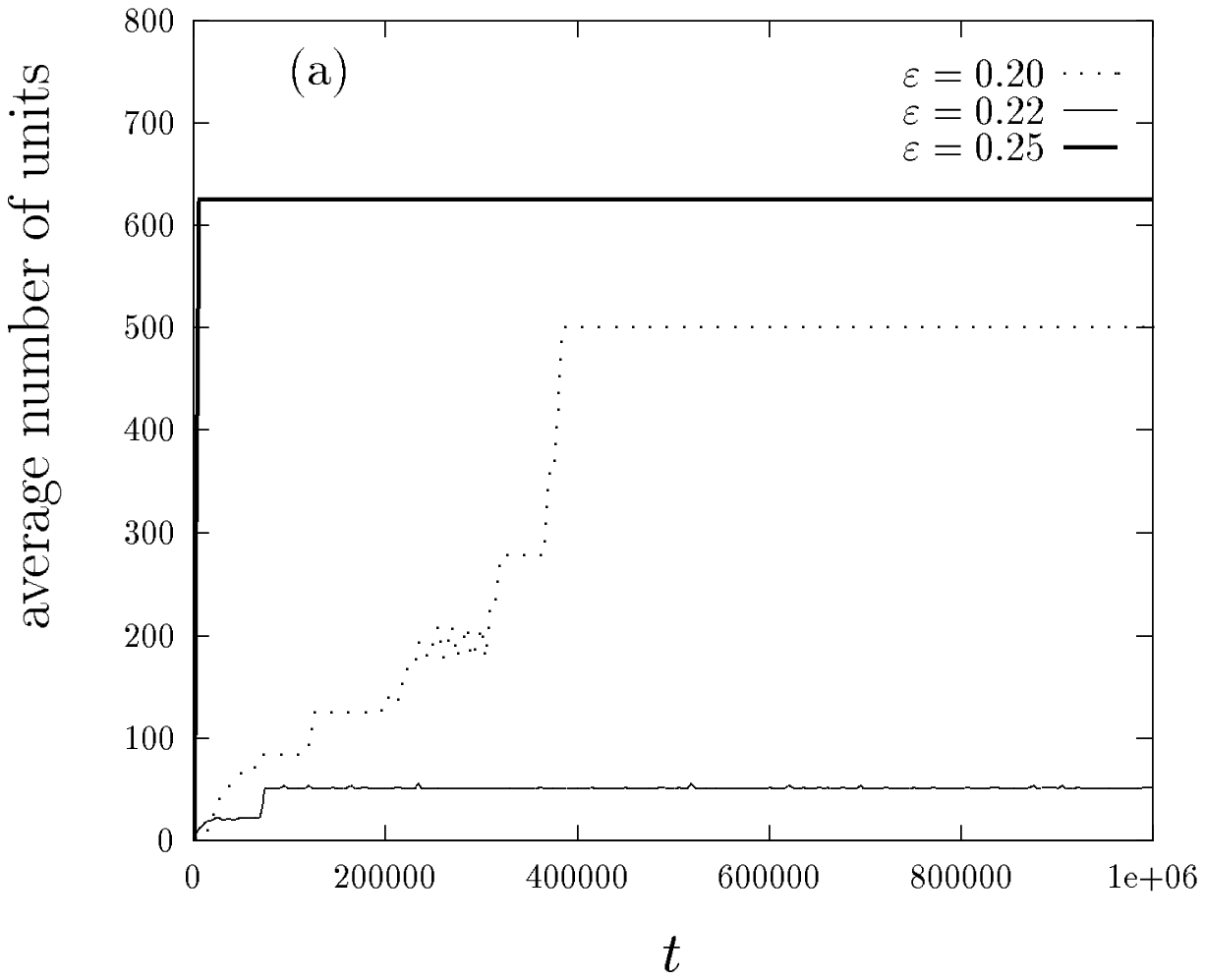}}
\vspace*{0.2cm}
\centerline{\epsfxsize=4.0truein\epsffile{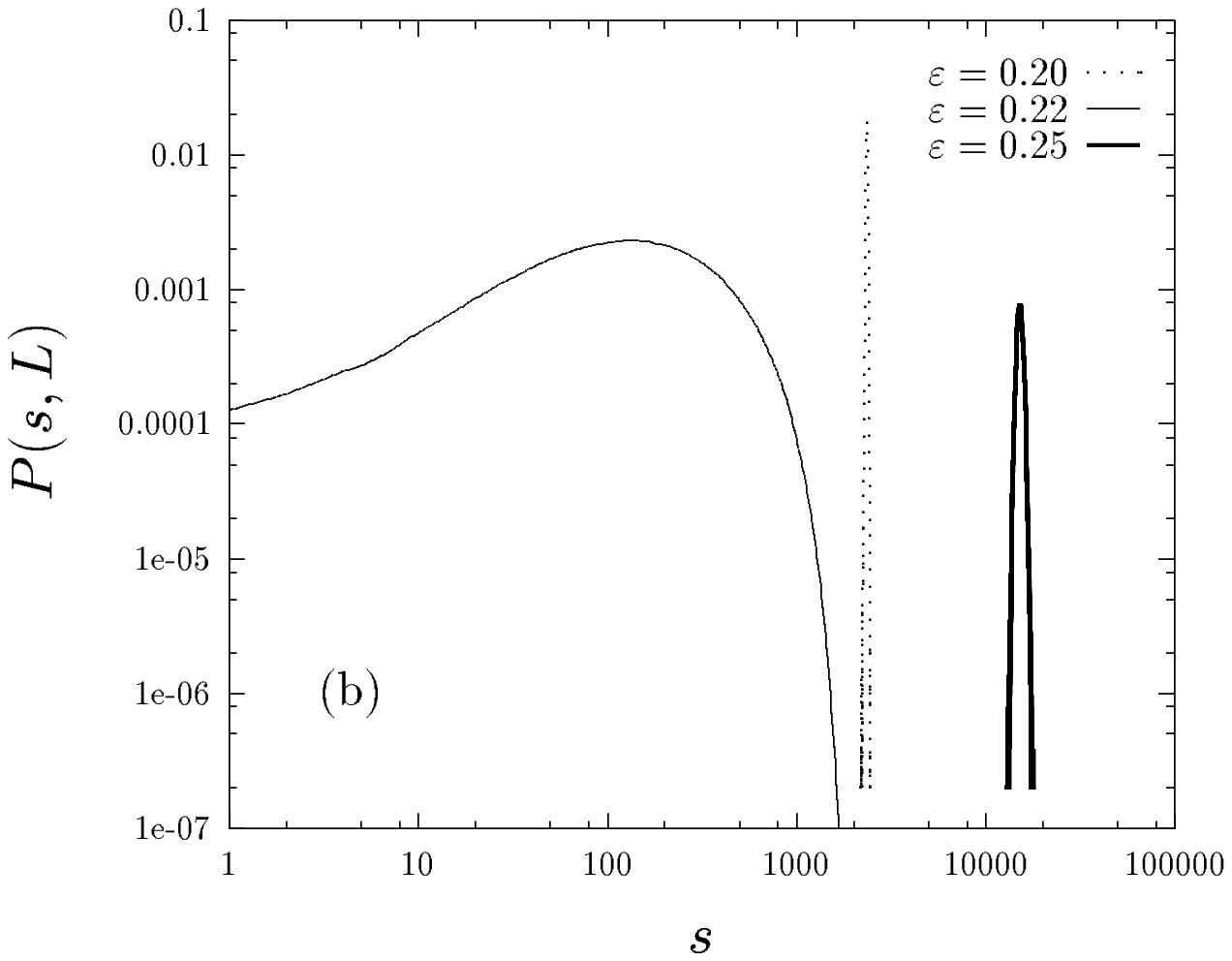}}
\vspace*{0.3cm}
\fcaption{(a) The average number of synchronized sites in the FF
model with $\varepsilon = 0.20, 0.22$, and $0.25$, respectively.
$E_c = 1$ and system size $L = 50$.  (b) The corresponding avalanche
distribution when the system has reached the final
partially-synchronized state. Time is measured in number of
avalanches.}
\label{epsilonfig}
\end{figure}

\section{Synchronization and Self-Organized Criticality}
\noindent
Up to now, we have seen the strong similarities between some models
displaying SOC and lattice models of IFO's showing different levels
of synchronization. In both types of systems, each unit has an
intrinsic dynamics leading to a threshold value. When this threshold
is reached, the unit interacts with its neighbors and it is reset.
This process is developed in a time scale much faster than that
associated to the natural driving of each unit. The features of the
couplings, the natural dynamics, and the reset mechanism define each
particular situation. In view of this analogy it is reasonable to
think that a general framework may be developed. We have shown that
by an appropriate choice of the relevant parameters of these models a
wide spectrum of collective behaviors can be observed. However, up to
now our attention has been focussed on systems which display either
SOC or any sort of phase locking. The goal of this final section is
to give a step forward by showing that there is an interesting group
of models, with quite different origins, which display both
behaviors.

\subsection{Integrate-and-fire oscillators}
\noindent
Let us start the discussion with a model studied in the previous
section.  We have seen that the FF model tends to develop, in the
stationary state, either assemblies of cells with the same phase
(when the ratio $E_c/\varepsilon$ is an integer) in which case the
big avalanches take place because a lot of units reach the threshold
simultaneously or a more complex behavior (when $E_c/\varepsilon$ is
not an integer) where avalanches of all sizes take place, but they
are not power-law distributed.  These behaviors are not robust to
noise.\cite{tesikim} In principle, the effect of the noise is to
prevent that two sites become critical simultaneously due to the
uniform driving. The nature of such noise can be based on the
existence of fluctuations: on the dynamic variables, on the
thresholds, or on the reset energy.\cite{tesikim,prl74.742} This
source of dynamical noise has to be distinguished from a quenched
source of noise, as for instance, a random distribution of
frequencies.\cite{futuro}

     The dramatic effect of noise on the features of the stationary
state of the model is observed in Fig.~\ref{ffpower}. When altering
the relaxation rule (\ref{ff}) slightly by adding a small random
number in the range $[0,0.001]$ to the reset
oscillator,\cite{tesikim} the system no longer settles into a state
with a few groups of oscillators with the same phase, but it goes
towards a new state in which the avalanche distribution follows a
power-law decay characteristic of SOC.

\begin{figure}[htbp]
\centerline{\epsfxsize=4.0truein\epsffile{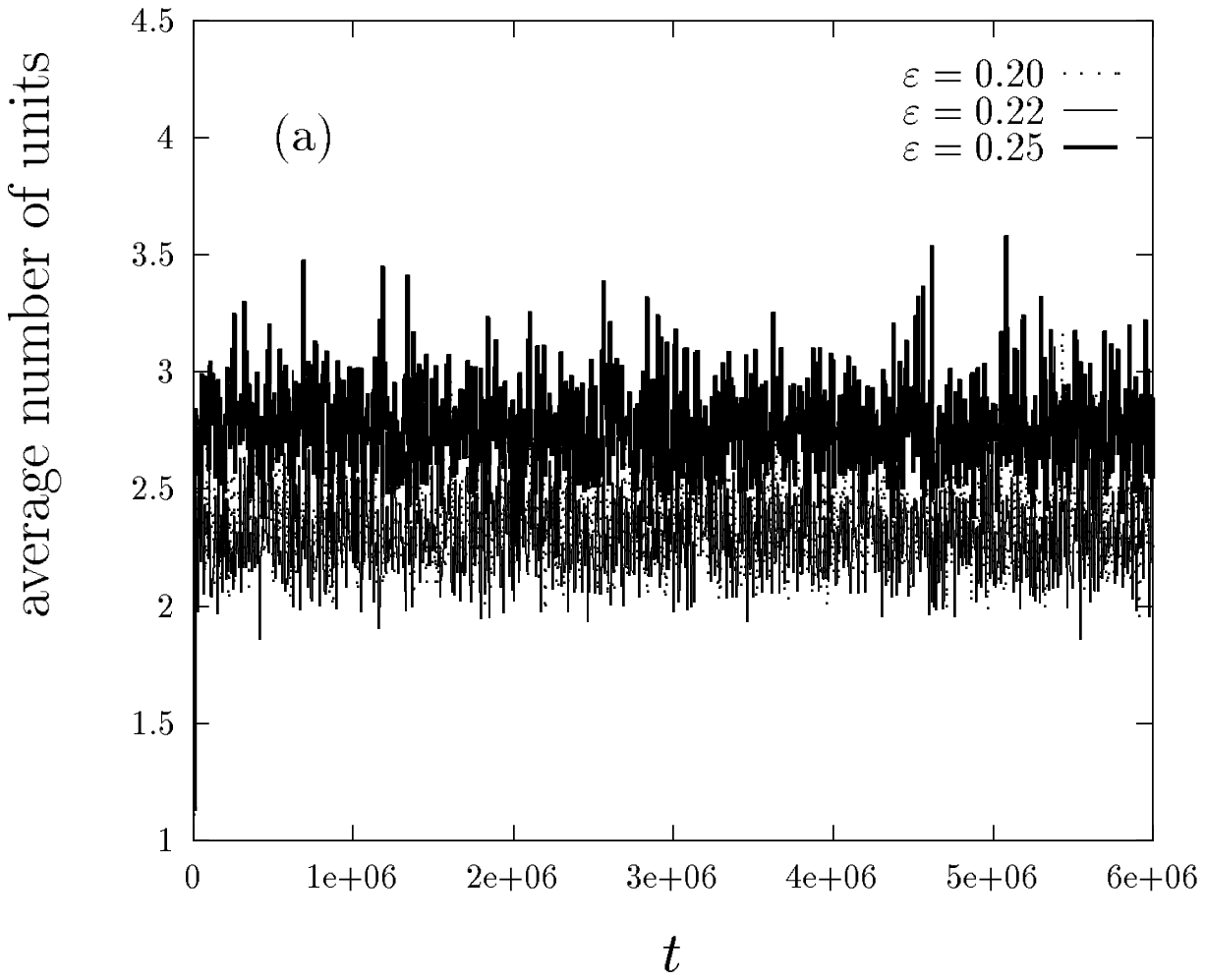}}
\vspace*{0.2cm}
\centerline{\epsfxsize=4.0truein\epsffile{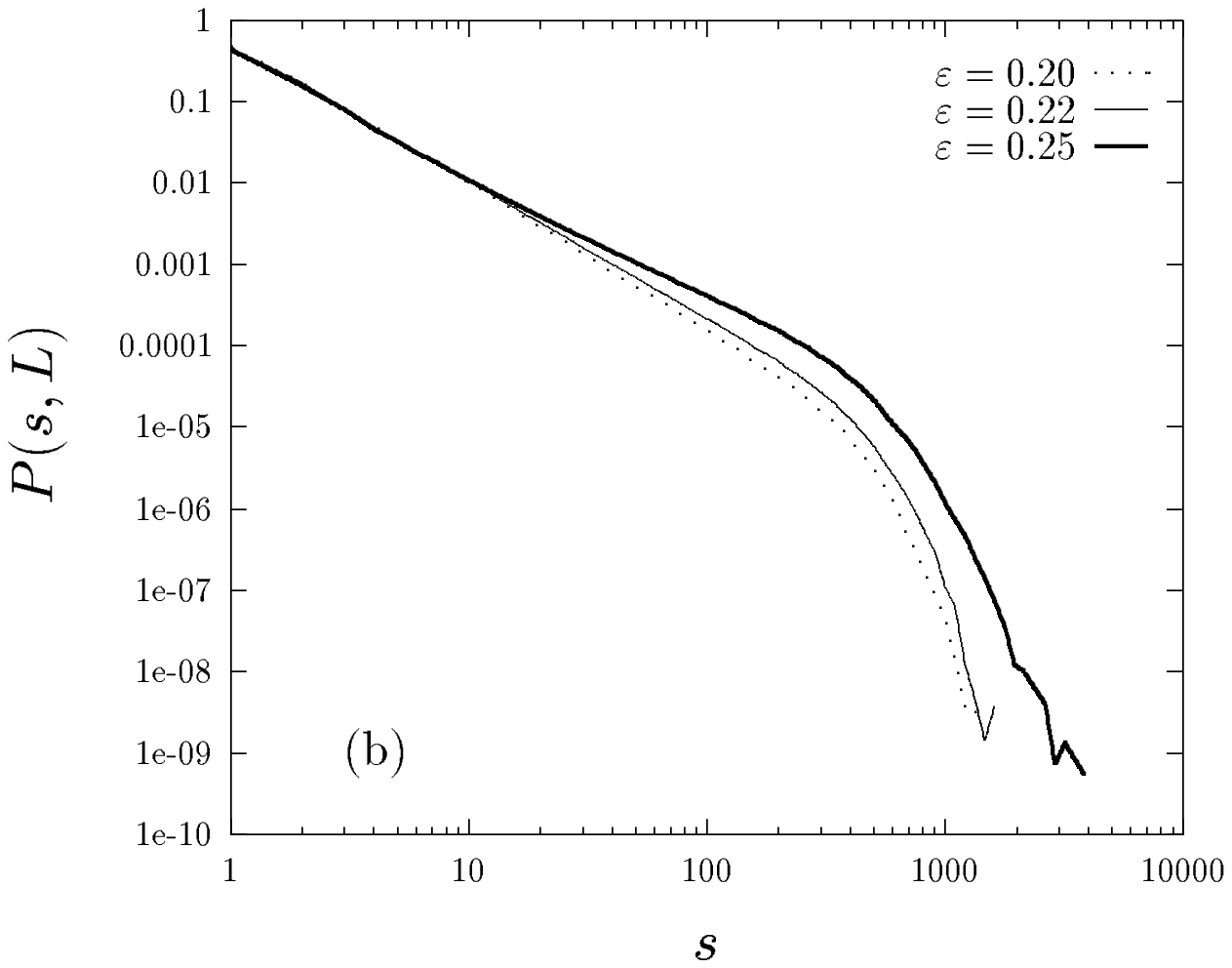}}
\vspace*{0.3cm}
\fcaption{The FF model ($L=50$ and $E_c = 1$) with dynamical noise.
(a) The average number of totally synchronized oscillators. As in
previous figures time is measured in terms of the number of
avalanches.  (b) The distribution of avalanche sizes are power laws
where the slopes change with $\varepsilon$.}
\label{ffpower}
\end{figure}

     Keeping in mind the results given in the previous section for
the MS model, it is not difficult to imagine that by increasing the
convexity of the driving (in the original FF model it is linear) SOC
fades out, even in the presence of a dynamical
noise,\footnote{Corral {\em et al.}\cite{prl74.118} 
introduced the noise by choosing at random
one of the cells that become critical simultaneously due to the
driving and making it the starting point of the avalanche, while the
energy of the rest of critical cells was decreased by a small random
amount.  However, different sources of dynamical noise give
similar 
results.}\, 
and for a certain degree of convexity a spatial structure
formed by large groups of oscillators with the same phase appears
spontaneously.\cite{prl74.118} In fact, the features of the model
allow to derive analytically an explicit condition, written in terms
of a critical convexity $\gamma_c$, which ensures that a state
characterized by the existence of relaxation oscillations of the size
of the system is a fixed point. However, this argument cannot ensure
that such fixed point is an attractor. We will show, by numerical
simulations, that it has a large basin of attraction.

     Let us assume that there is an avalanche sweeping the whole
lattice (for an arbitrary geometry). The idea is to derive a
condition that could ensure that the next event will reach the whole
population as well. Let $n$ denote the lattice coordination number
and $m$ the number of neighbors of the most unfavorable site (the
cell or group of cells with the smallest number of neighbors). These
sites have zero energy when the avalanche is over, and in order to
fire in the next avalanche, they have to satisfy\cite{prl74.118}
(for excitatory coupling), 
\begin{equation}
E\left( 1- \varphi (n \varepsilon )\right) +m \varepsilon  \geq E_c.
\end{equation}
In the case of the Peskin model the relations between the phase
and the energy are
\label{peskin}
\begin{equation}
E( \varphi )=C \left( 1-e^{-\gamma  \varphi } \right)
\end{equation}
and
\begin{equation}
\varphi (E)=\frac{1}{\gamma}\ln \frac{C}{C-E},
\end{equation}
where $C$ was already defined in (\ref{peskdri}). A simple
calculation yields, when $E_c = 1$, 
\begin{equation}
\varepsilon   \leq  \frac{C(m-n)+n}{mn}    
\label{epsilon}
\end{equation}
which is the relation between $ \varepsilon $ and $ \gamma $ that must
be satisfied for model (\ref{peskdri}) with interaction rules
(\ref{ff}) in order to show relaxation oscillations involving
all the cells. In the particular case of a regular $2D$ square
lattice with open bc, the most unfavorable sites are the
corners. Hence 
\begin{equation}
\varepsilon \leq \frac{2 - C}{4}.
\label{corners}
\end{equation}
This relation has been checked through simulations finding an
excellent agreement as can be seen in Fig.~\ref{phaseFF}. In region
A, the condition (\ref{corners}) is satisfied and therefore, the
system settles in a state where a macroscopic group of oscillators
whose size scales with the size of the lattice remain at the same
phase. In region B, the condition is not satisfied and relaxation
oscillations involving all the cells are not stable. The systems goes
towards an attractor of difficult classification whose
characteristics will be discussed later. In the limit of $\gamma =0$
we recover the features of the noisy FF model. Notice that a
convex driving is not enough to find relaxation oscillations. It is
necessary to go beyond $\gamma_c$. In fact, $\gamma_c$ tends to $\ln
2$ in the limit of $\varepsilon \rightarrow 0$.  

\begin{figure}[htbp]
\centerline{\epsfxsize=3.0truein \epsffile{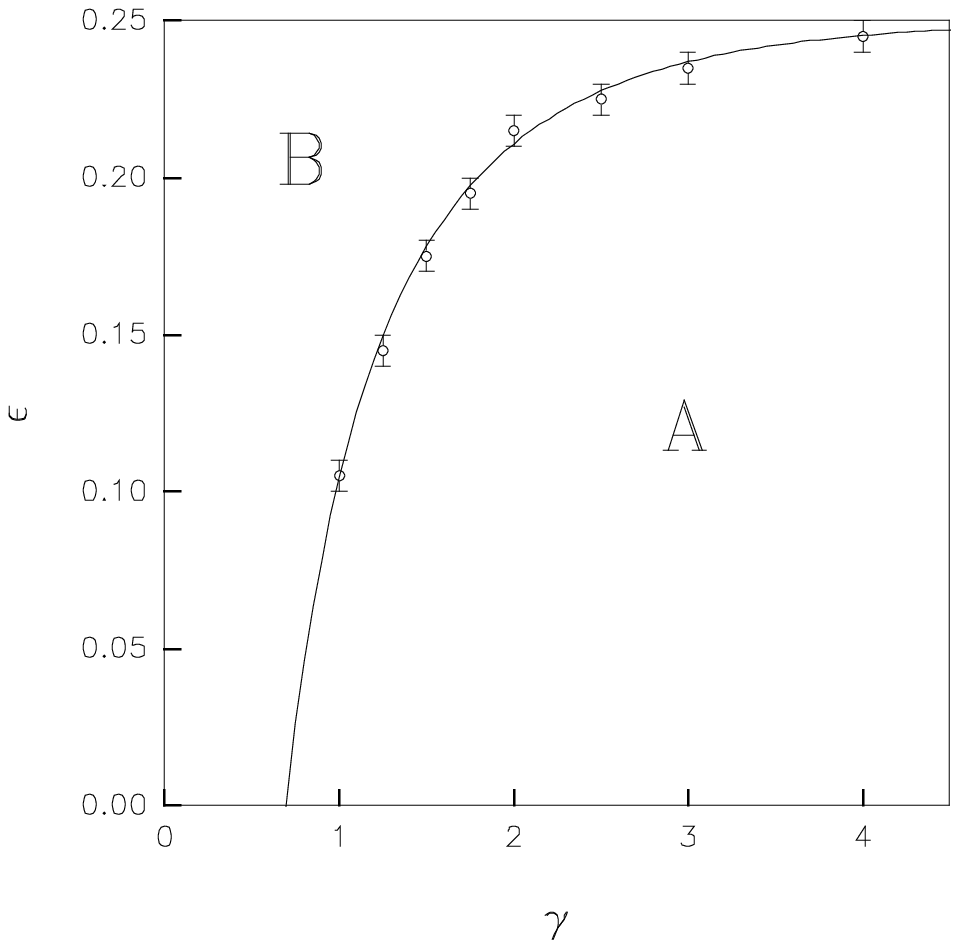}}
\vspace*{0.5cm}
\fcaption{Schematic phase diagram in terms of $ \gamma $ (the convexity
of the driving) and $ \varepsilon $ (the strength of the coupling) for
model (\ref{ff}) and units driven by (\ref{peskdri}). The
symbols correspond to the phase transition observed in the
simulations and the error bars are given by the standard
deviation over ten measures. The solid line corresponds to the
analytical result (\ref{corners}).} 
\label{phaseFF} 
\end{figure}

     When considering periodic bc in a $2D$ square lattice, Eq.
(\ref{epsilon}) reads 
\begin{equation}
\varepsilon  <  0.25
\label{bad.condition}
\end{equation}
thus favoring the appearance of relaxation oscillations in the sense
discussed above: it is a stable fixed point but, in principle, we
cannot ensure it is an attractor for the dynamics of the system.\footnote
{ For $\varepsilon \ge 0.25$ the model with
periodic bc has no dissipation and finally enters into a never-ending
avalanche.}\,
Indeed, numerical simulations show that this synchronized
state can be an attractor of the dynamics in some cases. 
This model has been also analyzed by Hopfield and
Herz\cite{prl.herz,nas.hopfield} in the context of computation by
networks of integrate-and-fire neurons. These authors study a
general model with interaction rules\cite{prl.herz} 
\begin{equation}
\left. \begin{array}{l}
E_{nn} \rightarrow E_{nn} + \varepsilon,  \\
E_{i,j} \rightarrow \Gamma (E_{i,j}-1),
\end{array}
\right. 
\label{herz}
\end{equation}
and a linear driving.  This model is equivalent to the FF model in
the $\Gamma =0$ limit and to the uniformly driven BTW when
$\Gamma =1$.  
As soon as every element has fired once,  
the system reaches a limit cycle in which 
each oscillator fires exactly once during one period 
of the oscillation, no matter the value of $\Gamma$.
For the BTW model it is possible to construct a Lyapunov function that
show this convergence.
However, for $\Gamma<1$, the volume of the attractor is greatly
reduced.  From numerical simulations on $2D$ square lattices they
construct a diagram in terms of the initial distribution of
energies and the coupling strength $\varepsilon$.
The behavior ranges
from a trivial fully synchronized state, when the initial
distribution of the oscillators is very narrow, 
to an exponential distribution of events, when the initial 
distribution is broad enough.
In the border line between both regions a power-law distribution
of events is reported.
The same
authors\cite{nas.hopfield} have also studied the limiting cases of
the BTW and FF models with units being driven nonlinearly.
Within the computational aspects of this work they perform simulations
from which one can conclude that all the models exhibit rapid
convergence to periodic solutions with locally synchronized clusters.
However, in the large time scale the behavior is quite different.
In this case, the BTW interaction rules lead to a global
synchronization whereas FF's show a partially synchronized behavior.
The existence of this nonglobal synchronized periodic state for the
FF model confirms that Eq. (\ref{bad.condition}) gives
the existence of a synchronized state 
(in the sense of a simultaneous firing of all the units)
that, however, it is
not necessarily a global attractor for the dynamics. 
These authors give conditions which show that if 
the strength of the coupling is strong enough,
a simple spatially nonuniform periodic solution
is also possible.

\subsection{Spring-block models}
\noindent
Now, we are going to analyze the effect of a nonlinear convex driving
in deterministic spring-block models, such as those discussed in the
context of earthquake dynamics, which display SOC with a linear
driving.\cite{prl68.1244,pra46.1829,pre47.2366} As in models of
IFO's, these nonlinearities tend to break SOC in favor of more
complex dynamical states. However, there are several differences that
must be remarked. In the OFC model (7) two cells very seldom reach
the threshold simultaneously and trigger an avalanche in two
different points of the lattice. There is no need to introduce a
dynamical noise in contrast with the FF model where the existence of
the noise is required to observe SOC.  Therefore, concerning the SOC
behavior, one might conclude that the existence of some kind of
noise, either in the way the system is driven or in the initial
conditions, is necessary. On the other hand, some authors have
reported power-law distributions for completely deterministic
automata.\cite{prl65.949,pra46.6288,p199a.254} However, for all these
deterministic models there is a single site with a special behavior;
this inhomogeneity, either in the initial conditions or in the
external loading, is what gives rise in these cases to the complex
behavior characteristic of SOC.

     For the OFC model it is also possible to deduce analytical
results giving information about the conditions required to observe a
transition between a regime where relaxation oscillations rule the
long time behavior and other dynamical states. However, the situation
here is more difficult and two equations are necessary. One ensures
that the avalanches will reach all the cells of the lattice, and the
other is based on the assumption that the configuration after a big
event always will be the same. The main difference with respect to
the FF model is that we have to replace $
\varepsilon $ by an effective value $\overline{ \varepsilon }$ since the
energy of a given site can be larger than $E_c=1$ when the avalanche
propagates through the lattice.\footnote{Actually, for a regular 2D
square lattice $\overline{ \varepsilon }$ is bounded between $ \varepsilon
$ and $ 2 \varepsilon $ since the energy at any site cannot exceed the
value of $2$.}\, 
Within a mean-field approximation we will assume that
this energy is the same for all cells, except the seed (the unit that
triggers the avalanche), its neighbors and the boundaries. Assuming
that any site fires at most once during an avalanche\footnote{ This
can only be ensured when $ \varepsilon
<0.22$.\cite{reply.kim}}\, the seed 
has a phase $\varphi (4\overline{
\varepsilon })$ when the avalanche has finished, and to fire again it
has to increment by an amount $1- \varphi (4\overline{ \varepsilon })$.
Since its neighbors are at $ \varphi (3\overline{ \varepsilon })$ the
condition to repeat permanently this situation is
\begin{equation}
4\overline{ \varepsilon } = 4  \varepsilon  \left[ E \left(
\varphi (3\overline{ \varepsilon })+1- \varphi (4\overline{
\varepsilon }) \right) + \varepsilon \right]. 
\label{epsilonbarra}
\end{equation}
This equality, in addition with (\ref{epsilon}), with $\varepsilon$
replaced by $\overline{ \varepsilon }$, gives the following relationship
between $\varepsilon$ and $\gamma $
\begin{equation}
\varepsilon \leq \frac{1}{16}\left[\sqrt{C^2-52C+164} - 
(C+6)\right].
\label{epsilondef}
\end{equation}
When this condition is fulfilled, only relaxation oscillations of the
size of the system can survive in the stationary state. The curve
corresponding to the equality has been plotted in Fig.~\ref{phaseOFC}
(solid line). It has been corroborated through simulations performed
in a system of size $64 \times 64$ with a fixed value of $\gamma$ and
increasing $ \varepsilon $ from below. For 10 different realizations
of the initial random conditions we look at the final state after $3
\cdot 10^6$ avalanches. The average value of $
\varepsilon $ where the transition is observed corresponds to the
circles in Fig.~\ref{phaseOFC} and the error bars correspond to the
standard deviation.  

\begin{figure}[htbp]
\centerline{\epsfxsize=3.0truein \epsffile{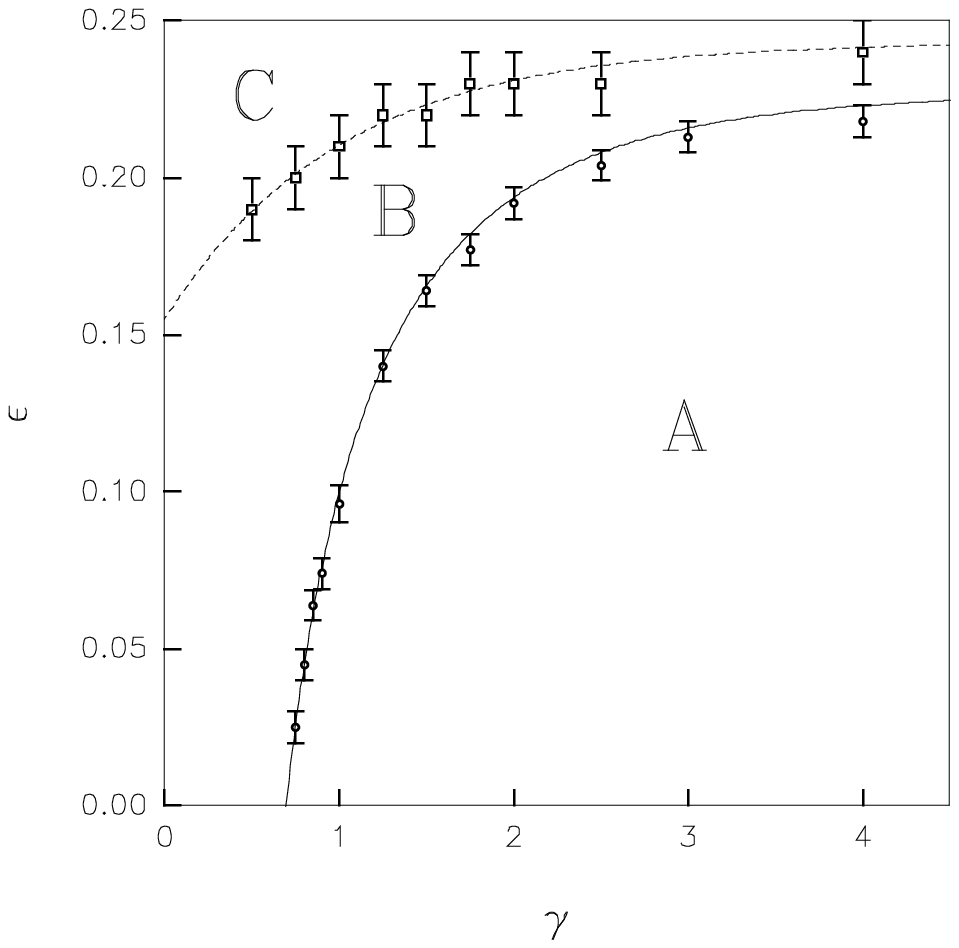}}
\vspace*{0.5cm}
\fcaption{Schematic phase diagram in terms of $ \gamma $ (the
convexity of the driving) and $ \varepsilon $ (the strength of the
coupling) for the OFC interaction rules with a driving given by
(\ref{peskdri}). The symbols correspond to the phase transitions
observed in the simulations. For the B-C transition the error
bars denote that above them we have always found SOC while below
them there is no power-law 
behavior. For the A-B transition the error bars are given by the
standard deviation over ten measures. The solid line corresponds to
the analytical result (\ref{epsilondef}) and the dashed line is an
exponential fit to the numerical data.} 
\label{phaseOFC}
\end{figure}

     Furthermore, to complement and illustrate the features of the
dynamical states observed in the phase diagram of the model, we have
plotted in Fig.  ~\ref{time} the time evolution of the avalanche size
as well as its return map, for a particular run. Thus, we can clearly
see that, in region A (Fig.~\ref{time}(a)) large avalanches appear
frequently until an avalanche sweeps the whole system and this state
is maintained. On the other hand, the return map shows the
evolution of the system at the fixed point. Let us remind that
in the theoretical approach we 
made the assumption that a whole system avalanche does exist. In the
simulations one observes that this hypothesis is indeed true. The
synchronization of the cells is not global but the number of cells
with the same phase is of the order of the size of the system. 

     Slightly above the curve given by Eq. (\ref{epsilondef}) an
avalanche sweeping all the lattice cannot repeat any more since the
next one will be unable to reach some of the cells in the boundaries
and these cells will be the starting point of future avalanches
(region B). This fact gives rise to a periodic behavior with a
discrete distribution of a few avalanche sizes that can
eventually be
broken only by the effect of the dynamical noise. This is indeed what
we have observed in the simulations; only some values of avalanche
sizes are present. The exact location is very sensitive to the
initial random configuration but they are always of order 1, $L$, and
$L^{2}$. In Fig.~\ref{time}(b) we show the time evolution of the
avalanche size in a typical run in this region after a transient
time. Here we can notice the appearance of the periodic state and a
small number of points in the return map.
     
\begin{figure}[htbp]
\epsfxsize=2.9truein
\epsffile{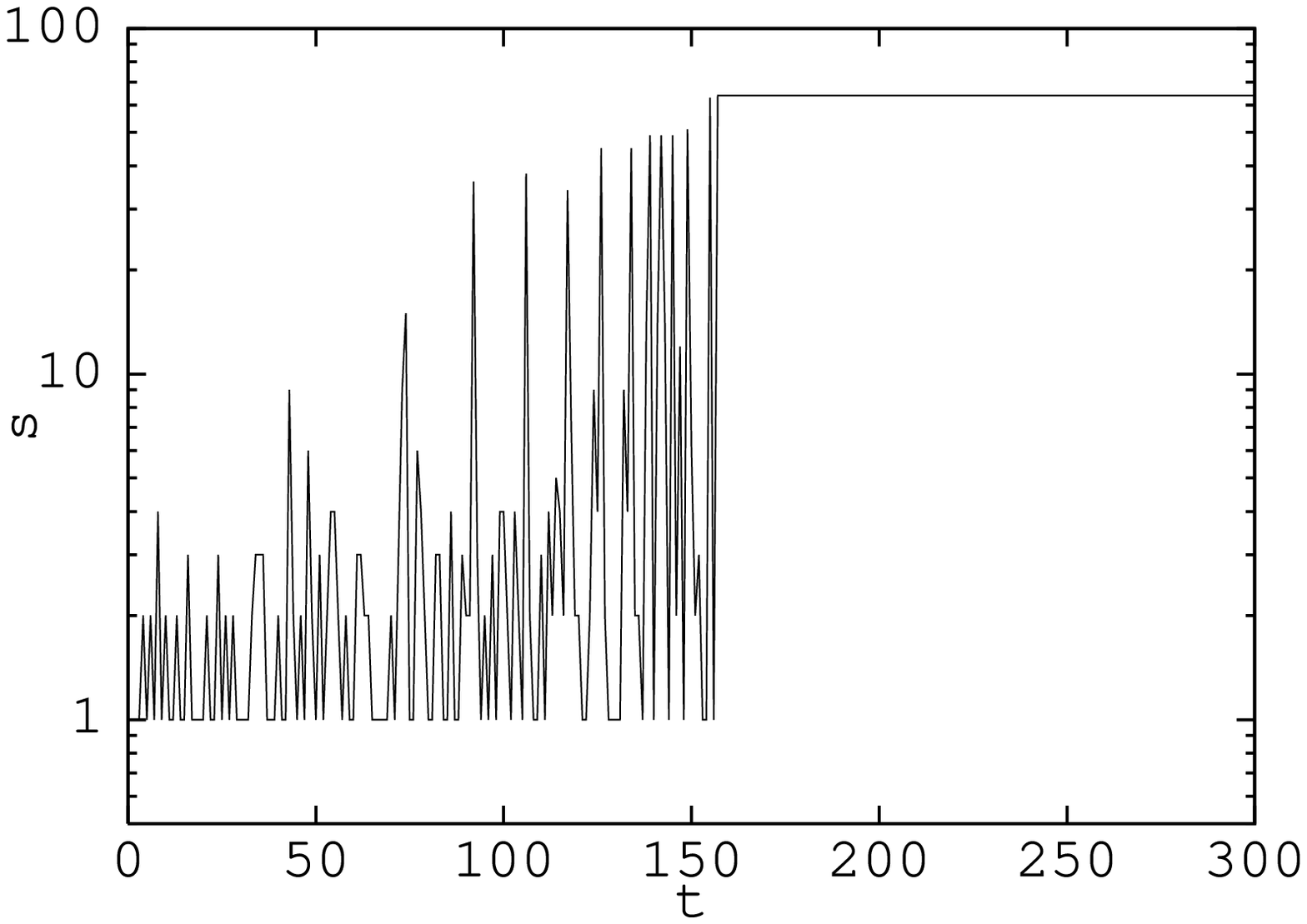}
\vskip -2.0truein
\hskip 2.7truein\epsfxsize=2.0truein
\epsffile{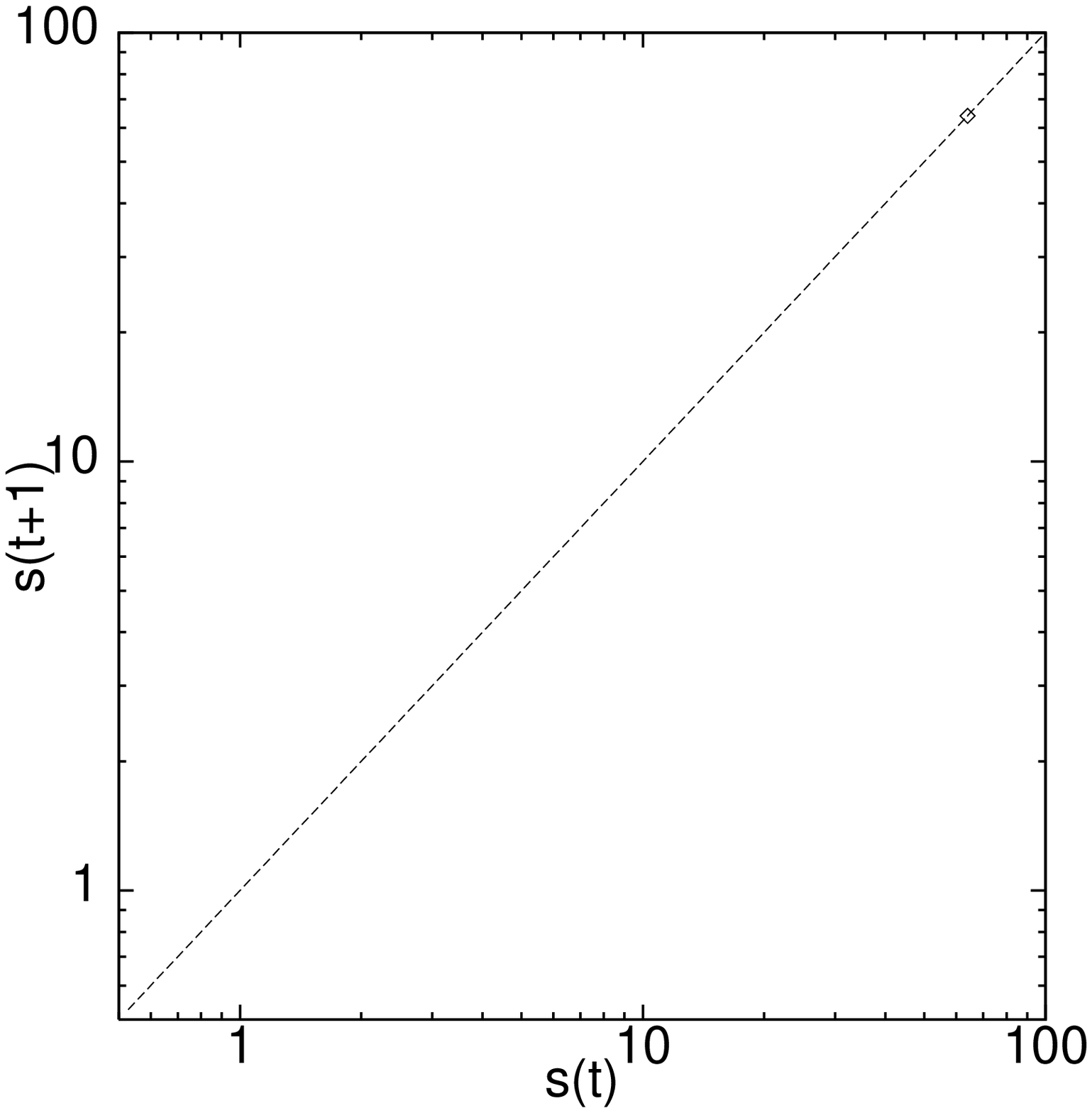}
\epsfxsize=2.9truein
\hskip 0.0truein\epsffile{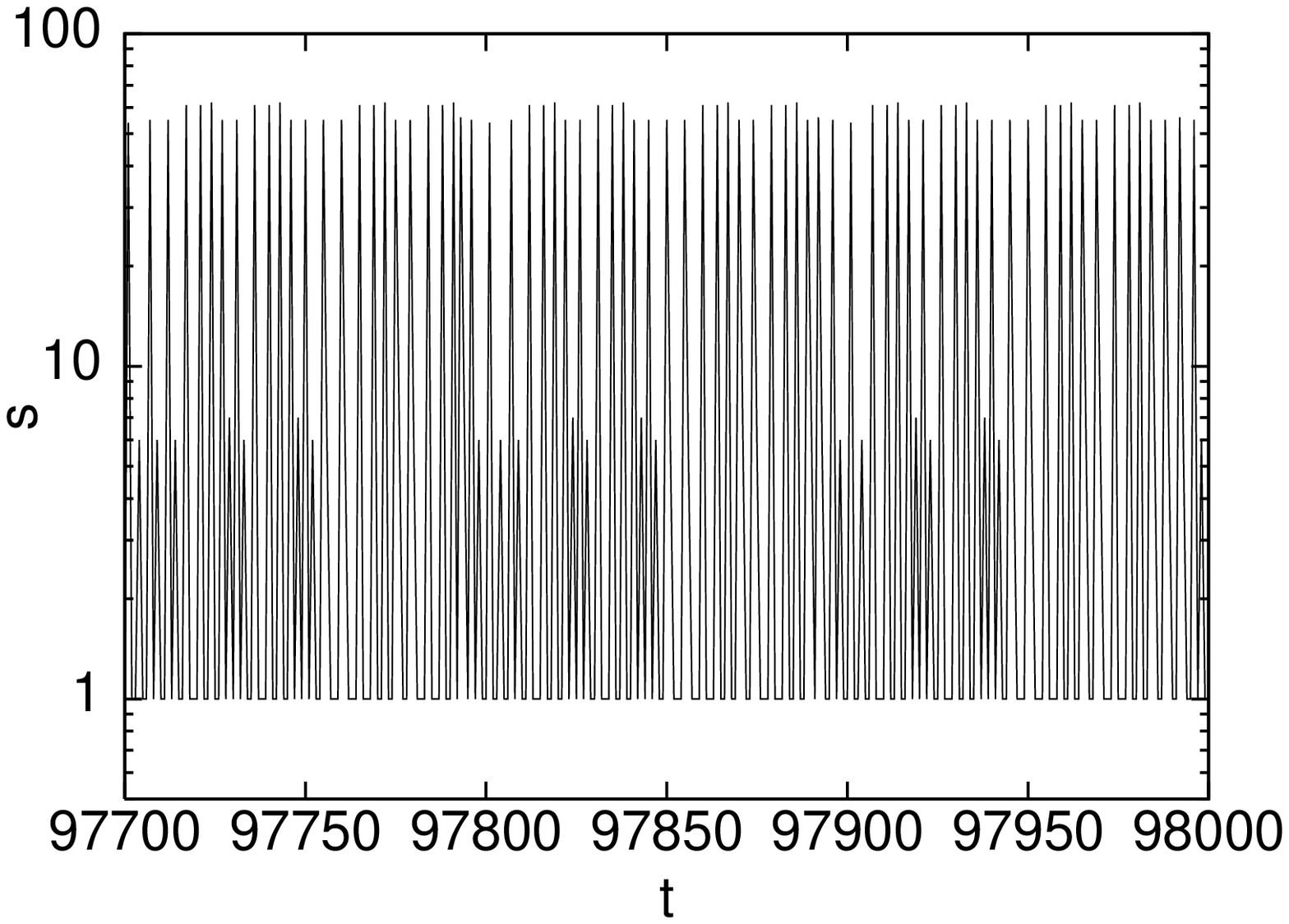}
\vskip -2.0truein
\hskip 2.7truein\epsfxsize=2.0truein
\epsffile{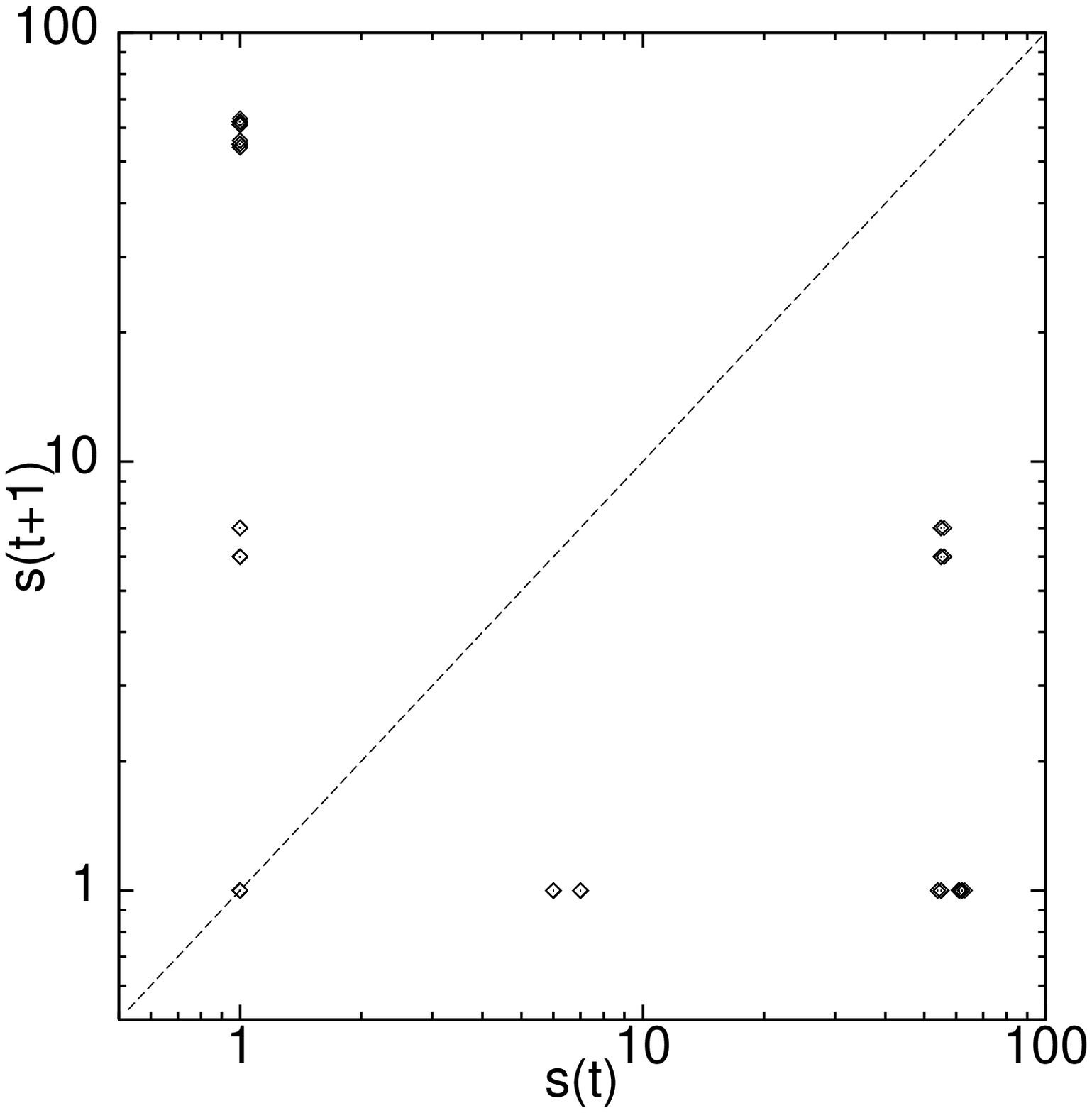}
\epsfxsize=2.9truein
\hskip 0.0truein\epsffile{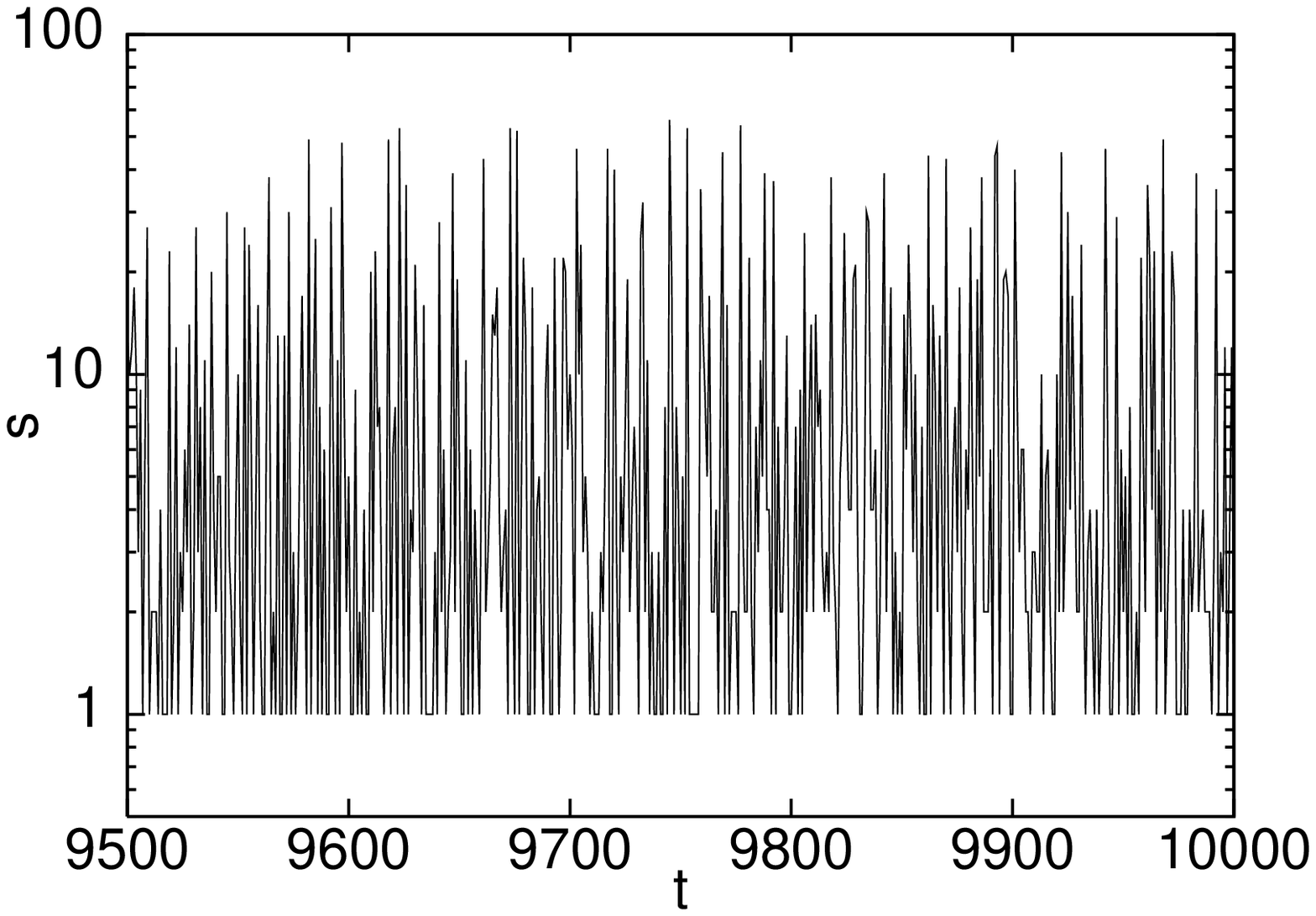}
\vskip -2.0truein
\hskip 2.7truein\epsfxsize=2.0truein
\epsffile{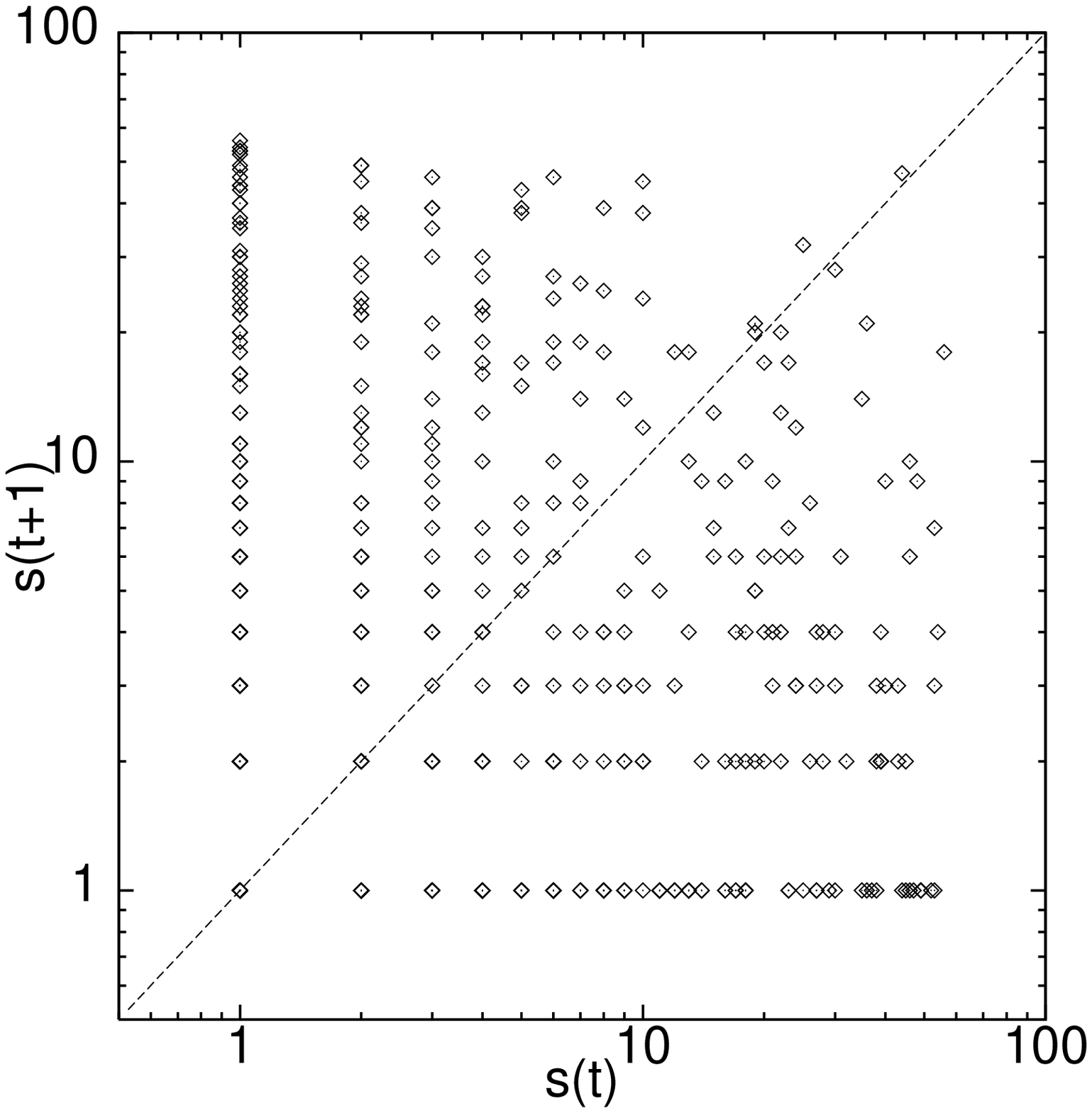}
\vspace*{0.2cm}
\fcaption{Left column: Time evolution of the avalanche size. 
Right column: return map of the avalanche size.
Top) in region A of
Fig.~\ref{phaseOFC} for $L=8$, middle) for region B (close to
region A), and bottom) for
region C.  
Each time step corresponds to an avalanche.}
\label{time}
\end{figure}

     By increasing $ \varepsilon $ we observe that the distribution
of avalanche sizes, $P(s)$, becomes continuous in the sense that all
bins of exponentially increasing width have at least one event, up to
some system size dependent cutoff. The transition from a sparse
distribution to a "continuous" one would require a careful
investigation. When $P(s)$ has this continuous appearance there are
peaks at some characteristic lengths, roughly at $L$, $2L$, $3L$,
$\ldots$, as can be seen in Fig.~\ref{dsb}, and no finite-size
scaling is possible.

     By increasing $ \varepsilon $ again we observe that the
intensity of the peaks decreases. Up to the system sizes and number
of avalanches we have studied, there exists a transition to a
stationary state with a power law, followed by an exponential
decay of $P(s)$. 
We have identified the transition from regions B to C (squares in
Fig.~\ref{phaseOFC}) by fixing $\gamma$ and increasing $\varepsilon$ up
to the appearance of a finite-size scaling in the distribution of
avalanche sizes.  The dashed line is an exponential fit of the
numerical results that we extrapolate to the linear driving case
($\gamma = 0$) and for large values of the convexity of the driving. 

\begin{figure}[htbp]
\centerline{\epsfxsize=4.0truein\epsffile{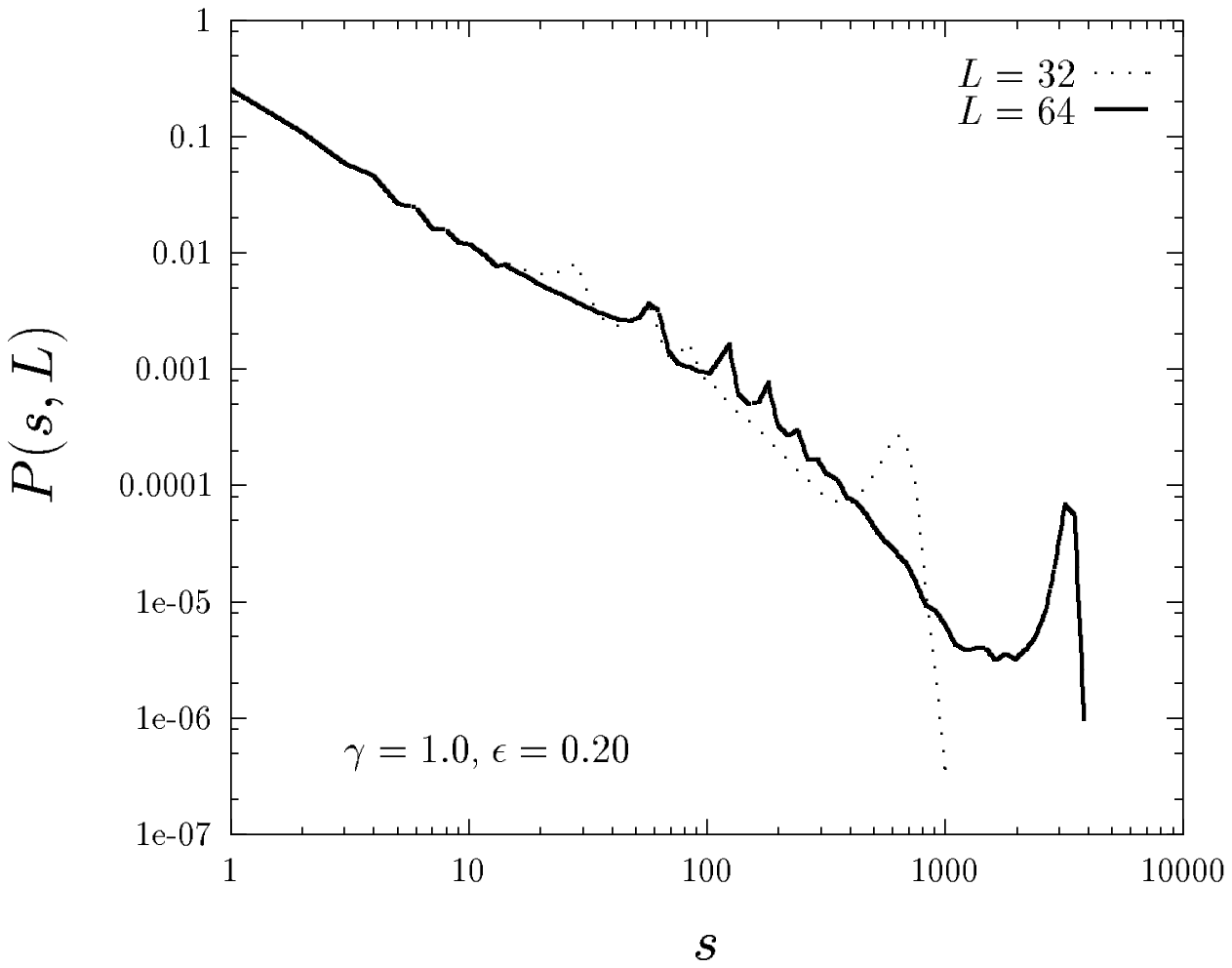}}
\vspace*{0.2cm}
\fcaption{Distribution of avalanche sizes in region B (close to
region C).}
\label{dsb}
\end{figure}

     The criticality of the SOC state stems from the lack of
characteristic time and length scales and for the apparent
unpredictability of the size of the next event. This can be
visualized in Fig.~\ref{time}(c) where no simple correlation can be
extracted from the time sequence. Just at the conservation line,
$ \varepsilon =0.25$, where the interaction between units is
stronger, we find SOC for a wide range of values of $\gamma$
with a power-law decay over more than three orders of magnitude
and the corresponding finite-size scaling as shown in
Fig.~\ref{dsc}.  

\begin{figure}[htbp]
\centerline{\epsfxsize=4.0truein\epsffile{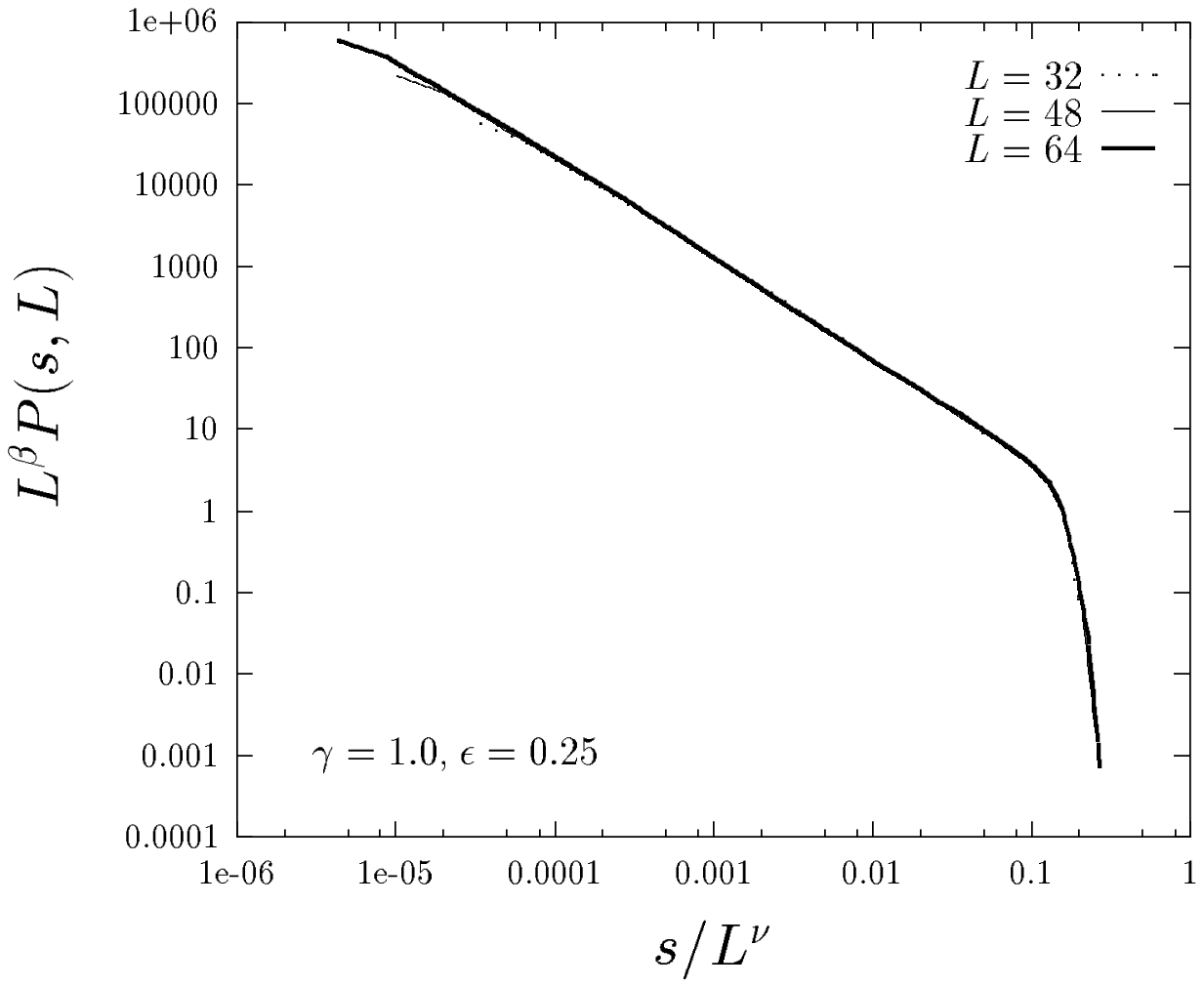}}
\vspace*{0.3cm}
\fcaption{Finite-size scaling ansatz of the distribution of
avalanche sizes in region C. In this case $\tau =2.2\pm 0.1$, 
$\beta =3.7 \pm 0.1$, and $\nu=3.0 \pm 0.1$}
\label{dsc}
\end{figure}

     Up to now we have studied a population of identical oscillators,
with intrinsic period $T=1$, but in a real system it is usual to find
it randomly distributed. We have consider a uniform distribution,
centered at $T=1$.\cite{futuro} For model (\ref{sbrelax}) and a
convex driving the results remain qualitatively the same for a width
in the distribution as large as $0.1 T$. However, if the width is
equal to the mean period $T$ the region where synchronization is
observed almost disappears as well as SOC which is constrained to a
narrow region near the line $\varepsilon=0.25$. The effect of a quenched
distribution of thresholds has been also analyzed by Janosi and
Kert\'{e}sz (uniform distribution)\cite{p200a.179} and by Torvund and
Fr{\o}yland (gaussian distribution)\cite{ps.torvund} for the linearly
driven OFC model, where it has been shown that the distribution of
avalanche sizes has an exponential decay for a sufficiently large
spread.

     A slightly different model with a uniform driving was analyzed
by Socolar, Grinstein, and Jayaprakash.\cite{pre47.2366} In this case
when a site reaches a critical value the interaction rules are
\begin{equation} 
\left. \begin{array}{l} 
E_{nn} \rightarrow \Lambda E_{nn} + \varepsilon E_{i,j}  \\ 
E_{i,j} \rightarrow 0.
\end{array} 
\right. 
\label{sgjrelax} 
\end{equation} 
In this model, $\Lambda \neq 1$ is a simple way to remove degeneracy,
though it does not model the elastic nonlinearities of the blocks and
springs model in detail. Rules (\ref{sgjrelax}) can be seen as a kind
of a state dependent transfer (ERC), which can be transformed into a
uniform transfer model with a nonlinear driving by means of the
appropriate transformation.\cite{pre51.738,PRL2} When open bc are
considered, $\Lambda =1$ corresponds to the original OFC model and in
an interval around this value there are indications of SOC behavior.
Nevertheless, for $\Lambda < 1$, the authors suggest the observed
behavior might correspond to a periodic state with either a long
transient or a long period.  For $\Lambda \geq 1$, the power law
decay in the avalanche size distribution observed is related to
apparently chaotic oscillations in the average energy
\begin{equation}
<E>=\frac{1}{L^{2}}\sum_{i,j} E_{i,j}.
\end{equation}
It is suggested that the boundary sites, which have a different cycle
than interior sites, might be the source of the criticality observed.
This issue has been analyzed in detail by Middleton and
Tang\cite{prl74.742} in a simple directed model; they show that the
slower growth rate of the boundary sites favors the creation of
synchronized clusters of all sizes. At a fixed time, the number of
clusters of size $d$ has a power-law tail that is independent of
$\varepsilon$, $n(d) \approx d^{-\sigma}$.  An avalanche can then
eventually cross a cluster boundary whenever both clusters have close
enough values of the energy, and this makes the avalanche size
distribution to be also a power law. 

     As we noticed for the FF model, also for the OFC model a very
important role is played by the bc. In particular, periodic bc break
completely the SOC behavior and can provide some hints on its origin
in the nonconservative continuously driven
models.\cite{prl74.742,pre47.2366,pre49.2436} For instance,
Socolar, Grinstein, and Jayaprakash obtain
\cite{pre47.2366} different behaviors, depending on the
values of $\Lambda$ in (\ref{sgjrelax}). Thus, for $\Lambda < 1$, the
system always settles in a periodic state, with transients that
become longer as $\Lambda$ approaches one. On the other hand, for
$\Lambda >1$ the most common states are periodic with a cycle
consisting of single avalanches that sweep the whole lattice. A
simple picture can be obtained by analyzing the return map of a
two-site model: for $\Lambda > 1$ both sites fire in the same
avalanche, whereas for $\Lambda <1$ there is a stable fixed point.
These results are in agreement with the ones obtained for a nonlinear
driving by means of the above mentioned transformation.\cite{PRL2}
Finally, the special case of $\Lambda =1$, shows a completely
different behavior from what is expected from the limits of the
previous cases: the system rapidly settles into a periodic state with
avalanches of size one which are marginally stable as those described
previously for the FF model.\cite{prl.herz,nas.hopfield}
Grassberger,\cite{pre49.2436} by means of very large-scale
simulations, realized that there exists a close relation between
temporal and spatial ordering; while time ordering is
associated with the periodic behavior and becomes stronger as
$\varepsilon$ decreases, spatial order is related to the
synchronization of neighboring sites which, in principle, seems to be
favored by a large interaction between units.  Concerning the SOC
behavior, one can conclude that open boundaries create an
inhomogeneity which prevents time ordering but spatial local
ordering is still maintained.  Along this line Middleton and
Tang\cite{prl74.742} have observed that by including some source of
dynamical noise the system with periodic bc is periodic in time,
after a brief transient time, and that these states are neutrally
stable.

     A different way to introduce the inhomogeneity that is
responsible for the SOC behavior has been followed by Torvund and
Fr{\o}yland.\cite{ps.torvund} They have considered a single site to
have a threshold larger than the threshold of the remaining sites, in
a system with periodic boundary conditions. These authors have found
some evidence for a SOC behavior when the larger threshold is above a
critical value. Nevertheless, also a non-simple periodic behavior can
be observed. This means that a more careful analysis would have to be
performed along this line.

\subsection{Other models}
\noindent
Forest-fire models are other systems where typical SOC behavior has
been observed.\cite{prl69.1629} The model is defined in a
$D$-dimensional lattice (usually $D=2$).  Each site of the lattice is
either occupied by a tree or empty. Trees grow according to a certain
function that depends on the age of the empty site. A tree burns if
it is struck by lightning with a small probability $f$. In this case,
the fire propagates through the neighbors burning the whole cluster
that immediately becomes empty.

     In this system the features of the stationary state depend
strongly on the specific form of the tree growth.
Drossel\cite{p.drossel.sy} has shown analytically (for a $1D$ system)
and through simulations that if the life-time distribution of empty
sites has a long-time tail a SOC state with non-universal exponents
is observed. On the other hand, for deterministic tree growth a
coherent temporal activity between trees in the same cluster comes
up. Therefore, we see again the relevance of the driving in the
collective properties of the model.

     Inspired by a forest-fire model, Clar, Drossel, and
Schwabl\cite{clar.noneq} have studied a nonequilibrium percolation
system which displays several behaviors ranging from SOC to
clusterization. The authors consider a population of objects (trees,
animals or others) distributed randomly over a lattice.  The density
of occupied sites is $\rho$. Then, a given site is chosen. If it is
not empty an ''explosion'' occurs affecting all the neighbors within
the same cluster, whose spatial position on the lattice is
redistributed (a group of animals dispersed by the action of a
predator could be another example). In these terms the density of the
system is a conserved quantity. 

     In the stationary state the model shows different regimes
depending on the density of occupied sites of the lattice. For
diluted networks ($\rho < \rho_1 = 0.41$) only small clusters of
units are developed. As a consequence, the average number of units
affected by an explosion and the size of the largest cluster
$s_{max}$ is small. For $\rho_1 < \rho < \rho_2=0.435$ the size of
the explosions diverge but more slowly than the size of the system.
They show that the relevant quantities diverge as a power law, with
an exponent that depends on the density $\rho$. This critical
behavior is characteristic of SOC. Finally, for larger values of the
density there is a region where the explosions have infinite size
(they scale with the size of the system) which is repeated
periodically in time.

     Related to this model we can notice the work of Csilling {\em et
al.}\cite{pre50.1083} about dynamics of populations. Each site on a
lattice denotes a population that evolves according to
\begin{equation}
N_{i,j}(t+1)=\lambda N_{i,j}(t)\left[ 1+aN_{i,j}(t) \right] ^{-\beta}
\label{janositime}
\end{equation}
until a critical population density is reached (overcrowding). In
this case a dispersal movement (migration) is triggered through
nearest neighbor local habitats
\begin{equation} 
\left. \begin{array}{l} 
N_{nn} \rightarrow N_{nn} + \Delta \frac{N_{i,j}-N_{sc}}{4}  \\ 
N_{i,j} \rightarrow N_{sc}.
\end{array} 
\right. 
\label{janosirelax} 
\end{equation} 
Notice that these rules are identical to OFC rules with
$N_{sc}=0$ and 
$\Delta = 4 \varepsilon$. This process may lead to a migration
avalanche, typical of the lattice models discussed so far. Without
interaction, a simple site behaves chaotically. However, by
increasing the interaction it reaches a noisy fixed point. Concerning
the lattice average, by lowering the threshold level, the collective
behavior becomes more pronounced, reflected by the appearance of
discrete frequencies on the power spectra; moreover, the time
evolution either becomes strictly periodic or it reaches a stable
fixed point. Finally, the distribution of avalanche sizes is
computed; whereas for a weak interaction the distribution is
exponential, for a stronger interaction a power law, characteristic
of SOC, was observed.

\section{Conclusions} 
\noindent
In this paper we have reviewed the collective behavior of a large
group of low dimensional systems characterized by two essential
features: a dynamical process to update the state of each member of
the system and a threshold condition which defines an interactive
process between members of the population. The coupling between units
is defined in terms of very simple local rules. Within this broad
framework we have considered different models ranging from cellular
automata where the state of each unit, described in terms of discrete
variables, is updated through a random process, to coupled dynamical
systems where the internal state is governed by differential
equations. In spite of the simplicity of the considered models, the
majority displays a complex behavior. In some cases they settle down
in stationary states without characteristic time or length scales
typical of self-organized criticality (SOC). 
In other cases, the attractors
are synchronized states with a partial or complete coherence in the
temporal activity of the members of a given population.  Most
interestingly, a few of them display both attractors (and others)
depending on the value of the parameters which define the model.

     There are several reasons which foster the study of these
models. Perhaps, the most relevant is to understand the origin of SOC
and identify the underlying mechanisms. Unfortunately, there is no
clear answer to this question and the problem remains open. SOC
behavior is very sensitive to the action of external factors, it can
be destroyed (and created) in many different ways and, 
{\em a priori}, it
is difficult to find the proper combination of ingredients that may
make a system to display it. It is well known that boundary
conditions play a relevant role. Big events (avalanches) are usually
generated near the boundaries and propagated towards the bulk, but
only under suitable conditions which, in general, imply open
boundaries. But apart from this important factor, there are other
components which also have a crucial effect on the features of the
stationary state. For some cellular automata such as the BTW model
conservation is the key. For the Zhang model either conservation or a
very tiny perturbation of the internal state of the cells. Some
continuous driven models display SOC even without conservation such
as the OFC model but additionally, in others such as the FF model a
dynamical noise is required. In the last two models the role of
memory effects (a trace of the initial state) can give the answer. 

     We have seen that in the continuous driven models the features
of the driving are also relevant. The original OFC and the noisy FF
models display SOC with constant or quasiconstant driving rate.
However, a non-constant driving rate, which implies a convex
(concave) relation between the state of each unit and time, leads to
new behaviors. In some cases the duality convexity/concavity lead to
full synchronization/anti-synchronization (phase locked state where
neighbors remain at a fixed phase difference).  Then SOC might be
understood as an intermediate behavior, balanced between these two
extreme situations. In other situations, local synchronization can be
responsible for a global SOC behavior. For these reasons, the
analysis of systems displaying a rich variety of attractor such as
those mentioned in the last section of the paper is of great
interest.

     Another important feature of the systems considered in this
review is the separation of time scales associated to the natural
dynamics of each unit and the interaction between them. This
separation, quite natural in some situations, seems to be mandatory
to observe SOC behavior. If this is true, systems where the
transmission of information between neighbors could take a finite
time might not display SOC. It would be interesting to pay more
attention at this point in the future and see whether it is possible
to observe power-law behavior in systems where both time scales are
overlapped.

     Finally, there are other aspects of interest not sufficiently
analyzed so far. The effect of different sources of noise (for
instance, the coupling with a thermal bath), the role played by
inhibitory rather than excitatory couplings, a careful analysis of
the role of memory effects are some of the topics that deserve
further research in the future.

\nonumsection{Acknowledgments}

     This work has benefited from discussions with several people,
mainly P.  Bak, L.F.  Abbott, C.  Van Vreeswijk, and A.V.M.  Herz.
Financial support from CICyT, project \# PB94-0897, and EC Grant \#
CHGE-CT92-0009 is also acknowledged.  K.C.  gratefully appreciates
the hospitality of the University of Barcelona.

\nonumsection{References}

} 


\begin{thebibliography}{000}

\bibitem{Mand82} B. B. Mandelbrot, 
{\nineit The Fractal Geometry of Nature} (Freeman, San
Francisco, 1982).

\bibitem{Feder88} J. Feder,
{\nineit Fractals} (Plenum Press, New York, 1988).

\bibitem{cmp7.103} W. H.  Press, {\nineit Comments  Astrophys.}
{\ninebf 7}, 103 (1978).

\bibitem{rmp53.497} P. Dutta and P. M.  Horn, 
{\nineit Rev.  Mod.  Phys.}  {\ninebf
53}, 497 (1981).

\bibitem{weissman} M. B. Weissman, {\nineit Rev. Mod. Phys.}
{\ninebf 60}, 537 (1988).

\bibitem{prl59.381} P.  Bak, C.  Tang, and K.  Wiesenfeld, 
{\nineit Phys.  Rev. Lett.} {\ninebf 59}, 381 (1987).

\bibitem{pra38.364} P.  Bak, C.  Tang, and K.  Wiesenfeld, 
{\nineit Phys.  Rev.} {\ninebf A38}, 364 (1988).

\bibitem{GR44} B.  Gutenberg and C.  F.  Richter, {\nineit Bull.
Seismol. Soc.  Amer.} {\ninebf 34}, 185 (1944).

\bibitem{prl62.2632} J. M.  Carlson and J. S.  Langer, 
{\nineit Phys.  Rev. Lett.} {\ninebf 62}, 2632 (1989).

\bibitem{prl68.1244} Z.  Olami, H. J. S.  Feder, and 
K.  Christensen, {\nineit Phys. Rev. Lett.} 
{\ninebf 68}, 1244 (1992).

\bibitem{prl69.1629} B.  Drossel and F.  Schwabl, 
{\nineit Phys. Rev. Lett.} {\ninebf 69}, 1629 (1992).

\bibitem{prl71.4083} P.  Bak and K.  Sneppen, 
{\nineit Phys. Rev. Lett.} {\ninebf 71}, 4083 (1993).

\bibitem{prl68.205} P. Bantay and I. M. Janosi, 
{\nineit Phys. Rev. Lett.} {\ninebf 68}, 2058 (1992).

\bibitem{prl67.919} O. Pla and F. Nori, 
{\nineit Phys. Rev. Lett.} {\ninebf 67}, 919 (1991).

\bibitem{prl67.1334} P. J.  Cote and L. V.  Meisel, 
{\nineit Phys. Rev. Lett.} {\ninebf 67}, 1334 (1991).

\bibitem{rmp66.657} J. M.  Carlson, J. S.  Langer, 
and B. E.  Shaw, {\nineit Rev. Mod. Phys.} {\ninebf 66}, 
657 (1994). 

\bibitem{ricerche} P.  Bak, K.  Chen, J. Scheikman, 
and M.  Woodford, {\nineit Ricerche Economiche } 
{\ninebf 47}, 3 (1993).

\bibitem{p184a.127} H.  Takayasu, H.  Miura, T.  Hirabayashi, 
and K. Hamada, {\nineit Physica } {\ninebf 184A}, 127 (1992).

\bibitem{prl62.40} H. M.  Jaeger, C. H. Liu, and S.  R.  Nagel, 
{\nineit Phys. Rev. Lett.} {\ninebf 62}, 40 (1989).

\bibitem{pra43.2720} P.  Evesque, {\nineit Phys. Rev.} 
{\ninebf A43}, 2720 (1991).

\bibitem{prl69.2431} M.  Bretz, J. B.  Cunningham, P. L.  
Kurczynski, and F. Nori, {\nineit Phys. Rev. Lett.} 
{\ninebf 69}, 2431 (1992).

\bibitem{pre47.2229} E.  Morales-Gamboa, J.  Lomnitz-Adler, V.
Romero-Roch\'{\i}n, R.  Chicharro-Serra, and R.  Peralta-Fabi,
{\nineit Phys.  Rev.} {\ninebf E47}, R2229 (1993).

\bibitem{prl65.1120} G. A.  Held, D. H.  Solina, D. T.  Keane, 
W. J.  Haag, P. M.  Horn, and G.  Grinstein, {\nineit Phys.  
Rev. Lett.} {\ninebf 65}, 1120 (1990).

\bibitem{ajp61.329} S. K.  Grumbacher, K. M.  McEwen, D. A.  
Halverson, D. T.  Jacobs, and J.  Lindner, 
{\nineit Am.  J.  Phys.} {\ninebf 61}, 329 (1993).

\bibitem{pre47.1401} J.  Rosendahl, M.  Veki\'c, 
and J.  Kelley, {\nineit Phys. Rev.} {\ninebf E47}, 1401 (1993).

\bibitem{prl73.537} J.  Rosendahl, M.  Veki\'c, and J. E.  
Rutledge, {\nineit Phys. Rev. Lett.} {\ninebf 73}, 537 (1994).

\bibitem{kim.nature} V.  Frette, K.  Christensen, A.  
Malthe-S{\o}renssen,
J.  Feder, T.  J{\o}ssang, and P.  Meakin, "Dynamics of
a One-Dimensional Rice Pile", {\nineit Nature} (in press).

\bibitem{newff} J. Feder, "The evidence for self-organized
criticality in sand-pile dynamics", preprint.

\bibitem{p186a.82} D.  Dhar, {\nineit Physica} 
{\ninebf 186A}, 82 (1992).

\bibitem{pra45.665} C. P. C.  Prado and Z.  Olami, 
{\nineit Phys. Rev.} {\ninebf A45}, 665 (1992).

\bibitem{rpp57.383} A.  Mehta and G. C.  Barker, 
{\nineit Rep. Prog. Phys.} {\ninebf 57}, 383 (1994).

\bibitem{p173a.22} L. Pietronero, P Tartaglia, and Y.-C. Zhang, 
{\nineit Physica} {\ninebf 173A}, 22 (1991).

\bibitem{prl64.1613} D.  Dhar, {\nineit Phys.  Rev.  Lett.} 
{\ninebf 64}, 1613 (1990).

\bibitem{jsp22.923} S. S.  Manna, L. B.  Kiss, and J.  
Kert\'{e}sz, {\nineit J. Stat.  Phys.} {\ninebf 22}, 923 (1990).

\bibitem{prl63.470} Y.-C.  Zhang, {\nineit Phys.  Rev.  Lett.} 
{\ninebf 63}, 470 (1989).

\bibitem{el26.177} A.  D\'{\i}az-Guilera, {\nineit Europhys.
Lett.} {\ninebf 26}, 177 (1994).

\bibitem{pre51.1711} A.  Vespignani, S.  Zapperi, and L.  
Pietronero, {\nineit Phys. Rev.} {\ninebf E51}, 1711 (1995).

\bibitem{pra45.2211} J.  Lomnitz-Adler, L.  Knopoff, and G.
Mart\'{\i}nez-Mekler, {\nineit Phys.  Rev.} {\ninebf A45},
2211 (1992).

\bibitem{granada} A.  D\'{\i}az-Guilera, in 
{\nineit 3rd Granada Lectures in Computational Physics}, 
P.L.  Garrido and L.  Marro eds., (Springer-Verlag, Berlin, 
1995), p. 115.

\bibitem{adg.fractals} A.  D\'{\i}az-Guilera, 
{\nineit Fractals} {\ninebf 1}, 963 (1993).

\bibitem{prl74.2511} E. T.  Lu, {\nineit Phys. Rev. Lett.} 
{\ninebf 74}, 2511 (1995).

\bibitem{23DRG} A. Corral and A. D\'{\i}az-Guilera, 
"Symmetries and Fixed Point Stability of Stochastic 
Differential Equations Modeling Self-Organized Criticality", 
preprint.

\bibitem{prl62.1813} T.  Hwa and M.  Kardar, {\nineit Phys.  
Rev. Lett.} {\ninebf 62}, 1813 (1989).

\bibitem{prl64.1927} G.  Grinstein, D.-H.  Lee, and S.  
Sachdev, {\nineit Phys. Rev. Lett.} {\ninebf 64}, 1927 (1990).

\bibitem{pra45.7002} T.  Hwa and M.  Kardar, 
{\nineit Phys.  Rev.} {\ninebf A45}, 7002 (1992).

\bibitem{BK67} R.  Burridge and L.  Knopoff, 
{\nineit Bull.  Seismol.  Soc. Am.} {\ninebf 57}, 3411 (1967).

\bibitem{pra46.1829} K.  Christensen and Z.  Olami, 
{\nineit Phys. Rev.} {\ninebf A46}, 1829 (1992).

\bibitem{comment.ofc} W.  Klein and J. Rundle, 
{\nineit Phys. Rev. Lett.} {\ninebf 71}, 1288 (1993).

\bibitem{reply.kim} K.  Christensen, 
{\nineit Phys. Rev. Lett.} {\ninebf 71}, 1289 (1993).

\bibitem{pre49.2436} P.  Grassberger, {\nineit Phys. Rev.}
{\ninebf E49}, 2436 (1994).

\bibitem{tesikim} K.  Christensen, Ph.D.  Thesis, 
University of Aarhus, Denmark, 1992.

\bibitem{pre48.3361} K. Christensen and Z. Olami, 
{\nineit Phys.  Rev.} {\ninebf E48}, 3361 (1993).

\bibitem{winfree} A. T.  Winfree, 
{\nineit The Geometry of Biological Time}
(Springer-Verlag, New York, 1980).

\bibitem{qrb63.265} J.  Buck, {\nineit Q. Rev. Biol.} 
{\ninebf 63}, 265 (1988).

\bibitem{s82.151} H. M.  Smith, {\nineit Science} 
{\ninebf 81}, 151 (1935).

\bibitem{pt.feb94.40} J. J.  Hopfield, 
{\nineit Phys.  Today} {\ninebf 47}, 40 (1994).

\bibitem{bc54.29} C.  Von der Malsburg and W.  Schneider, 
{\nineit Biol. Cybern.} {\ninebf 54}, 29 (1986).

\bibitem{bc60.121} R.  Eckhorn, R.  Bauer, W.  Jordan, 
M.  Brosch, W. Kruse, M.  Munk, and H.  Reitboeck, 
{\nineit Biol. Cybern.} {\ninebf 60}, 121 (1988).

\bibitem{pnas86.1698} C. M.  Gray and W.  Singer, 
{\nineit Proc.  Natl.  Acad. Sci. USA} {\ninebf 86}, 1698 (1989).

\bibitem{n388.334} C. M.  Gray, P.  Konig, A. K.  Engel, and 
W.  Singer, {\nineit Nature} {\ninebf 388}, 334 (1989).

\bibitem{jmb12.13} M.  Kawato, {\nineit J. Math. Biol.} 
{\ninebf 12}, 13 (1981).

\bibitem{bc43.157} N. Ikeda, {\nineit Biol. Cybern.} 
{\ninebf 43}, 157 (1982).

\bibitem{jmb24.291} C. Torras, {\nineit J. Math. Biol.} 
{\ninebf 24}, 291 (1986).

\bibitem{Kurabook} Y. Kuramoto, 
{\nineit Chemical Oscillations, Waves, and Turbulence} 
(Springer-Verlag, Berlin, 1984).

\bibitem{ptp79.223} Y. Kuramoto, {\nineit Prog. Theor. Phys. 
Suppl.} {\ninebf 79}, 223 (1984).

\bibitem{ptp75.1319} S. Shinomoto and Y. Kuramoto, 
{\nineit Prog.  Theor. Phys.} {\ninebf 75}, 1319 (1986).

\bibitem{jsp63.613} S. H.  Strogatz and R. E.  Mirollo, 
{\nineit J.  Stat.  Phys.} {\ninebf 63}, 613 (1991).

\bibitem{jsp67.313} L. L. Bonilla, J. C. Neu, and R. Spigler, 
{\nineit J. Stat. Phys.} {\ninebf 67}, 313 (1992).

\bibitem{prl68.1073} H. Daido, {\nineit Phys. Rev. Lett.} 
{\ninebf 68}, 1073 (1992).

\bibitem{el26.79} A. Arenas and C. J. P\'{e}rez, 
{\nineit Europhys. Lett.} {\ninebf 26}, 79 (1994).
\bibitem{ptp88.1213} H.  Daido, {\nineit Prog.  Theor.  Phys.  } {\ninebf 88},
1213 (1992).

\bibitem{ptp89.929} H.  Daido, {\nineit Prog.  Theor.  Phys.  } {\ninebf 89}, 929
(1993).

\bibitem{ptp79.39} H.  Sakaguchi, {\nineit Prog.  Theor.  Phys.  } {\ninebf 79},
39 (1988).

\bibitem{jsp70.921} L. L.  Bonilla, C. J.  P\'{e}rez, and J. M.  Rub\'{\i},
{\nineit J.  Stat.  Phys.  } {\ninebf 70}, 921 (1993).

\bibitem{jpa23.3835} L. F.  Abbott, {\nineit J.  Phys.  A:  Math.  Gen.  } {\ninebf
23}, 3835 (1990).

\bibitem{schuster} H. G. Schuster and P. Wagner,
{\nineit Biol. Cibern.} {\ninebf 64}, 77 (1990); {\nineit ibid.}
{\ninebf 64}, 83 (1990).

\bibitem{cn4.193} D. Hansel, G. Mato, and C. Meunier,
{\nineit Concepts Neurosci.} {\ninebf 4}, 193 (1993).

\bibitem{jtb16.15} A. T.  Winfree, {\nineit J.  Theoret.  Biol.  } {\ninebf 16}, 15
(1967).

\bibitem{jmb22.1} G. B.  Ermentrout, {\nineit J.  Math.  Biol.  } {\ninebf 22}, 1
(1985).

\bibitem{sjam50.1645} R. E.  Mirollo and S. H.  Strogatz, {\nineit SIAM J.  Appl.
Math.  } {\ninebf 50}, 1645 (1990).

\bibitem{peskin} C. S.  Peskin, {\nineit Mathematical Aspects of Heart
Physiology}, Courant Institute of Mathematical Sciences (New
York University, New York, 1975), p. 268.

\bibitem{jppg9.620} L.  Lapicque, {\nineit J.  Physiol.  Pathol.
Gen.} {\ninebf 9}, 620 (1907).

\bibitem{Tuck} H. C. Tuckwell,
{\nineit Introduction to Theoretical Neurobiology} (Cambridge
University Press, Cambridge, 1988).

\bibitem{prl74.4189} S.  Bottani, {\nineit Phys.  Rev.  Lett.  } {\ninebf 74},
4189 (1995).

\bibitem{p50d.15} Y.  Kuramoto, {\nineit Physica} {\ninebf 50D}, 15 (1991).

\bibitem{pre49.2668} C.-C.  Chen, {\nineit Phys.  Rev.} {\ninebf E49}, 2668
(1994).

\bibitem{pre48.1483} L. F.  Abbott and C.  van Vreeswijk, 
{\nineit Phys.  Rev.} {\ninebf E48}, 1483 (1993).

\bibitem{n4.259} A.  Treves, {\nineit Network } {\ninebf 4}, 259 (1993).

\bibitem{prl71.1280}
M. Tsodyks, I. Mitkov, and H. Sompolinsky, {\nineit Phys. Rev. Lett.}
{\ninebf 71}, 1280 (1993).

\bibitem{pre51.738} W.  Gerstner, {\nineit Phys. Rev.} {\ninebf E51}, 738
(1995).

\bibitem{jcn1.313} C.  van Vreeswijk, L. F.  Abbott, and G. B.  Ermentrout,
{\nineit J.  Comp.  Neur.} {\ninebf 1}, 313 (1994).

\bibitem{prl74.1570}
U. Ernst, K. Pawelzik, and T. Geisel, {\nineit Phys. Rev. Lett.}
{\ninebf 74}, 1570 (1995).

\bibitem{nc4.84} X. J.  Wang and J.  Rinzel, {\nineit Neural Comp.} 
{\ninebf 4}, 84 (1992).

\bibitem{PRL2} A.  Corral, C. J.  P\'{e}rez, A.  D\'{\i}az-Guilera, and A.
Arenas, {\nineit Phys. Rev. Lett.} {\ninebf 75}, 3697 (1995).

\bibitem{ptp92.1039} H.  Sakaguchi, {\nineit Prog.  Theor.  Phys.  } {\ninebf 92},
1039 (1994).

\bibitem{chinchon} Q. Zhilin, H. Gang, M. Benkun, and T. Gang,
{\nineit Phys. Rev.} {\ninebf E50}, 163 (1994).

\bibitem{physd} A. D\'{\i}az-Guilera, A. Arenas, A. Corral,
and C. J. P\'{e}rez, "Stability of Spatio-Temporal Structures in a
Lattice Model of Pulse-Coupled Oscillators", preprint.

\bibitem{prl66.2669} H. J. S.  Feder and J.  Feder, {\nineit Phys.  Rev.  Lett.
} {\ninebf 66}, 2669 (1991).

\bibitem{prl74.742} A. A.  Middleton and C.  Tang, {\nineit Phys.  Rev.  Lett.
} {\ninebf 74}, 742 (1995).

\bibitem{futuro} A.  Corral, C. J.  P\'{e}rez, A.  D\'{\i}az-Guilera, and A.
Arenas, (in preparation).

\bibitem{prl74.118} A.  Corral, C. J.  P\'{e}rez, A.  D\'{\i}az-Guilera, and
A.  Arenas, {\nineit Phys.  Rev.  Lett.} {\ninebf 74}, 118
(1995).

\bibitem{prl.herz} A. V. M.  Herz and J. J.  Hopfield, 
{\nineit Phys.  Rev.  Lett.} {\ninebf 75}, 1222 (1995).

\bibitem{nas.hopfield} J. J.  Hopfield and A. V. M.  Herz, 
{\nineit Proc.  Natl. Acad.  Sci.  USA} {\ninebf 92}, 6655 (1995).

\bibitem{pre47.2366} J. E. S.  Socolar, G.  Grinstein, 
and C.  Jayaprakash,
{\nineit Phys.  Rev.} {\ninebf E47}, 2366 (1993).

\bibitem{prl65.949} K.  Wiesenfeld, J.  Theiler, and B.  McNamara, {\nineit
Phys.  Rev.  Lett.  } {\ninebf 65}, 949 (1990).

\bibitem{pra46.6288} M. S.  Vieira, 
{\nineit Phys.  Rev.} {\ninebf A46}, 6288
(1992).

\bibitem{p199a.254} K.  Nagel and H. J.  Herrmann, 
{\nineit Physica} {\ninebf 199A}, 254 (1993).

\bibitem{p200a.179} I. M.  Janosi and J.  Kertesz, 
{\nineit Physica} {\ninebf 200A}, 179 (1992).

\bibitem{ps.torvund} F.  Torvund and J.  Fr{\o}yland, 
"Strong Ordering
by Non-Uniformity of Thresholds in a Coupled Map Lattice",
{\nineit Phys.  Scripta} (in press).

\bibitem{p.drossel.sy} B.  Drossel, "Self-Organized Criticality and
Synchronization in the Forest-Fire Model", preprint.

\bibitem{clar.noneq} S.  Clar, B.  Drossel, and F.  Schwabl, 
{\nineit Phys. Rev. Lett.} {\ninebf 75}, 2722 (1995).

\bibitem{pre50.1083} A.  Csilling, I. M.  Janosi, G.  Pasztor, and I.
Scheuring, {\nineit Phys.  Rev.} {\ninebf E50}, 1083 (1994).

    
\end{thebibliography}
\end{document}